BEng in Industrial Technologies Engineering

2024-2025

*BEng Thesis*

# Novel Numerical Methods for Accurate Space Thermal Analysis

Bernat Frangi

**Supervisor:** Antonio Soria Verdugo
*Full Professor, Universidad Carlos III de Madrid*

**Co-supervisor:** David Criado Pernía
*CEO, Radian Space S.L.*

Madrid, November 2025

*This page intentionally left blank.*

# uc3m | ESCUELA POLITÉCNICA SUPERIOR

# Novel Numerical Methods for Accurate Space Thermal Analysis

Enforcing View Factors and Modeling Diffuse Reflectivity

**Bernat Frangi**

**Supervisor:** Antonio Soria Verdugo
*Full Professor, Universidad Carlos III de Madrid*

**Co-supervisor:** David Criado Pernía
*CEO, Radian Space S.L.*

*Universidad Carlos III de Madrid*

Escuela Politécnica Superior

Department of Thermal and Fluids Engineering

BEng in Industrial Technologies Engineering

*BEng Thesis*

Madrid, November 2025

*This page intentionally left blank.*

**Novel Numerical Methods for Accurate Space Thermal Analysis**

*This page intentionally left blank.*

# Abstract

Accurate thermal analysis is crucial for modern spacecraft, driving demand for reliable modeling tools. This research advances space thermal modeling by improving the simulation accuracy and efficiency of radiative heat transfer, the dominant mode of heat exchange in space.

To this end, we incorporate diffuse reflectivity using the Gebhart method, which computes radiative exchange factors (REFs) from geometric view factors. The view factors, obtained via MCRT, require post-processing to mitigate statistical errors. Critically, existing correction schemes cannot simultaneously enforce closure and reciprocity for open systems. This research addresses this gap by proposing two novel enforcement methods: (i) a least-squares optimization with non-negativity rectification (NNR) and small positive value avoidance (SPVA), and (ii) an iterative enforcement algorithm.

To ensure consistency across different discretization levels, this work also introduces the multi-node surface model relations to formalize the connection between sub-face, face, and node representations of view factors and REFs.

A simple case study demonstrates a substantial reduction in mean absolute error (MAE): the least-squares method achieves an 81% MAE reduction, while the iterative method offers the best balance of accuracy (56% MAE reduction) and computational efficiency. A second case study shows that including diffuse reflections decreases the steady-state temperature of a plate by $4°C$, reinforcing that reflected radiation reduces net absorption.

This work introduces and validates computationally efficient methods for integrating diffuse reflectivity into space thermal analyses and for consistently coupling multi-node surface radiative models. The results enable more accurate and robust thermal predictions for spacecraft systems.

*This page intentionally left blank.*

# Acknowledgements

I would like to express my sincere gratitude to the individuals and organizations who provided invaluable support and guidance throughout the development of this thesis.

I am deeply grateful to Antonio Soria, my supervisor at the *Universidad Carlos III de Madrid* (Department of Thermal and Fluids Engineering). His expertise and insightful guidance were invaluable, greatly shaping the quality of this thesis.

I am profoundly grateful to David Criado, CEO of *Radian Space S.L.*, who served as my co-supervisor. He not only proposed the interesting problem that forms the core of this thesis but also generously provided access to the *Radian* software package, which was essential for developing and testing the methods presented herein. His helpful discussions and guidance along the way were instrumental to my progress.

I also wish to thank my father, Alejandro Frangi, for his continuous support, encouragement from the beginning, and for his valuable research insights, drawing on his experience as a high-impact researcher.

Finally, my heartfelt thanks go to my mother Silvia and my seven siblings (Nuria, Pau, Inés, Aina, Alex, Silvia and Leo) for their unwavering belief in me and their constant support.

*This page intentionally left blank.*

# Contents

**Appendices**

# List of Figures

*This page intentionally left blank.*

# List of Tables

*This page intentionally left blank.*

# Acronyms

**AHF**    Absorbed Heat Flux

**CAGR**    Compound Annual Growth Rate

**DHF**    Direct Heat Flux

**ESA**    European Space Agency

**FE**    Finite Element

**GR**    Radiative Coupling Coefficient
**GTE**    Global Thermodynamic Equilibrium

**IR**    Infrared

**LPN**    Lumped Parameter Network
**LTE**    Local Thermodynamic Equilibrium

**MAE**    Mean Absolute Error
**MCRT**    Monte Carlo Ray Tracing

**NNR**    Non-negativity Rectification

**REF**    Radiative Exchange Factor
**RHS**    Right-hand Side

**SPVA**    Small Positive Value Avoidance

**TB**    Thermal Balance
**TE**    Thermodynamic Equilibrium
**TVAC**    Thermal Vacuum

*This page intentionally left blank.*

# 1
# Introduction

## 1.1 Market and regulatory context

The small satellite industry in Europe is rapidly growing, driven by miniaturization, falling costs, and increasing demand for Earth observation, IoT connectivity, and communications. The European small satellite market is projected to grow from approximately USD 1.23 billion in 2024 to over USD 3 billion by 2033—a robust Compound Annual Growth Rate (CAGR) of around 10–11% [1, 2]. Leading European players include *Airbus SE*[1] (Netherlands, France), *Thales Alenia Space*[2] (France), and *OHB SE*[3] (Germany), recognized for their extensive small-satellite offerings [3]. Other prominent contributors in Europe comprise *Spire Global, Inc.*[4] (Glasgow, Luxembourg, Munich), *ICEYE Ltd.*[5] (Finland), *Aerospacelab NV/SA*[6] (Belgium), *NanoAvionics Corp*[7] (Lithuania), *ISISPACE*[8] (Netherlands), *Alén Space*[9] (Spain), and *FOSSA Systems*[10] (Spain)—all specializing in satellite buses, manufacturing, and IoT solutions. The European Space Agency (ESA) plays a strategic role by funding programs, fostering development, and enabling innovation across the industry [2, 4–6].

Global forums for small satellite innovation, such as the *SmallSat Conference*[11], help in bringing together industry, academia, and government to share advances and foster collaboration. They aim to accelerate the transition of new technologies into practical applications, from communications to Earth observation. By driving knowledge exchange and networking, they have a major impact on the small satellite market—shaping standards, opening commercial opportunities, and fueling the sector's rapid growth and democratization of space access.

---

[1] https://airbus.com/
[2] https://thalesaleniaspace.com/
[3] https://ohb.de/
[4] https://spire.com/
[5] https://iceye.com/
[6] https://aerospacelab.com/
[7] https://nanoavionics.com/
[8] https://isispace.nl/
[9] https://alen.space/
[10] https://fossa.systems/
[11] https://smallsat.org/

As the small satellite sector evolves, new regulations are put in place to ensure mission safety, reliability, and sustainability in increasingly crowded orbital environments. Some of these specifically affect the thermal subsystem, such as the *ECSS-E-ST-31 thermal control standards* [7] and the *Thermal Engineering* section of the *ESA IOD CubeSat Guidelines*, which requires a 15 K modeling uncertainty and strict reporting of thermal analyses, as well as mandatory Thermal Vacuum (TVAC) and Thermal Balance (TB) testing in extreme temperature ranges defined based on thermal analysis [8]. To comply with such requirements, developers must perform rigorous simulations that predict spacecraft temperature behavior under diverse orbital and environmental conditions. This drives the need for advanced thermal analysis software capable of accurately modeling heat transfer, subsystem interactions, and material properties to validate compliance before launch.

## 1.2 Space thermal analysis software

Several specialized tools are available to meet these demands in space thermal analysis. Among the most widely used are *ESATAN-TMS*[12], developed under ESA oversight, and the *Thermica*[13] suite, both well-established in European missions for their advanced radiation and orbital heat-transfer capabilities. In the United States, *Thermal Desktop*[14] integrated with *SINDA/FLUINT* is the industry standard, offering powerful finite-difference modeling and multi-physics coupling. Recently, *Radian*[15] has emerged as the first truly cloud-based thermal analysis platform, developed by *Radian Space SL*, enabling engineers to perform end-to-end simulations directly from a web browser, import CAD models, and leverage scalable computing resources without local hardware constraints [9]. *Radian* also includes a reusable databank of materials, orbits, attitudes, and components, and has been successfully used in over 60 satellite missions with strong correlation between predicted and telemetry-reported thermal performance [10]. These software packages form the backbone of thermal subsystem design and verification, supporting compliance with evolving regulatory frameworks and enabling reliable mission performance.

---

[12] https://www.esatan-tms.com/
[13] https://airbus.com/products-services/space/space-customer-support/systema/thermica-suite
[14] https://www.ansys.com/products/fluids/ansys-thermal-desktop
[15] https://radian.systems/

## 1.3 Problem statement

The increasing complexity of satellite missions and the growing demand for high-performance thermal control systems necessitate ever more advanced analysis capabilities. Among these, the accurate modeling of radiative heat transfer is particularly critical, as radiation—along with conduction—constitutes a dominant mode of heat exchange in space. Capturing effects such as diffuse reflectivity is essential to correctly represent surface interactions and achieve reliable thermal predictions. Equally important is the consistent enforcement of reciprocity and closure conditions on view factors, ensuring that radiative exchange relationships remain physically meaningful. In this work, the focus is placed on incorporating the modeling of diffuse reflectivity and implementing enforcers in the radiative heat transfer calculations within *Radian*, thereby enhancing the accuracy and reliability of satellite thermal simulations.

## 1.4 Objectives and scope

Based on the problem statement, the main objectives of this work are:

1. To enhance the modeling of surface interactions in *Radian* by incorporating diffuse reflectivity into the radiative heat transfer calculations.
2. To ensure the physical consistency of view factors and REFs by implementing methods that enforce closure and reciprocity conditions in radiative exchange modeling.

These objectives align with the overarching aim of this work: to increase the accuracy and reliability of thermal simulations for space applications, in response to the growing complexity of satellite missions and the regulatory requirements for rigorous verification. From these main objectives, the following secondary objectives are derived:

1. To develop and implement original methods for simultaneously enforcing closure and reciprocity for view factors in open systems[16].
2. To formally extend Gebhart's formulation and matrix method for computing REFs to open systems.
3. To develop a framework for efficiently handling and switching between different levels of subdivision in a multi-node surface model[17].
4. To ensure the preservation of closure and reciprocity relations when applying these transformations across different levels of the multi-node model.

---

[16] Currently, many of these methods are only properly developed for enclosed systems and are not fit for direct application in open-space environments.

[17] This is a specific requirement for the implementation of the proposed methods in *Radian*, which uses a multi-node surface representation for improved efficiency.

5. To develop applied examples that demonstrate the practical relevance and effectiveness of the proposed methods.

These objectives entail a comprehensive enhancement of the radiative heat transfer modeling capabilities in *Radian*, covering both theoretical foundations and practical implementations. They also involve a substantial amount of original development not found in the existing literature, with these novel contributions highlighted throughout the text. Together, they aim to advance the accuracy and reliability of thermal simulations for satellite applications.

# 2
# Notions of Heat Transfer and Radiation

## 2.1 What is heat transfer?

While thermodynamics focuses on the initial and final states of a thermal system and examines how systems exchange energy with their surroundings through work and heat, it does not describe the mechanisms by which these interactions occur. The study of these mechanisms is the domain of **heat transfer**. In heat transfer, we analyze the different *modes* of heat transmission, which represent the fundamental processes through which heat is exchanged in nature.

What does heat transfer mean in a physical sense? In Bergman et al. [11], heat transfer is simply defined as "thermal energy in transit due to a spatial temperature difference." When a temperature difference exists within a system, a temperature gradient is established, driving the flow of heat from the hotter region to the colder one.

In the following sections, we will summarize the three fundamental modes of heat transfer: conduction, convection and radiation. We will briefly study the mechanisms for each of these modes and define their corresponding rate equations[1]. Further attention will be paid later to radiation heat transfer, which is the main focus of this work.

## 2.2 Conduction

Conduction is a mode of heat transfer that occurs due to exchange of energy between particles at an atomic or molecular level. Energy is transferred from more energetic

---

[1] Rate equations quantify each of these transfer modes and allow us to compute the amount of heat that is transferred per unit time.

molecules to less energetic ones.

In the case of gases, we associate the temperature at a particular region of the gas with the energy of the particles in that region. This energy is due to the random motion of the particles in the gas, as well as their internal rotational and vibrational motions. If the gas has no bulk motion, the main mechanism of heat transfer is conduction[2], which takes place due to the collisions between molecules as they move, causing energy and momentum to transfer between them [11].

Conduction in liquids happens similarly to gases, but the particles are more tightly packed. In solids, the particles are even closer together, and the mechanism of conduction can happen in two ways: for non-conductors of electricity, heat is conducted by lattice waves caused by atomic motion; for good conductors of electricity, the principal contribution is due to the motion of free electrons [12].

Note that conduction is a mode of heat transfer that requires the presence of a material medium in order to occur, be it a solid, liquid or gas. Thus, if two systems are separated by a vacuum, heat transfer by conduction between those systems is not possible.

### 2.2.1 Conduction rate equation

The rate equation for conduction is known as **Fourier's Law**, which was first published by Joseph Fourier, in his 1822 book titled *Théorie analytique de la chaleur* [13]. Its general form, considering an isotropic medium with constant thermal conductivity, $k$, is given by **Definition 1**.

> **Definition 1: Fourier's Law**
>
> The rate $q''_{\text{cond}}$ at which heat is conducted through an isotropic medium with constant conductivity $k$, per unit time and per unit surface, depends on the spatial distribution of the temperature $T$ and is given by:
>
> $$q''_{\text{cond}} = -k \nabla T$$

In Cartesian coordinates, Fourier's law can be written as:

$$q''_{\text{cond}} = -k \left( \boldsymbol{i} \frac{\partial T}{\partial x} + \boldsymbol{j} \frac{\partial T}{\partial y} + \boldsymbol{k} \frac{\partial T}{\partial z} \right) \quad (2.1)$$

Where $\boldsymbol{i}$, $\boldsymbol{j}$ and $\boldsymbol{k}$ are the unit vectors in the $x$, $y$ and $z$ directions, respectively.

The minus sign in Fourier's law captures the fact that heat is transferred from the hotter regions to the colder ones.

---

[2] We forget about radiation for now.

## 2.2.2 One-dimensional conduction and thermal resistance

In the particular case of one-dimensional conduction[3], Fourier's law is simplified. Taking, the $x$ direction as the direction of transfer, **Equation 2.1** becomes:

$$q''_{\text{cond}} = -k\frac{dT}{dx} \qquad (2.2)$$

Under *steady state* conditions, a plane wall has a linear temperature distribution in its interior, meaning that the temperature $T$ varies linearly with the position $x$ along thickness of the wall from $T_1$ to $T_2$. In this case, the heat transfer rate from 1 to 2 by conduction can be expressed as:

$$q''_{\text{cond}} = -k\frac{T_2 - T_1}{L} \qquad (2.3)$$

where $L$ is the thickness of the wall [11].

This expression, which appears many times in conduction scenarios for space applications, gives rise to the concept of *thermal resistance*, $R_{\text{th}}$, which is defined as the ratio between the temperature difference and the heat transfer rate per unit time $q_{\text{cond}} = q''_{\text{cond}}A$:

$$R_{\text{th}} = \frac{T_1 - T_2}{q''_{\text{cond}}A} = \frac{\Delta T}{q_{\text{cond}}} = \frac{L}{kA} \qquad (2.4)$$

Knowing the temperature difference $\Delta T$ and computing the thermal resistance $R_{\text{th}}$, the conduction heat transfer rate can be easily computed as:

$$q_{\text{cond}} = \frac{\Delta T}{R_{\text{th}}} \qquad (2.5)$$

The advantage of this approach is that it allows us to work in analogy with electrical circuits, where the thermal resistance plays a role similar to that of an electrical resistance, the temperature difference is analogous to the voltage difference, and the heat transfer rate is analogous to the electrical current. This analogy is particularly useful for combining multiple thermal resistances in series or parallel [11].

In full analogy with electrical circuits, series thermal resistances are added, while parallel thermal resistances are combined using the reciprocal formula:

$$\frac{1}{R_{\text{th,eq}}} = \sum_{i=1}^{N} \frac{1}{R_{\text{th},i}} \qquad (2.6)$$

---

[3] Meaning that heat is transferred in a single Cartesian direction

## 2.3 Convection

Convection is a mode of heat transfer that occurs due to the combined effect of conduction with bulk fluid motion, and it is realized via mobile fluid particles, which are "portions" of a fluid that move together due to advection [14]. The movement of the fluid particles implies the transfer of heat. Convection processes are usually classified into two types: natural convection and forced convection.

In natural convection, the fluid motion is due to buoyancy forces. When the temperature of a fluid particle increases with respect to that of its surrounding fluid, its density decreases. As a consequence, buoyancy forces appear that cause the fluid particle to rise, leaving behind a void that is filled with colder fluid particles.

In forced convection, the fluid is forced to move by an external source, such as a pump or a fan. Mixed convection is also possible, where both natural and forced convection occur at the same time.

Convection is closely linked to conduction, as conduction is the primary mechanism by which fluid particles receive heat initially. Typically, the fluid flows along a solid surface or an interface with another fluid, where the relative velocity of the fluid with respect to the solid or the other fluid is zero. At this interface, conduction occurs first, heating the fluid directly adjacent to the surface or interface. This heated fluid then rises, initiating convection [14].

Note that, as convection requires the presence of a fluid, it cannot occur in a vacuum. Since space has a negligible atmosphere, convection heat transfer in space applications is generally not important and completely neglected, unless for fluid tanks and reentry phenomena.

### 2.3.1 Convection rate equation

The rate equation for convection is given by **Newton's Law of Cooling**, first published anonymously in 1701 by Isaac Newton in "Scala Graduum Caloris. Calorum Descriptiones & Figna" [15]. We present the law in **Definition 2**

> **Definition 2: Newton's Law of Cooling**
>
> The rate $q''_{conv}$ at which heat is transferred by convection between a fluid and a solid surface (or liquid interface) per unit time and per unit surface is given by:
>
> $$q''_{conv} = h\,(T_s - T_\infty)$$

where $T_\infty$ is the temperature of the fluid at a distance from the solid surface (or liquid interface) where its effect is negligible, $T_s$ is the temperature of the solid surface (or liquid interface), and $h$ is the convective heat transfer coefficient.

The *convective heat transfer coefficient h* is a measure of the ability of a fluid to transfer heat to a solid surface or liquid interface. It can depend on many things such as the fluid properties, the flow conditions, and the geometry of the solid surface or liquid interface. Discussing the calculation of $h$ is beyond the scope of this work.

## 2.4 Radiation

Radiation is the emission of electromagnetic radiation by all matter that is at nonzero absolute temperature, and it is due to a combination of electronic, molecular and lattice oscillations of the emitting material [16].

Radiation differs from the other two modes of heat transfer in several ways. Unlike conduction and convection, radiation does not require a material medium to propagate, as it consists of electromagnetic waves. Additionally, radiation occurs independently of the surroundings of the emitting body. Any object with a temperature above absolute zero emits radiation, regardless of whether nearby objects are hotter or colder. If these objects intercept the emitted radiation, they may absorb it, depending on their own properties and the properties of the radiation, but not directly on their temperature. Only the net radiative heat transfer depends on the temperature difference between the emitting body and its surroundings, as the surroundings also emit radiation in return [17].

### 2.4.1 Radiation heat fluxes

**Surface emissive power**

Consider a surface emitting thermal radiation. The rate at which radiation is emitted by the surface is known as its *surface emissive power*, and it is governed by the **Stefan-Boltzmann Law**. This law is named after Josef Stefan, who first discovered it empirically in 1879 based on measurements by John Tyndall [18, 19], and after Ludwig Boltzmann, who derived it theoretically in 1884 [20]. The law is stated in **Definition 3** for the special case of a black body.

> **Definition 3: Black Body Surface Emissive Power (Stefan-Boltzmann Law)**
>
> The rate $E_b$ at which heat is radiated by a black body per unit time and per unit surface is given by:
> $$E_b = \sigma T^4$$
> where $\sigma$ is the *Stefan-Boltzmann constant*, equal to $5.67 \times 10^{-8}\,\mathrm{W\,m^{-2}\,K^{-4}}$, and $T$ is the absolute temperature of the black body.

The black body will be discussed in more detail further on, but for now let it suffice to say that it is a theoretical body capable of absorbing and emitting the maximum possible amount of radiation for a given temperature. Therefore, the expression in **Definition 3** gives the theoretical maximum emissive power that a surface can emit.

However, real surfaces do not emit radiation as efficiently as black bodies. To account for this, we introduce the *emissivity* of a surface, $\varepsilon$, which takes values between $0$ and $1$ and gives a measure of how well a surface emits radiation compared to a black body. We can then define in **Definition 4** the surface emissive power for a real surface [11].

> **Definition 4: Real Body Surface Emissive Power**
>
> The rate $E$ at which heat is radiated by a real body per unit time and per unit surface is given by:
> $$E = \varepsilon E_b = \varepsilon \sigma T^4 \qquad (2.7)$$
> where $\varepsilon$ is the emissivity of the surface of the real body, $\sigma$ is the *Stefan-Boltzmann constant*, equal to $5.67 \times 10^{-8}\,\mathrm{W\,m^{-2}\,K^{-4}}$, and $T$ is the absolute temperature of the real body.

Setting $\varepsilon = 1$, one recovers the surface emissive power for a black body. Emissivity and real surfaces will be discussed in more detail later. For now, we will simply remark that emissivity generally depends on the material properties of the surface and on the wavelength and angle of incidence of the radiation.

**Irradiation**

Radiation can also be emitted by surrounding objects or the environment and directed *onto* the surface of the body under study. The rate at which this radiation is incident on a unit area of the surface is known as *irradiation*, as shown in **Definition 5** [11].

> **Definition 5: Irradiation**
>
> The rate $G$ at which heat (energy) is incident onto a surface, per unit time and per unit area, is called *irradiation*.

Some or all of the irradiation $G$ may be absorbed by the body that is receiving the radiation. The parameter that quantifies how much of the incident radiation is absorbed is known as the *absorptivity*, $\alpha$. It is equal to the ratio between the absorbed irradiation and the total incident irradiation, and takes values between $0$ and $1$. We can then define the absorbed irradiation in **Definition 6**.

> **Definition 6: Absorbed Irradiation**
>
> The absorbed irradiation $G_{\text{abs}}$ is the portion of the irradiation $G$ incident on a surface that is finally absorbed by the surface. It is given by:
>
> $$G_{\text{abs}} = \alpha G \tag{2.8}$$
>
> where $\alpha$ is the absorptivity of the surface.

The portion of irradiation that is not absorbed by the surface is either reflected or transmitted, meaning it does not contribute to the surface's heating. These processes are quantified by parameters analogous to absorptivity: the *reflectivity* $\rho$ and the *transmissivity* $\tau$. Like absorptivity, these parameters range between $0$ and $1$.

Since energy is conserved, the sum of the absorbed, reflected and transmitted irradiation must be equal to the total incident irradiation. Then, we have the relationship

$$\alpha + \rho + \tau = 1 \tag{2.9}$$

Just like emissivity, all these properties will be discussed later in more detail. However, we will remark that they all generally depend on the material properties of the surface and on the wavelength and angle of incidence of the irradiation. For a black body, $\alpha = 1$, and $\rho = \tau = 0$.

**Radiosity**

Another common radiative heat flux is the radiosity, presented in **Definition 7** [11].

> **Definition 7: Radiosity**
>
> The radiosity $J$ is defined as the total radiation energy leaving a surface per unit area and per unit time. It is given by:
>
> $$J = E + \rho G \tag{2.10}$$
>
> where $E$ is the self-emitted surface emissive power, $\rho$ is the reflectivity of the surface and $G$ is the irradiation incident on the surface.

### 2.4.2 Radiation rate equation

The rate of heat transfer by radiation between two bodies (one of them may represent the environment) is simply given by the net radiative power transferred between them. For example, focusing on the rate of heat transfer by radiation from a body to the environment, we obtain **Definition 8** [11]:

> **Definition 8: Radiation Rate Equation (Real Body and Environment)**
>
> The rate $q''_{\text{rad}}$ at which heat is transferred by radiation from a real surface at temperature $T$ to the environment, per unit time and per unit surface, is given by:
>
> $$q''_{\text{rad}} = \varepsilon \sigma T^4 - \alpha \sigma T_\infty^4 \tag{2.11}$$
>
> where $\varepsilon$ is the emissivity of the surface, $\alpha$ is the absorptivity of the surface, $\sigma$ is the *Stefan-Boltzmann constant*, equal to $5.67 \times 10^{-8}\,\text{W}\,\text{m}^{-2}\,\text{K}^{-4}$, and the environment is considered to be a black body at temperature $T_\infty$ (it has unit emissivity).

When we consider the radiation heat transfer between a single body and the environment, we assume that the entirety of the radiation emitted by the environment is incident onto the body, and vice versa. This is a reasonable assumption, since the environment totally surrounds the body, and it is also valid when we are studying a system with only two bodies if one of them forms an "enclosure" around the other. However, when more bodies come into play, this is no longer the case, and we need to take into account the fractions of the total radiation emitted by each surface that reach each of the other surfaces. Then, the expression in **Definition 8** gets slightly more complicated.

The discussion of radiative heat transfer between multiple surfaces is left for **Section 2.4.12**, when we will talk about view factors.

### 2.4.3 Directional nature of radiation

One key property of surfaces is how they distribute emitted radiation across the various possible directions. A *diffuse* surface, such as a black body, emits radiation uniformly in all directions, while most real surfaces may exhibit preferential emission in specific directions [17]. This behavior is influenced by factors such as surface roughness [21].

Since radiation can also be incident from multiple directions, the directionality of emission is closely related to the surface's ability to absorb radiation from different angles.

These directional effects, illustrated in **Figure 2.1**, play a crucial role in radiative heat transfer and are addressed through the concept of *radiance*.

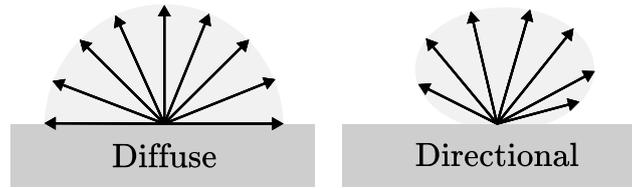

**Figure 2.1:** *Angular distribution of emitted power from an ideal diffuse surface (left), such as a black body, and from a real, mildly directional surface (right).*

### 2.4.4 Emitted radiance and surface emissive power

The quantity that characterizes the amount of radiation emitted by a surface in a particular direction is known as the *emissive radiation intensity*, or simply the *emitted radiance*, denoted $I_e$ [17]. The simplest and most common way of specifying the direction is by means of spherical coordinates, as shown in **Figure 2.2**.

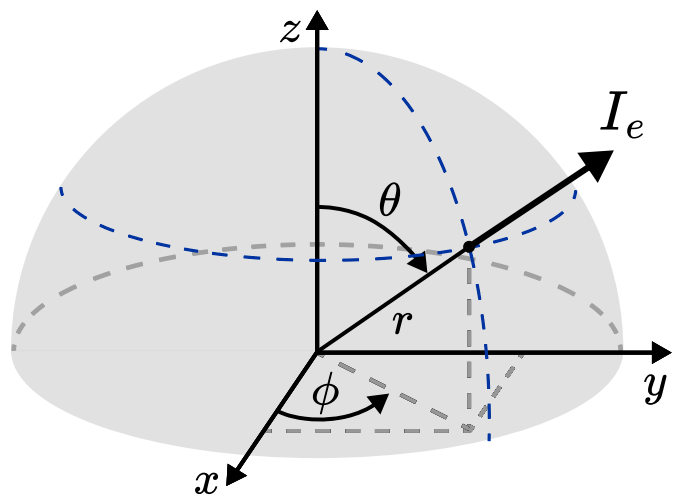

**Figure 2.2:** *Direction of emitted radiation defined in spherical coordinates $\theta$ (zenith angle) and $\phi$ (azimuth angle).*

Another important quantity in radiation heat transfer is the *surface emissive power*, denoted $E$, which is the total rate at which radiation is emitted by a surface per unit area and per unit time (recall **Section 2.4.1**).

Before properly defining emitted radiance and surface emissive power, let us introduce the *solid angle*, which will be a useful tool in what follows.

**Solid angle**

The *flat angle $\theta$* is commonly measured in *radians* (rad). As illustrated in **Figure 2.3a**, one radian is defined as the angle subtended by a circular sector whose arc length equals the radius of the circle [22]. The differential of the flat angle can then be defined as $d\theta = \frac{ds}{r}$, where $ds$ is the differential arc length and $r$ is the radius of the circle. We can compute the total flat angle $\Theta$ in the circle by integrating this differential over the circle's perimeter:

$$\Theta = \int_{\text{circle}} d\theta = \int_0^{2\pi r} \frac{ds}{r} = 2\pi \text{ rad} \tag{2.12}$$

In two dimensions, the flat angle precisely quantifies the proportion of the circle that the sector occupies. For example, a semicircular sector subtends an angle of $\pi$ radians, consistently representing half ($\pi/\Theta = 1/2$) of the complete circle, independently of the circle's radius.

The *solid angle* $\omega$ represents this same concept, but for three dimensions, as illustrated in **Figure 2.3b**. It characterizes the fraction of the sphere that a surface element on the sphere represents. The most commonly used unit for the solid angle is the *steradian* (sr). One steradian is defined as the solid angle subtended by a surface (of any shape) on a sphere that has an area equal to the square of the sphere radius [23].

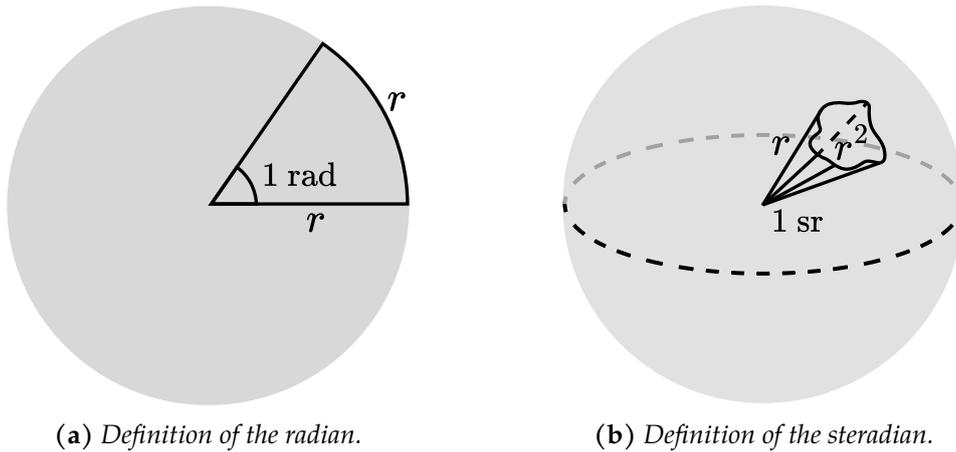

(**a**) *Definition of the radian.* (**b**) *Definition of the steradian.*
**Figure 2.3:** *The flat and the solid angle.*

As we did for the flat angle, we can define the differential of the solid angle as $d\omega = \frac{dS}{r^2}$, where $dS$ is the differential spherical surface and $r$ is the radius of the sphere. In spherical coordinates, this is equal to $d\omega = \sin\theta d\theta d\phi$. The total solid angle in the sphere, denoted $\Omega$, can be computed using spherical coordinates as:

$$\Omega = \int_{\text{sphere}} d\omega = \int_0^{2\pi} d\phi \int_0^{\pi} d\theta \, \sin\theta = 4\pi \text{ sr} \tag{2.13}$$

With this useful tool in hand, we are now ready to define *emitted radiance* and *surface emissive power*.

**Emitted Radiance**

Consider the surface elements $dA$ and $dS$ in **Figure 2.4**. We want to determine the rate at which radiation emitted by $dA$ traverses $dS$. This quantity is what we call the *emitted radiance* or the *emissive radiation intensity*, denoted $I_e$ (the $e$ for "emitted" is included to avoid confusion with photometric quantities) and defined in **Definition 9**.

> **Definition 9: Emitted Radiance**
>
> The emitted radiance[a] $I_e(\theta, \phi)$ is the rate $dQ$ at which radiant energy is emitted per unit time in the $(\theta, \phi)$ direction, per unit area of the emitting surface $dA$ normal to this direction, per unit solid angle $d\omega$ about this direction. That is:
>
> $$I_e(\theta, \phi) = \frac{dQ}{dA \cos\theta \cdot d\omega} \tag{2.14}$$
>
> [a] Sometimes also known as *emissive radiation intensity*.

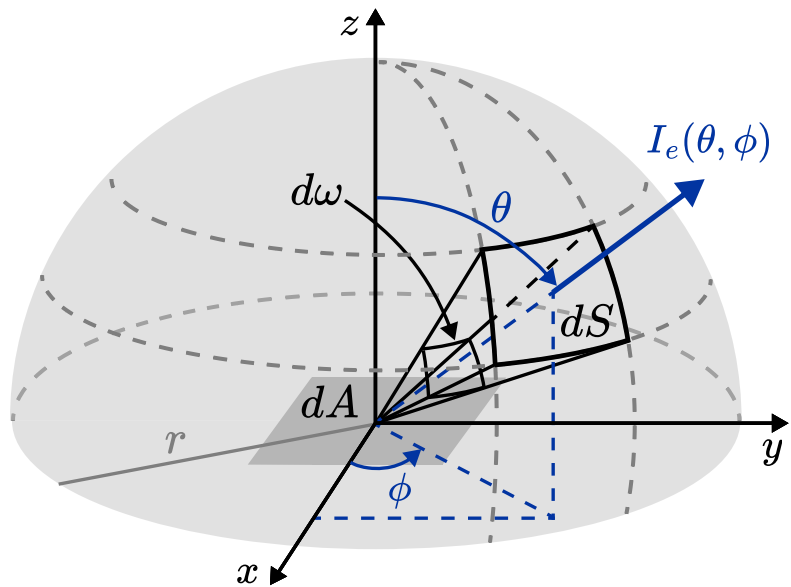

**Figure 2.4:** *Emitted radiance in the direction defined by the zenith angle $\theta$ and the azimuth angle $\phi$ for a surface element $dA$.*

Note that, in this definition, we consider only the component of the emitting area $dA$ that is normal to the direction of radiant emission defined by $(\theta, \phi)$. We "project" the area $dA$ as shown in **Figure 2.5**. The projected area is equal to $dA \cos\theta$, which is the "effective" area that an observer situated at $dS$ would see.

**Spectral emitted radiance**

**Definition 9** gives the total emitted radiance $I_e(\theta, \phi)$ in a particular direction $(\theta, \phi)$. However, in many cases, we are interested in the emitted radiance at a specific wave-length $\lambda$. This is captured by the *spectral emitted radiance*, denoted $I_{e,\lambda}$ and defined in **Definition 10**.

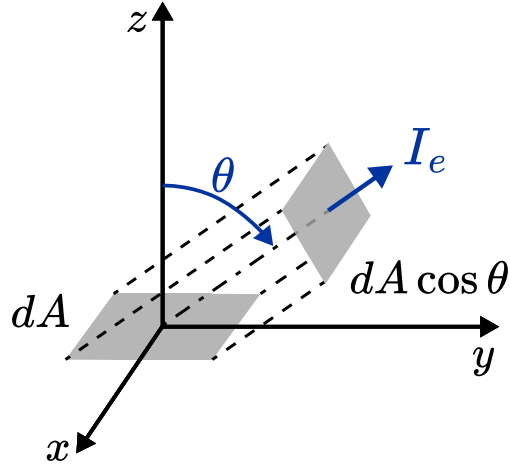

**Figure 2.5:** *Normal component of dA with respect to the direction of radiant emission.*

> **Definition 10: Spectral Emitted Radiance**
>
> The spectral emitted radiance[a] $I_{e,\lambda}(\theta, \phi, \lambda)$ is the rate $dQ$ at which radiant energy is emitted per unit time in the $(\theta, \phi)$ direction and at the wavelength $\lambda$, per unit area of the emitting surface $dA$ normal to this direction, per unit solid angle $d\omega$ about this direction, and per unit wavelength interval $d\lambda$ around $\lambda$. That is:
>
> $$I_{e,\lambda}(\theta, \phi, \lambda) = \frac{dQ}{dA \cos\theta \cdot d\omega \cdot d\lambda} \qquad (2.15)$$
>
> ---
> [a] Sometimes also known as *spectral emissive radiation intensity*.

The spectral emitted radiance is related to the total emitted radiance through:

$$I_e(\theta, \phi) = \int_0^\infty I_{e,\lambda}(\theta, \phi, \lambda) d\lambda \qquad (2.16)$$

**Hemispherical surface emissive power**

From **Definition 9**, we can write the differential form of the *surface emissive power* (recall **Section 2.4.1**) in the direction $(\theta, \phi)$ as:

$$dE = \frac{dQ}{dA} = I_e(\theta, \phi) \cos\theta \, d\omega = I_e(\theta, \phi) \cos\theta \sin\theta \, d\theta \, d\phi \qquad (2.17)$$

Then, for a flat emitting surface, the total power per unit area emitted by the surface in all directions is given by the integral of $dE$ over the hemisphere, as stated in **Definition 11**.

### Definition 11: Hemispherical Surface Emissive Power

The total hemispherical surface emissive power $E$ is the radiation energy emitted by a flat surface per unit area and per unit time. It is given by:

$$E = \int_0^{2\pi} d\phi \int_0^{\pi/2} d\theta \, I_e(\theta, \phi) \cos\theta \sin\theta \tag{2.18}$$

In reality, the emitted radiance $I_e(\theta, \phi)$ is directional. That is, it depends on the direction of emission $(\theta, \phi)$. However, in many cases we approximate real surfaces as being diffuse [17]. In this case $I_e(\theta, \phi) \equiv I_e$ is constant, and the hemispherical surface emissive power $E$ is simply:

$$E = I_e \int_0^{2\pi} d\phi \int_0^{\pi/2} d\theta \, \cos\theta \sin\theta = \pi I_e \tag{2.19}$$

**Spectral hemispherical surface emissive power**

Again, in some cases we are interested in the surface emissive power at a specific wavelength $\lambda$. In this case, we can define the *spectral surface emissive power* in the direction $(\theta, \phi)$, using **Definition** 10. In differential form, we have:

$$dE_\lambda = \frac{dQ}{dA \, d\lambda} = I_{e,\lambda}(\theta, \phi, \lambda) \cos\theta \, d\omega = I_{e,\lambda}(\theta, \phi, \lambda) \cos\theta \sin\theta \, d\theta \, d\phi \tag{2.20}$$

Then, for a flat emitting surface, the total power per unit area emitted by the surface in all directions at wavelength $\lambda$, per unit wavelength interval $d\lambda$ around $\lambda$, is given by the integral of $dE_\lambda$ over the hemisphere, as stated in **Definition** 12.

### Definition 12: Spectral Hemispherical Surface Emissive Power

The total spectral hemispherical surface emissive power $E_\lambda(\lambda)$ is the total radiation energy emitted by a flat surface per unit area and per unit time at a specific wavelength $\lambda$, per unit wavelength interval $d\lambda$ around $\lambda$. It is given by:

$$E_\lambda(\lambda) = \int_0^{2\pi} d\phi \int_0^{\pi/2} d\theta \, I_{e,\lambda}(\theta, \phi, \lambda) \cos\theta \sin\theta \tag{2.21}$$

Again, for diffuse surfaces, we have $I_{e,\lambda}(\theta, \phi, \lambda) \equiv I_{e,\lambda}(\lambda)$, and the spectral hemispherical surface emissive power $E_\lambda(\lambda)$ is simply:

$$E_\lambda(\lambda) = I_{e,\lambda}(\lambda) \int_0^{2\pi} d\phi \int_0^{\pi/2} d\theta \, \cos\theta \sin\theta = \pi I_{e,\lambda}(\lambda) \tag{2.22}$$

The spectral hemispherical surface emissive power is related to the total hemispherical surface emissive power through:

$$E = \int_0^\infty E_\lambda(\lambda) d\lambda \qquad (2.23)$$

### 2.4.5 Incident radiance and irradiation

**Incident radiance**

Everything discussed in **Section 2.4.4** for emissive radiance and surface emissive power can be readily applied to incident radiance and irradiation. Instead of talking about the emissive radiance $I_e$, we talk about the *incident radiance*, denoted $I_i$. The difference between $I_e$ and $I_i$ is that $I_e$ depends directly on the properties of the surface, since it concerns *emission* which comes from the surface; while $I_i$ can depend on many factors, since in general it comes from an unknown source of radiation completely unrelated to the intercepting surface.

Incident radiance is defined in **Definition 13**.

> **Definition 13: Incident Radiance**
>
> The incident radiance[a] $I_i(\theta, \phi)$ is the rate $dQ$ at which radiant energy is incident on a surface element $dA$ per unit time from the $(\theta, \phi)$ direction, per unit area of the intercepting surface normal to this direction, per unit solid angle $d\omega$ about this direction. That is:
> $$I_i(\theta, \phi) = \frac{dQ}{dA \cos\theta \cdot d\omega} \qquad (2.24)$$
>
> [a] Sometimes also known as *incident radiation intensity*.

Again, we consider only the component of the intercepting area $dA$ that is normal to the direction of incident radiance defined by $(\theta, \phi)$.

**Spectral incident radiance**

Just like we did for emissive radiance, we can define the *spectral incident radiance*, denoted $I_{i,\lambda}$, which is defined in **Definition 14**.

> **Definition 14: Spectral Incident Radiance**
>
> The spectral incident radiance[a] $I_{i,\lambda}(\theta, \phi, \lambda)$ is the rate $dQ$ at which radiant energy is incident on a surface element $dA$ per unit time from the $(\theta, \phi)$ direction and at the wavelength $\lambda$, per unit area of the intercepting surface $dA$ normal to this direction, per unit solid angle $d\omega$ about this direction, and per unit wavelength interval $d\lambda$ around $\lambda$. That is:

$$I_{i,\lambda}(\theta, \phi, \lambda) = \frac{dQ}{dA \cos \theta \cdot d\omega \cdot d\lambda} \tag{2.25}$$

[a] Sometimes also known as *spectral incident radiation intensity*.

The spectral incident radiance is related to the total incident radiance through:

$$I_i(\theta, \phi) = \int_0^\infty I_{i,\lambda}(\theta, \phi, \lambda) d\lambda \tag{2.26}$$

**Hemispherical irradiation**

From **Definition 13**, we can write the differential form of the *irradiation* (recall **Section 2.4.1**) from the direction $(\theta, \phi)$ as:

$$dG = \frac{dQ}{dA} = I_i(\theta, \phi) \cos \theta \, d\omega = I_i(\theta, \phi) \cos \theta \sin \theta \, d\theta \, d\phi \tag{2.27}$$

Then the total irradiation incident on a flat surface from all directions is given by the integral of $dG$ over the hemisphere, as stated in **Definition 15**.

**Definition 15: Hemispherical Irradiation**

The total hemispherical irradiation $G$ is the radiation energy incident on a flat surface per unit area and per unit time. It is given by:

$$G = \int_0^{2\pi} d\phi \int_0^{\pi/2} d\theta \, I_i(\theta, \phi) \cos \theta \sin \theta \tag{2.28}$$

If the irradiation is diffuse, we have $I_i(\theta, \phi) \equiv I_i$, and the hemispherical irradiation $G$ is simply:

$$G = I_i \int_0^{2\pi} d\phi \int_0^{\pi/2} d\theta \, \cos \theta \sin \theta = \pi I_i \tag{2.29}$$

**Spectral hemispherical irradiation**

We can also define the *spectral irradiation*, which is the irradiation at a specific wavelength $\lambda$. Using **Definition 14**, we can write its differential form as:

$$dG_\lambda = \frac{dQ}{dA \, d\lambda} = I_{i,\lambda}(\theta, \phi, \lambda) \cos \theta \, d\omega = I_{i,\lambda}(\theta, \phi, \lambda) \cos \theta \sin \theta \, d\theta \, d\phi \tag{2.30}$$

Then, the total irradiation incident on a flat surface from all directions at wavelength $\lambda$, per unit wavelength interval $d\lambda$ around $\lambda$, is given by the integral of $dG_\lambda$ over the hemisphere, as stated in **Definition 16**.

> **Definition 16: Spectral Hemispherical Irradiation**
>
> The total spectral hemispherical irradiation $G_\lambda(\lambda)$ is the total radiation energy incident on a flat surface per unit area and per unit time at a specific wavelength $\lambda$, per unit wavelength interval $d\lambda$ around $\lambda$. It is given by:
>
> $$G_\lambda(\lambda) = \int_0^{2\pi} d\phi \int_0^{\pi/2} d\theta \, I_{i,\lambda}(\theta,\phi,\lambda) \cos\theta \sin\theta \tag{2.31}$$

For diffuse irradiation, we have $I_{i,\lambda}(\theta,\phi,\lambda) \equiv I_{i,\lambda}(\lambda)$, and the spectral hemispherical irradiation $G_\lambda(\lambda)$ is simply:

$$G_\lambda(\lambda) = I_{i,\lambda}(\lambda) \int_0^{2\pi} d\phi \int_0^{\pi/2} d\theta \, \cos\theta \sin\theta = \pi I_{i,\lambda}(\lambda) \tag{2.32}$$

The spectral hemispherical irradiation is related to the total hemispherical irradiation through:

$$G = \int_0^\infty G_\lambda(\lambda) d\lambda \tag{2.33}$$

### 2.4.6  Exiting radiance and radiosity

In **Definition 7**, we defined *radiosity* as the total radiation leaving a surface per unit area and per unit time, which includes both self-emitted and reflected radiation. Following the same procedure as for emitted radiance in, we can also define *reflected radiance*, denoted $I_r$, and the *spectral reflected radiance*, denoted $I_{r,\lambda}$. The sum of the reflected radiance $I_r$ and the emitted radiance $I_e$ gives the total *exiting radiance* leaving the surface, which is commonly denoted $I_{r+e}$; or $I_{r+e,\lambda}$ when we talk about the total *spectral exiting radiance* leaving the surface.

$$I_{r+e}(\theta,\phi) = I_e(\theta,\phi) + I_r(\theta,\phi) \tag{2.34}$$

$$I_{r+e,\lambda}(\theta,\phi,\lambda) = I_{e,\lambda}(\theta,\phi,\lambda) + I_{r,\lambda}(\theta,\phi,\lambda) \tag{2.35}$$

From $I_r$ and $I_{r,\lambda}$, respectively, we can define the *surface reflected power*, denoted $R$, and the *spectral surface reflected power*, denoted $R_\lambda$. The sum of the surface emissive power $E$ and the surface reflected power $R$ gives the total power leaving the surface per unit area, which is the radiosity $J$. The same is true for the spectral versions of these quantities.

$$J = E + R \tag{2.36}$$

$$J_\lambda(\lambda) = E_\lambda(\lambda) + R_\lambda(\lambda) \tag{2.37}$$

### 2.4.7 Net radiance and net surface radiative flux

Having defined all the relevant radiances and surface powers, we can now determine the net radiance and the net radiative heat flux per unit area (or surface radiative flux) exiting from a surface. Net radiance is defined in **Definition 17**.

> **Definition 17: Net Radiance**
>
> The net radiance[a] $I_{\text{net}}$ is the rate $dQ$ at which radiant energy exits a surface element $dA$ per unit time from the direction $(\theta, \phi)$, per unit area of $dA$ normal to this direction, per unit solid angle $d\omega$ about this direction. In terms of the emitted, reflected and incident radiances, it can be written as:
>
> $$I_{\text{net}}(\theta, \phi) = I_e(\theta, \phi) + I_r(\theta, \phi) - I_i(\theta, \phi) \tag{2.38}$$
>
> [a] Sometimes also known as the *net radiation intensity*.

The spectral counterpart of the net radiance is the net spectral radiance, defined in **Definition 18**.

> **Definition 18: Net Spectral Radiance**
>
> The net spectral radiance[a] $I_{\text{net},\lambda}$ is the rate $dQ$ at which radiant energy exits a surface element $dA$ per unit time from the direction $(\theta, \phi)$ and at wavelength $\lambda$, per unit area of $dA$ normal to this direction, per unit solid angle $d\omega$ about this direction, and per unit wavelength interval $d\lambda$ around $\lambda$. In terms of the emitted, reflected and incident spectral radiances, it can be written as:
>
> $$I_{\text{net},\lambda}(\theta, \phi, \lambda) = I_{e,\lambda}(\theta, \phi, \lambda) + I_{r,\lambda}(\theta, \phi, \lambda) - I_{i,\lambda}(\theta, \phi, \lambda) \tag{2.39}$$
>
> [a] Sometimes also known as the *net spectral radiation intensity*.

The net surface radiative flux can then be defined as in **Definition 19**.

> **Definition 19: Net Hemispherical Surface Radiative Flux**
>
> The net hemispherical surface radiative flux $q''_{\text{rad}}$ is the net radiation energy exiting a flat surface per unit area and per unit time. It is given by:

$$q''_{\text{rad}} = \int_0^{2\pi} d\phi \int_0^{\pi/2} d\theta \, I_{\text{net}}(\theta, \phi) \cos\theta \sin\theta \qquad (2.40)$$

This can be expanded in terms of each of the radiances as:

$$\begin{aligned} q''_{\text{rad}} &= \int_0^{2\pi} d\phi \int_0^{\pi/2} d\theta \, I_{\text{net}}(\theta, \phi) \cos\theta \sin\theta \\ &= \int_0^{2\pi} d\phi \int_0^{\pi/2} d\theta \, \left( I_e(\theta, \phi) + I_r(\theta, \phi) - I_i(\theta, \phi) \right) \cos\theta \sin\theta \\ &= E + R - G = J - G \end{aligned} \qquad (2.41)$$

The net spectral surface radiative flux is defined in **Definition 20**.

**Definition 20: Net Spectral Hemispherical Surface Radiative Flux**

The net spectral hemispherical surface radiative flux $q''_{\text{rad},\lambda}$ is the net radiation energy exiting a flat surface at a frequency $\lambda$ per unit area, per unit time and per unit wavelength interval $d\lambda$ around $\lambda$. It is given by:

$$q''_{\text{rad},\lambda}(\lambda) = \int_0^{2\pi} d\phi \int_0^{\pi/2} d\theta \, I_{\text{net},\lambda}(\theta, \phi, \lambda) \cos\theta \sin\theta \qquad (2.42)$$

Again, we can obtain:

$$\begin{aligned} q''_{\text{rad},\lambda}(\lambda) &= \int_0^{2\pi} d\phi \int_0^{\pi/2} d\theta \, I_{\text{net},\lambda}(\theta, \phi, \lambda) \cos\theta \sin\theta \\ &= \int_0^{2\pi} d\phi \int_0^{\pi/2} d\theta \, \left( I_{e,\lambda}(\theta, \phi, \lambda) + I_{r,\lambda}(\theta, \phi, \lambda) - I_{i,\lambda}(\theta, \phi, \lambda) \right) \cos\theta \sin\theta \\ &= E_\lambda(\lambda) + R_\lambda(\lambda) - G_\lambda(\lambda) = J_\lambda(\lambda) - G_\lambda(\lambda) \end{aligned}$$
$$(2.43)$$

Integrating $q''_{\text{rad},\lambda}(\lambda)$ over all wavelengths we can obtain $q''_{\text{rad}}$:

$$q''_{\text{rad}} = \int_0^\infty q''_{\text{rad},\lambda}(\lambda) d\lambda \qquad (2.44)$$

## 2.4.8 Black body radiation

The preceding sections explored concepts of emitted radiance, incident radiance, irradiation, and radiosity, along with derivations for emitted and incident surface power. However, these expressions have been presented only in general terms of radiances, without addressing how these radiances are actually determined. This section examines the specific case of a *black body*, which provides a natural foundation for broader generalization.

**The black body**

Any body at a temperature $T$ above absolute zero emits radiation. The amount of radiation emitted by a body in a specific direction and at a specific wavelength depends on many factors, such as the material and the surface properties [17]. The same thing can be said for absorption of incident radiation. In order to study the radiation properties of a general body, it is useful to first introduce the concept of the *black body*, which is defined in **Definition 21**.

> **Definition 21: Black Body**
>
> A *black body* is an idealized physical object exhibiting the following properties [11]:
>
> - It absorbs all incident radiation, regardless of wavelength or direction of incidence.
> - For a given temperature and wavelength, no surface can emit more energy than a black body.
> - It is a diffuse emitter, meaning it emits radiation uniformly in all directions.

In other words, the black body is a perfect emitter and absorber of radiation. As such, it is very useful as a standard against which other surfaces can be compared.

The black body spectral emitted radiance at thermal equilibrium, determined first by Planck [24], is given by **Definition 22**.

> **Definition 22: Black Body Spectral Emitted Radiance**
>
> The spectral emitted radiance of a black body at equilibrium at temperature $T$ is given by:
> $$I_{b,\lambda}(\lambda, T) = \frac{2hc^2}{\lambda^5} \frac{1}{\exp\left(\frac{hc}{\lambda k_B T}\right) - 1} \quad (2.45)$$
>
> where $h = 6.62607015 \times 10^{-34}$ $J \cdot s$ is *Planck's constant*, $c = 299792458$ $m/s$ is the speed of light in vacuum, and $k_B = 1.38065 \times 10^{-23}$ $J/K$ is *Boltzmann's constant*.

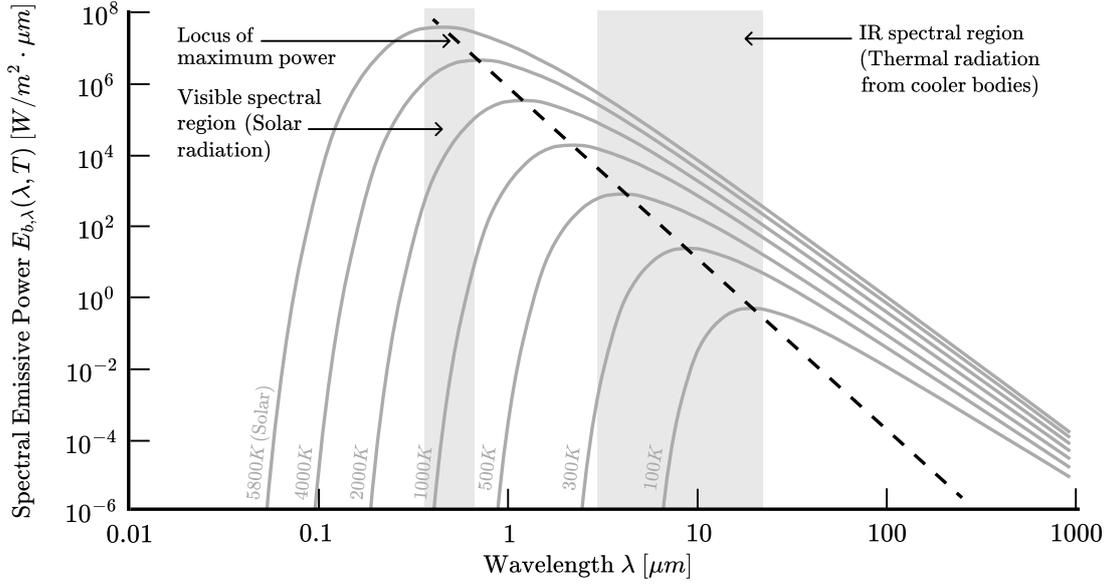

**Figure 2.6:** *Spectral emissive power of a black body for different temperatures. Note that, since a black body is (by definition) diffuse, the spectral emissive power is just the black body spectral emitted radiance multiplied by π. Based on a similar figure in [17].*

Combining **Definition 22** with **Equation 2.22**, we can express the spectral hemispherical surface emissive power of a black body, which depends only on temperature and wavelength, as:

$$E_{b,\lambda}(\lambda, T) = \pi I_{b,\lambda}(\lambda, T) = \frac{2\pi h c^2}{\lambda^5} \frac{1}{\exp\left(\frac{hc}{\lambda k_B T}\right) - 1} \quad (2.46)$$

The *spectral hemispherical surface emissive power of a black body*, also known simply as *spectral emissive power of a black body*, is plotted in **Figure 2.6** for different temperatures. Integrating over all wavelengths, we can work towards obtaining the total hemispherical surface emissive power of a black body:

$$E_b(T) = \int_0^\infty E_{b,\lambda}(\lambda, T) d\lambda = 2\pi h c^2 \int_0^\infty \frac{1}{\exp\left(\frac{hc}{\lambda k_B T}\right) - 1} \frac{d\lambda}{\lambda^5} \quad (2.47)$$

Using $\lambda = \frac{2\pi c}{\omega}$ and $d\lambda = -\frac{2\pi c}{\omega^2} d\omega$, we have:

$$\frac{d\lambda}{\lambda^5} = \frac{-\frac{2\pi c}{\omega^2} d\omega}{\left(\frac{2\pi c}{\omega}\right)^5} = -\frac{\omega^3}{(2\pi c)^4} d\omega \quad (2.48)$$

Then, we can change the variable of integration[4] and use the *reduced Planck constant*

---

[4] Note that the minus in **Equation 2.48** is cancelled by changing the order of the integration limits, which are now $\omega \in [\infty, 0]$, back to $\omega \in [0, \infty]$.

$\hbar = h/2\pi$ to obtain:

$$E_b(T) = \frac{\hbar}{(2\pi c)^2} \int_0^\infty \frac{\omega^3}{\exp\left(\frac{\hbar \omega}{k_B T}\right) - 1} \, d\omega \tag{2.49}$$

One last change of variable $\omega = \frac{k_B T}{\hbar} x$ and $d\omega = \frac{k_B T}{\hbar} dx$ gives us:

$$E_b(T) = \frac{k_B^4 T^4}{(2\pi c)^2 \hbar^3} \int_0^\infty \frac{x^3}{e^x - 1} \, dx \tag{2.50}$$

The integral $\int_0^\infty \frac{x^3}{e^x - 1} \, dx$ is tricky, but doable. The result, given in [25], is:

$$\int_0^\infty \frac{x^3}{e^x - 1} \, dx = \frac{\pi^4}{15} \tag{2.51}$$

Thus, we can finally express the total hemispherical surface emissive power of a black body as:

$$E_b(T) = \frac{k_B^4 T^4}{(2\pi c)^2 \hbar^3} \cdot \frac{\pi^4}{15} = \frac{\pi^2 k_B^4}{60 \hbar^3 c^2} T^4 \tag{2.52}$$

Recalling **Section 2.4.1**, we can see that this agrees with the *Stefan-Boltzmann Law* from **Definition 3**, and we can define the *Stefan-Boltzmann constant* $\sigma$ as:

$$\sigma \equiv \frac{\pi^2 k_B^4}{60 \hbar^3 c^2} = 5.67 \times 10^{-8} \, \frac{W}{m^2 \cdot K^4} \tag{2.53}$$

So that we recover:

$$E_b(T) = \sigma T^4 \tag{2.54}$$

### 2.4.9 Real body radiation and material properties

Having defined the ideal black body, and having obtained the expressions for its emitted radiance and its spectral emissive power, we are now ready to consider real surfaces.

As we already established, a black body emits the maximum amount of radiation that any object can emit at the same temperature, for any direction and wavelength. It also absorbs all incident radiation, regardless of wavelength and angle of incidence. For this reason, it is useful to use the black body as a reference when studying the properties of real surfaces. The way we do that is by means of two analogous properties called the emissivity and the absorptivity, which are ratios of the power emitted or absorbed by the real body to the power emitted or absorbed by the ideal black body.

Note that these ratios in general depend on the wavelength and the direction considered [11]. For instance, snow and white paint are very bad absorbers of visible light, and that is why they appear white. However, they are very good absorbers of infrared (IR) radiation, so for those wavelengths they are essentially black [17].

**Emissivity**

The *emissivity* $\varepsilon$ of a surface is the ratio of the emitted radiation from the surface to the emitted radiation from a black body at the same temperature. It takes values between $0 \leq \varepsilon \leq 1$ and gives a measure of how well a surface emits radiation compared to a black body, which is the ideal emitter with $\varepsilon = 1$.

As we mentioned already, although emissivity is often given as a constant value, it is in fact a function of the temperature, the wavelength and the angle of incidence. **Definition 23** defines the emissivity of a surface in the most general way.

> **Definition 23: Spectral Directional Emissivity**
>
> The spectral directional emissivity $\varepsilon_{\lambda,\theta}(\lambda, \theta, \phi, T)$ of a surface is defined as the ratio of the radiance emitted by the surface at temperature $T$, at a specified wavelength $\lambda$ in a specified direction $(\theta, \phi)$ to the radiance emitted by a black body at the same temperature, at the same wavelength and in the same direction. It is given by:
>
> $$\varepsilon_{\lambda,\theta}(\theta, \phi, \lambda, T) = \frac{I_{e,\lambda}(\theta, \phi, \lambda, T)}{I_{b,\lambda}(\lambda, T)} \qquad (2.55)$$

Note that in this definition we have included the temperature dependence $T$ in the emission radiances, which we omitted in the previous sections for simplicity. Note also that the black body radiance $I_{b,\lambda}(\lambda, T)$ is independent of the direction $(\theta, \phi)$, since a black body is a diffuse emitter.

Similarly, we can define the *spectral hemispherical emissivity* $\varepsilon_\lambda(\lambda, T)$ as in **Definition 24**. This value gives us information about the emissive behavior of the surface at different wavelengths and will be very important in following chapters, when we will need to deal with radiation in different frequency bands, namely visible solar radiation and IR thermal radiation from cooler bodies.

> **Definition 24: Spectral Hemispherical Emissivity**
>
> The spectral hemispherical emissivity $\varepsilon_\lambda(\lambda, T)$ of a surface is defined as the ratio of the spectral emitted radiation of the surface at temperature $T$ and wavelength $\lambda$ to the spectral emitted radiation of a black body at the same temperature and wavelength, regardless of direction. It is given by:

$$\varepsilon_\lambda(\lambda, T) = \frac{E_\lambda(\lambda, T)}{E_{b,\lambda}(\lambda, T)} \tag{2.56}$$

It is common to assume the *gray surface* approximation, which considers a constant emissivity $\varepsilon$ for all wavelengths, in conjunction with the *diffuse surface* approximation, which assumes that the emissivity is independent of the angle of incidence. In this case, we talk about the *total hemispherical emissivity* $\varepsilon(T)$ of a surface, defined in **Definition 25** in terms of the emitted radiation over all directions and all wavelengths. This is the emissivity referred to at the beginning of this section simply as *emissivity* $\varepsilon$, neglecting even the temperature dependence.

**Definition 25: Total Hemispherical Emissivity**

The total hemispherical emissivity[a] $\varepsilon(T)$ of a surface is the ratio of the total emitted radiation emitted by the surface at temperature $T$ to the total emitted radiation of a black body at the same temperature. It is given by:

$$\varepsilon(T) = \frac{E(T)}{E_b(T)} = \frac{\int_0^\infty \varepsilon_\lambda(\lambda, T) E_{b,\lambda}(\lambda, T) \, d\lambda}{\sigma T^4} \tag{2.57}$$

[a] Commonly referred to simply as *emissivity* $\varepsilon$, when we neglect even the temperature dependence.

**Absorptivity**

*Absorptivity* $\alpha$ is defined as the ratio of the absorbed incident radiation, which we will denote $A$, to the total incident radiation $G$. Since black bodies absorb all incident radiation, it can also be thought of as a measure of how well the body absorbs compared to a black body at the same temperature. Absorptivity takes values between $0 \leq \alpha \leq 1$, with the black body having $\alpha = 1$.

As with emissivity, absorptivity is not a constant value. Thus, we can define the *spectral directional absorptivity* (**Definition 26**), the *spectral hemispherical absorptivity* (**Definition 27**) and the *total hemispherical absorptivity* (**Definition 28**) of a surface in analogy to the definitions of emissivity.

**Definition 26: Spectral Directional Absorptivity**

The spectral directional absorptivity $\alpha_{\lambda,\theta}(\lambda, \theta, \phi, T)$ of a surface is defined as the ratio of the radiance absorbed by the surface at temperature $T$, at a specified wavelength $\lambda$ and in a specified direction $(\theta, \phi)$ to the total radiance incident on the surface at the specified temperature, at the same wavelength and in the same di-

rection. It is given by:

$$\alpha_{\lambda,\theta}(\theta, \phi, \lambda, T) = \frac{I_{i,\lambda}^{\text{abs}}(\theta, \phi, \lambda, T)}{I_{i,\lambda}(\theta, \phi, \lambda)} \quad (2.58)$$

where $I_{i,\lambda}^{\text{abs}}(\theta, \phi, \lambda, T)$ is the portion of the radiance corresponding to the absorbed irradiation.

Note that in **Definition 26** $I_{i,\lambda}(\theta, \phi, \lambda)$ does not depend on the temperature $T$, since the incident radiation is completely independent of the surface and its properties or characteristics. It does, however, depend on the direction and the wavelength, as these are properties of the incident radiation itself.

**Definition 27: Spectral Hemispherical Absorptivity**

The spectral hemispherical absorptivity $\alpha_\lambda(\lambda, T)$ of a surface is defined as the ratio of the spectral power absorbed by the surface at temperature $T$ and wavelength $\lambda$ to the irradiation incident on the surface at that same wavelength, regardless of direction. It is given by:

$$\alpha_\lambda(\lambda, T) = \frac{A_\lambda(\lambda, T)}{G_\lambda(\lambda)} = \frac{A_\lambda(\lambda, T)}{A_{b,\lambda}(\lambda)} \quad (2.59)$$

Where $A_\lambda(\lambda, T)$ is the spectral power absorbed by the surface at temperature $T$ and wavelength $\lambda$ and $A_{b,\lambda}(\lambda)$ is the spectral power absorbed by a black body at the same temperature and wavelength.

Since a black body absorbs all incident radiation, $A_{b,\lambda}(\lambda)$ does not depend on its temperature.

**Definition 28: Total Hemispherical Absorptivity**

The total hemispherical absorptivity[a] $\alpha(T)$ of a surface is the ratio of the total radiation absorbed by the surface at temperature $T$ to the total irradiation incident on the surface. It is given by:

$$\alpha(T) = \frac{A(T)}{G} = \frac{A(T)}{A_b(T)} = \frac{\int_0^\infty \alpha_\lambda(\lambda, T) G_\lambda(\lambda) \, d\lambda}{G} \quad (2.60)$$

where $A(T)$ is the total power absorbed by the surface at temperature $T$ and $A_b(T)$ is the power absorbed by a black body at the same temperature.

[a] Commonly referred to simply as *absorptivity* $\alpha$, when we neglect even the temperature dependence.

As in the case of emissivity, it is common to assume the *gray surface* and the *diffuse surface* approximations, which assume that the absorptivity is independent of the wavelength and the angle of incidence. In these cases, we use the definition of the *total hemispherical absorptivity* $\alpha(T)$ and even shorten this to *absorptivity* $\alpha$, neglecting the dependence on temperature.

**The gray surface and radiative properties of surfaces at different spectral regions**

We have already mentioned that emissivity and absorptivity, as well as other properties of surfaces, generally depend on the wavelength and the angle of incidence of the radiation under consideration. However, in practice, it is often useful to consider the *gray surface* approximation, which assumes that the emissivity and absorptivity are constant values in the spectral region of interest (see **Definition 29**). In other words, we assume that the emissivity and absorptivity are independent of the wavelength. This allows us to simplify the already complex calculations of radiative heat transfer.

> **Definition 29: Gray Surface**
>
> A *gray surface* is a surface whose emissivity and absorptivity are independent of the wavelength. This means that:
>
> - Its spectral hemispherical emissivity is constant and equal to its total hemispherical emissivity: $\varepsilon_{\lambda,\theta}(\lambda, T) = \varepsilon(T)$.
> - Its spectral hemispherical absorptivity is constant and equal to its total hemispherical absorptivity: $\alpha_{\lambda,\theta}(\lambda, T) = \alpha(T)$.
> - Similar definitions can be made for the directional versions of emissivity and absorptivity.

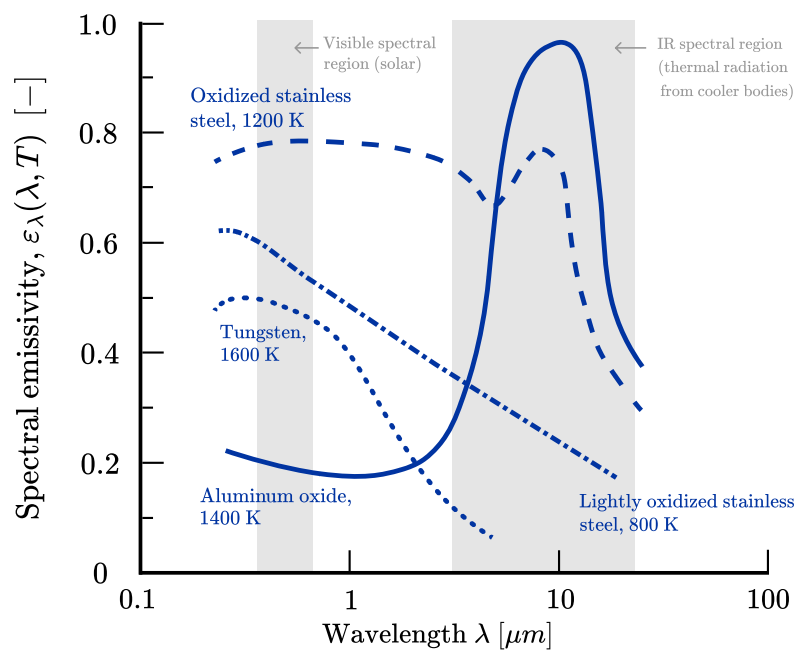

**Figure 2.7:** *Real spectral emissivity of some important metals, taken from* [17].

Taking aluminum oxide from **Figure 2.7** as an example, we may consider it as a gray surface with emissivity around $\varepsilon_{\text{vis}} \approx 0.2$ in the visible region and $\varepsilon_{\text{IR}} \approx 0.8$ in the IR region.

Many commercial software packages, such as *Radian* and *ESATAN-TMS*, allow the user to define the emissivity and absorptivity of surfaces in these two spectral regions, which are the most important for space applications. These are able to account for both the visible radiation that comes from the Sun (directly or through albedo) and the thermal radiation emitted by the spacecraft, the Earth and other nearby bodies, which is mostly in the IR. It is worth noting, however, that many of these software packages employ the notation $\alpha$ for both the absorptivity and the emissivity of surfaces in the visible range (since bodies mostly absorb visible radiation, from the Sun), and the notation $\varepsilon$ for the emissivity in the IR range (since bodies mostly emit thermal radiation in the IR). This can be done by virtue of Kirchhoff's law of thermal radiation, which will be discussed in **Section 2.4.11**.

### 2.4.10 Diffuse and specular reflection

In **Section 2.4.3**, we briefly mentioned that surfaces can distribute emitted and reflected radiation in different ways among all the possible directions, depending on their surface properties such as roughness. In terms of reflection, there are two extreme ways in which a surface can reflect incident radiation: through *diffuse reflection* (**Definition 30**) and through *specular reflection* (**Definition 31**) [17].

> **Definition 30: Diffuse Reflection**
>
> In *diffuse reflection*, the surface reflects incident radiation uniformly in all directions. Thus, the reflected radiance is independent of the angle of incidence and is the same in all directions.

> **Definition 31: Specular Reflection**
>
> In *specular reflection*, the surface reflects incident radiation in a single direction, which is determined by the angle of incidence and the surface normal. The reflected radiance is therefore dependent on the angle of incidence and is concentrated in a single direction.

The fact that we can see surfaces from different angles is precisely due to their diffuse behavior. Without diffuse reflection, we would only be able to see surfaces from very specific angles. Translating this now to radiative heat transfer, it is clear that being able to model diffuse reflection is crucial for accurately predicting the radiative heat transfer between surfaces. That is the purpose of this work: to add the feature of diffuse reflection to the *Radian* software package, which at the time of starting this thesis only supports specular reflection.

Generally, real surfaces exhibit a combination of both diffuse and specular reflectivity. To account for this, some software packages allow the user to assign both a diffuse and a specular component of the reflectivity to surfaces.

### 2.4.11 Kirchhoff's law of thermal radiation

It will be useful to first talk about the concept of *thermodynamic equilibrium* and the principle of *detailed balance*.

**Thermodynamic equilibrium and detailed balance**

A system is in *thermodynamic equilibrium* (TE) when it is in thermal, mechanical and chemical balance [26]. Macroscopically, this translates into a uniform temperature, pressure and chemical potential throughout the system, as well as zero net heat transfer. Microscopically, this means that every microscopic process is time-reversible, that is, that for every microscopic process there exists a reverse process that can occur with the same probability. If this was not the case, there would be an imbalance in the system, violating the condition of TE [27].

These microscopic implications lead to the concept of *detailed balance*, which in the context of radiation states that, in TE, the energy radiated and absorbed by a body must be equal for every infinitesimal surface element, in every direction and polarization state, and across every wavelength interval, resulting in zero net energy transfer [28].

**Kirchhoff's law in thermodynamic equilibrium**

One consequence of TE is *Kirchhoff's law of thermal radiation* [29] which is stated in **Definition 32**.

> **Definition 32: Kirchhoff's Law of Thermal Radiation**
>
> For a body in thermodynamic equilibrium, the absorptivity $\alpha(\lambda)$ and the emissivity $\varepsilon(\lambda)$ at each wavelength $\lambda$ are equal:
>
> $$\alpha(\lambda) = \varepsilon(\lambda) \tag{2.61}$$

Where does this come from? Consider a body in thermodynamic equilibrium with its surroundings. This means that the environment is radiating back at the body with a Planckian distribution (that of a black body) for the same temperature $T$ as the body. The power per unit area and per unit wavelength interval around $\lambda$ that the body is receiving from the environment is given by $E_{b,\lambda}(\lambda, T)$. Of this total incident power, the body absorbs an amount equal to $\alpha(\lambda) \cdot E_{b,\lambda}(\lambda, T)$, where $\alpha(\lambda)$ is the absorptivity of the body at wavelength $\lambda$. Similarly, the power per unit area and per unit wavelength interval around $\lambda$ that the body is emitting is given by $\varepsilon(\lambda) \cdot E_{b,\lambda}(\lambda, T)$, where $\varepsilon(\lambda)$ is the emissivity of the body at wavelength $\lambda$.

Since the body is in TE, it is not exchanging any net power with its surroundings, meaning that the power absorbed by the body must be equal to the power emitted by the body at each wavelength $\lambda$:

$$\alpha(\lambda) \cdot E_{b,\lambda}(\lambda, T) = \varepsilon(\lambda) \cdot E_{b,\lambda}(\lambda, T) \tag{2.62}$$

This means that the absorptivity and emissivity of the body at each wavelength $\lambda$ must be equal, since in general $E_{b,\lambda}(\lambda, T) \neq 0$ for any wavelength $\lambda$.

In fact, under TE, Kirchhoff's law of thermal radiation holds *in detail*, meaning that the absorptivity and emissivity are equal not only for a specific wavelength $\lambda$, but also for every infinitesimal surface element, in every direction and polarization state, and across every wavelength interval.

Kirchhoff's law of thermal radiation is the reason why many times $\varepsilon$ and $\alpha$ are used interchangeably in the context of thermal radiation.

**Kirchhoff's law in local thermodynamic equilibrium**

A quick look at the definition of Kirchhoff's law of thermal radiation in TE is enough to see that the conditions for its validity are very strict. For many interesting applications, such as spacecraft thermal control, we are almost always dealing with systems that do not have a uniform temperature distribution, thus violating the condition of TE.

Here, it is useful to make the distinction between *global thermodynamic equilibrium* GTE, which is what we have called just TE so far, and *local thermodynamic equilibrium* (LTE). In GTE, temperature and other intensive properties are uniform throughout the system and constant over time. In contrast, LTE allows these properties to vary in space and time, but slowly enough so that each point can still be considered locally in TE [30]. That is, each point in the system is in thermodynamic equilibrium with its local radiative field. This is a much more realistic condition for many practical applications, such as spacecraft thermal control, and it is sufficient to be able to apply Kirchhoff's law of thermal radiation for each point in the system [31].

### 2.4.12 Radiation between surfaces

Up to this point, we have mainly discussed radiation exchange between a single surface and its surroundings. This could be useful, for example, to obtain the average temper-ature of a satellite, as a whole, in outer space. However, we generally require a more detailed analysis, which involves breaking down the satellite into many individual surfaces.

This breakdown introduces additional challenges. For instance, the radiation emitted by one surface may be absorbed not only by a single other surface but also fractionally by multiple surfaces, depending on their visibility and properties. To tackle this issue, we must take into account the concept of *view factors*.

The medium through which radiation is exchanged plays also a crucial role, as it can influence the radiation transfer between surfaces. Luckily, in the context of spacecraft thermal control, the medium is almost always vacuum, which means that radiation exchange occurs solely between the surfaces themselves, without any involvement from the medium. This simplifies the analysis.

**The view factor**

The radiative exchange between two surfaces is dependent on their relative positions and orientations. For example, two parallel surfaces will exchange radiation differently than two surfaces that are perpendicular to each other. Likewise, two closely spaced surfaces will exchange radiation differently than two surfaces that are far apart. To account for these geometric factors, we introduce the concept of *view factors* (**Definition 33**), which quantify the fraction of radiation emitted by one surface that is directly received by another surface. Other names for view factors include *configuration factors*, *shape factors*, *form factors*, *geometric factors* and *angle factors* [17, 32].

> **Definition 33: View Factor**
>
> The *view factor* $F_{i \to j}$, or simply $F_{ij}$, from surface $i$ to surface $j$ is a purely geometric quantity defined as the fraction of the total radiation emitted by surface $i$ that strikes surface $j$ directly.

Note that view factors tell us the fraction of radiation that *arrives* at a surface from another surface. However, the radiation that *arrives* at a surface does not necessarily need to be *absorbed* by that surface. This will depend on the absorptivity of the receiving surface, which is not considered in the calculation of view factors. Furthermore, view factors do not account for visibility between surfaces due to reflection or transmission, since these depend on the reflectivity and the transmissivity.

View factors generally assume diffuse emission, meaning that they account for the maximum possible visibility between surfaces.

**The view factor integral**

After giving the conceptual definition of view factors, the natural question is: how do we actually compute them? That is the question we will answer in this section.

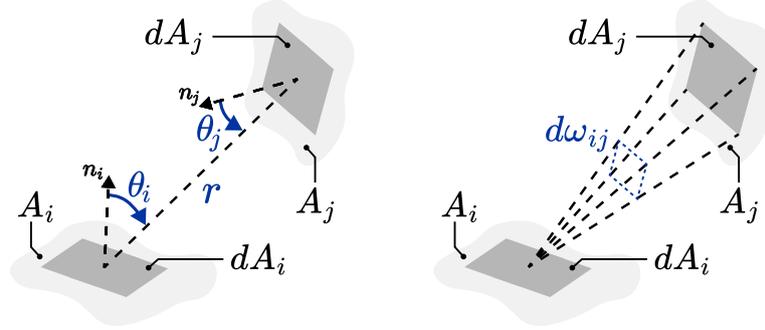

**Figure 2.8:** *Definition of the view factor between two elementary surfaces $dA_i$ and $dA_j$.*

To begin with, consider the two arbitrarily oriented surfaces $A_i$ and $A_j$ shown on **Figure 2.8**. The relative position and orientation of these surfaces are given by the distance $r$ and the angles $\theta_i$ and $\theta_j$ (measured with respect to the surface normals $\boldsymbol{n_i}$ and $\boldsymbol{n_j}$). Taking now a pair of elementary surfaces $dA_i$ and $dA_j$ on $A_i$ and $A_j$, respectively, and following **Definition 9**, we find that the exiting radiance[5] from $dA_i$ in the direction ($\boldsymbol{n_i}$, $\theta_i$), denoted $dQ_{ij}$, is given by:

$$dQ_{ij} = I_{r+e, dA_i}(\boldsymbol{n_i}, \theta_i) \cdot dA_i \, \cos \theta_i \cdot d\omega_{ij} \qquad (2.63)$$

Noting that $d\omega_{ij} = \cos \theta_j \, dA_j / r^2$, we obtain:

$$dQ_{ij} = I_{r+e, dA_i}(\boldsymbol{n_i}, \theta_i) \cdot \frac{\cos \theta_i \cos \theta_j \, dA_i \, dA_j}{r^2} \qquad (2.64)$$

The total power leaving the elementary surface $dA_i$ via both reflection and emission in all directions is just the radiosity $J_i$ which, considering a diffuse surface, is given by:

$$J_i = \pi \, I_{r+e, dA_i}(\boldsymbol{n_i}, \theta_i) \qquad (2.65)$$

Substituting this expression into **Equation 2.64**, we obtain:

$$dQ_{ij} = J_i \cdot \frac{\cos \theta_i \cos \theta_j}{\pi r^2} \cdot dA_i \, dA_j \qquad (2.66)$$

We can now integrate over the two surfaces $A_i$ and $A_j$ to obtain the total power leaving the surface $A_i$ and striking the surface $A_j$:

$$Q_{ij} = \int_{A_j} \int_{A_i} dQ_{ij} = \int_{A_j} \int_{A_i} J_i \cdot \frac{\cos \theta_i \cos \theta_j}{\pi r^2} \cdot dA_i \, dA_j \qquad (2.67)$$

---

[5] Recall from **Section 2.4.6** that this is the total radiance leaving the emitting surface.

Since the total power leaving the surface $A_i$ in all directions is given by $Q_i = J_i \cdot A_i$, the view factor integral from **Definition 34** can be obtained by computing $F_{ij} = Q_{ij}/Q_i$.

> **Definition 34: View Factor Integral**
>
> The view factor integral $F_{ij}$ between two surfaces $A_i$ and $A_j$ is given by:
>
> $$F_{ij} = \frac{1}{A_i} \int_{A_j} \int_{A_i} \frac{\cos \theta_i \cos \theta_j}{\pi r^2} \cdot dA_i \, dA_j \quad (2.68)$$

It is common to organize the view factors $F_{ij}$ into an $N \times N$ *view factor matrix*, where $i$ and $j$ represent the row and column indices, respectively, and $N$ denotes the total number of surfaces in the enclosed system. In the case of an open system, an additional imaginary surface is introduced to represent the environment (typically outer space), effectively closing the system. Consequently, the matrix expands to $N \times (N + 1)$, with the last column indicating the view factors to the environment.

**View factor relations**

We have already obtained the integral expression for $F_{ij}$. It is then simple to find the expression for the view factor $F_{ji}$:

$$F_{ji} = \frac{1}{A_j} \int_{A_i} \int_{A_j} \frac{\cos \theta_j \cos \theta_i}{\pi r^2} \cdot dA_j \, dA_i \quad (2.69)$$

Comparing **Equation 2.68** and **Equation 2.69**, it is easy to deduce the **Definition 35** by equating the integrals.

> **Definition 35: Reciprocity Relation**
>
> For any two surfaces $A_i$ and $A_j$ in a system, the following reciprocity relation holds:
>
> $$A_i F_{ij} = A_j F_{ji} \quad (2.70)$$

Another important property arises from the conservation of energy, which requires that all power emitted by each of the surfaces in an enclosure[6] be intercepted by some surface in the same enclosure. No "fraction of power" is lost, so the sum of all fractions from a surface $A_i$ must be equal to 1 (see **Definition 36**).

---

[6] We include also the surfaces of the enclosure itself.

**Definition 36: Summation Rule**

For a surface $A_i$ in a system of $N$ surfaces, the following summation rule[a] holds:

$$\sum_{j=1}^{N} F_{ij} = 1 \qquad (2.71)$$

[a] Also known as the *closure rule*.

In the particular application of spacecraft thermal analysis, when we are often in an outer space scenario, an imaginary surface is added to represent outer space and "close" the system to form an enclosure. This way, the *summation rule* still holds.

**Radiation between black surfaces**

The net radiative heat transfer $q_{ij}$ from surface $i$ to surface $j$ is the net rate at which heat emitted[7] by surface $i$ gets to surface $j$. It is given by the difference of the radiosities of the two surfaces, properly weighted by the areas and view factors between them:

$$q_{ij} = A_i F_{ij} J_i - A_j F_{ji} J_j = A_i F_{ij} \left( J_i - J_j \right) \qquad (2.72)$$

In **Equation 2.72**, we have used the reciprocity relation $F_{ji} A_j = F_{ij} A_i$. The direction of the heat transfer is such that a positive value of $q_{ij}$ means that heat flow is from surface $i$ to surface $j$, while a negative value means that heat flow is from surface $j$ to surface $i$. Again, using the reciprocity relation, we can write this as:

$$q_{ij} = -A_j F_{ji} \left( J_j - J_i \right) = -q_{ji} \qquad (2.73)$$

When working with black bodies, we have no reflection, thus the problem is simplified since heat leaves each surface only due to emission. That is, for a black body, $J = E_b = \sigma T^4$. Then, the net radiative heat transfer rate between surfaces $i$ and $j$ is given by:

$$q_{ij} = \sigma A_i F_{ij} \left( T_i^4 - T_j^4 \right) \qquad (2.74)$$

The final form of the resulting rate equation is given in **Definition 37**, where we have divided by the area to obtain the rate per unit area of the emitting surface $i$.

**Definition 37: Radiation Rate Equation (Between Black Surfaces)**

The rate $q''_{ij}$ at which heat is transferred by radiation from one black surface at temperature $T_i$ to another black surface at temperature $T_j$, per unit time and per

---

[7] By direct emission or by reflection.

unit area of the emitting surface $i$, is given by:

$$q''_{ij} = \sigma F_{ij} \left(T_i^4 - T_j^4\right) \tag{2.75}$$

where $F_{ij}$ is the view factor from surface $i$ to surface $j$ and $\sigma$ is the *Stefan-Boltzmann constant*, equal to $5.67 \times 10^{-8}\,\text{W}\,\text{m}^{-2}\,\text{K}^{-4}$.

The net transfer of radiation energy per unit time from surface $i$, denoted $q_i$, can now be obtained considering all surfaces $j = 1, \ldots, N$ in the system:

$$q_i = \sigma A_i \sum_{j=1}^{N} F_{ij} \left(T_i^4 - T_j^4\right) \tag{2.76}$$

**Radiation between real surfaces**

In general, surfaces are not black bodies, so we must take into account that radiation may leave them by both emission and reflection. This complicates things a bit further, although the basic principles remain the same. For example, **Equation 2.72** still holds, but now we must consider the radiosity of a real surface as depicted in **Figure 2.9**, which is here given for a particular surface $i$:

$$J_i = \varepsilon_i E_{b,i} + \rho_i G_i \tag{2.77}$$

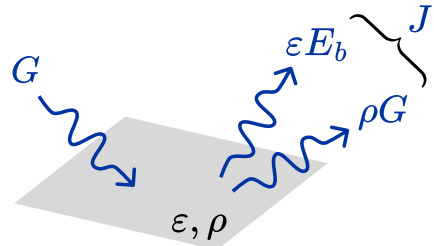

**Figure 2.9:** *Radiation exchange on a real surface.*

With this expression, we can see that the net radiative heat transfer $q_{ij}$ from surface $i$ to surface $j$ is given by:

$$\begin{aligned}
q_{ij} &= A_i F_{ij} \left(J_i - J_j\right) \\
&= A_i F_{ij} \left(\varepsilon_i E_{b,i} + \rho_i G_i - \varepsilon_j E_{b,j} - \rho_j G_j\right) \\
&= A_i F_{ij} \left(\varepsilon_i E_{b,i} - \varepsilon_j E_{b,j} + \rho_i G_i - \rho_j G_j\right) \\
&= A_i F_{ij} \sigma \left(\varepsilon_i T_i^4 - \varepsilon_j T_j^4\right) + A_i F_{ij} \left(\rho_i G_i - \rho_j G_j\right)
\end{aligned} \tag{2.78}$$

**Equation 2.78** makes it clear that the situation is now different from that of a system of black bodies, but it is not really very useful as it is. Our aim is to find the net transfer of radiation power from a real surface $i$ in terms of just the temperatures, the view factors and the areas, as we did for black bodies. We can simplify the problem by considering isothermal[8], diffuse, gray and opaque[9] surfaces with uniform radiosity and irradiation. We consider also LTE, so that $\varepsilon = \alpha$. Under these assumptions, the radiosity of a particular real surface $i$ in the system can be written as[10]:

$$J_i = \varepsilon_i E_{b,i} + (1 - \varepsilon_i) G_i = \varepsilon_i \left( E_{b,i} - G_i \right) + G_i \tag{2.79}$$

The irradiation $G_i$ is the sum of the irradiation from each of the other surfaces in the system $j = 1, \ldots, N$:

$$G_i = \frac{1}{A_i} \sum_{j=1}^{N} A_j F_{ji} J_j = \frac{1}{A_i} \sum_{j=1}^{N} A_i F_{ij} J_j = \sum_{j=1}^{N} F_{ij} J_j \tag{2.80}$$

Combining these expressions with **Equation 2.41**, we obtain a system of equations where the unknowns are $q_i$, $G_i$ and $J_i$ for each surface $i$ in the system:

$$\begin{cases} J_i = \varepsilon \left( E_{b,i} - G_i \right) + G_i \\ G_i = \sum_{j=1}^{N} F_{ij} J_j \\ q_i = A_i \left( J_i - G_i \right) \end{cases} \tag{2.81}$$

Substituting $G_i = J_i - q_i/A_i$ from the second equation into the first equation, we obtain[11]:

$$J_i = E_{b,i} - \left( \frac{1}{\varepsilon_i} - 1 \right) \frac{q_i}{A_i} \tag{2.82}$$

We can then substitute the first equation into the last, and use **Equation 2.82** to obtain the expression for the net radiative heat transfer rate from surface $i$ to the rest of the system [33]:

$$\begin{aligned} q_i &= A_i \left( J_i - G_i \right) \\ &= A_i \left( \varepsilon_i \left( E_{b,i} - G_i \right) + G_i - G_i \right) \\ &= A_i \varepsilon_i \left( E_{b,i} - G_i \right) \\ &= A_i \varepsilon_i \left( E_{b,i} - \sum_{j=1}^{N} F_{ij} J_j \right) \\ &= A_i \varepsilon_i \left( E_{b,i} - \sum_{j=1}^{N} F_{ij} \left( E_{b,j} - \left( \frac{1}{\varepsilon_j} - 1 \right) \frac{q_j}{A_j} \right) \right) \end{aligned} \tag{2.83}$$

---

[8] This will make sense especially as we will discretize systems into isothermal nodes in **Section 3.3**.
[9] This means that $\tau = 0$.
[10] Note that, for a black body, $\varepsilon_i = 1$, so the radiosity reduces to $J_i = E_{b,i}$, as expected.
[11] This expression is valid for $\varepsilon_i \neq 0$, since for $\varepsilon_i = 0$ we have a completely reflective surface with $J_i = G_i$ and $q_i = 0$.

We can develop this further to obtain a similar expression to **Equation 2.76** for the net radiative heat transfer rate from surface $i$ to the rest of the system as a summation of terms that can be attributed to the radiative heat transfer from surface $i$ to each of the other surfaces $j = 1, \ldots, N$ in the system. Using the summation rule from **Definition 36**, we can put all the terms into the sum:

$$\begin{aligned} q_i &= A_i \varepsilon_i \left( \left( \sum_{j=1}^{N} F_{ij} \right) E_{b,i} - \sum_{j=1}^{N} F_{ij} \left( E_{b,j} - \left( \frac{1}{\varepsilon_j} - 1 \right) \frac{q_j}{A_j} \right) \right) \\ &= \sum_{j=1}^{N} F_{ij} A_i \varepsilon_i \left( E_{b,i} - E_{b,j} + \left( \frac{1}{\varepsilon_j} - 1 \right) \frac{q_j}{A_j} \right) \\ &= \sum_{j=1}^{N} F_{ij} A_i \varepsilon_i \left( \sigma \left( T_i^4 - T_j^4 \right) + \left( \frac{1}{\varepsilon_j} - 1 \right) \frac{q_j}{A_j} \right) \end{aligned} \qquad (2.84)$$

We can identify each term in the sum as the net radiative heat transfer rate from surface $i$ to surface $j$, as given in **Definition 38**.

---

**Definition 38: Radiation Rate Equation (Between Real Surfaces)**

The rate $q_{ij}''$ at which heat is transferred by radiation from one real surface[a] at temperature $T_i$ to another real surface at temperature $T_j$, per unit time and per unit area of the emitting surface $i$, is given by:

$$q_{ij}'' = \varepsilon_i F_{ij} \left( \sigma \left( T_i^4 - T_j^4 \right) + \left( \frac{1}{\varepsilon_j} - 1 \right) \frac{q_j}{A_j} \right) \qquad (2.85)$$

where $F_{ij}$ is the view factor from surface $i$ to surface $j$, $\varepsilon_i$ and $\varepsilon_j$ are the emissivities of the surfaces $i$ and $j$, respectively, $A_j$ is the area of surface $j$, $q_j$ is the net radiative heat transfer rate from surface $j$ to the rest of the system and $\sigma$ is the *Stefan-Boltzmann constant*, equal to $5.67 \times 10^{-8}\,\mathrm{W\,m^{-2}\,K^{-4}}$.

[a] Under the assumptions of isothermal, diffuse, gray and opaque surfaces in LTE.

---

It is clear to see that, for a system of black bodies, the expression in **Definition 38** reduces to that in **Definition 37**. Furthermore, if we write **Equation 2.85** in terms of the reflectivity $\rho_j = 1 - \varepsilon_j$, we can more clearly see how the importance of the reflected term changes with the reflectivity of the surface $j$:

$$q_{ij}'' = \sigma \varepsilon_i F_{ij} \left( T_i^4 - T_j^4 \right) + \frac{\rho_j}{1 - \rho_j} \frac{\varepsilon_i F_{ij} q_j}{A_j} \qquad (2.86)$$

For zero reflectivity $\rho_j = 0$, we have a black body, and the second term vanishes. As the reflectivity $\rho_j$ increases, the second term becomes more significant, scaling roughly[12] as $\rho_j$. Note that the first term also decreases as the reflectivity increases, and at the limit of $\rho_j = 1$, the surface is completely reflective, and it does not participate in the heat transfer by emission or absorption, as we already mentioned, so $q_j = 0$.

## 2.5 Multimode heat transfer

*Multimode heat transfer* refers to heat transfer processes in which more than one mode of heat transfer is important. In the context of spacecraft thermal control, the most common modes of heat transfer are conduction and radiation, and these are usually the only modes of heat transfer considered in most space thermal analysis software packages.

If conduction and radiation are considered, the total heat transfer rate between two surfaces is just the sum of the separate heat transfer rates due to conduction and radiation.

## 2.6 Heat diffusion equation

The main objectives when solving heat transfer problems are to determine the temperature distribution in a system and the heat transfer rates between surfaces. The temperature distribution can be obtained by solving the *heat diffusion equation*, and the heat transfer rates can be obtained by applying the relevant rate equations once the temperature distribution is known.

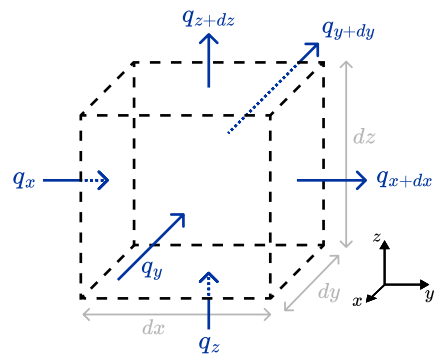

**Figure 2.10:** *Heat going in and coming out of an infinitesimal volume element $dV = dx\,dy\,dz$.*

We will now obtain the heat diffusion equation in three dimensions. Consider a homogeneous and incompressible[13] medium with no advection[14]. Suppose the temperature distribution is a function $T(x, y, z)$ of the three Cartesian coordinates. Consider an infinitesimal cube of volume $dV = dx\,dy\,dz$ as shown in **Figure 2.10**. The conduction heat rates going into the volume $dV$ in the Cartesian directions are indicated by $q_x$, $q_y$ and $q_z$, and they can be related to the heat rates at the opposite faces of the cube by a first order Taylor expansion, neglecting higher order terms [11]:

$$q_{x+dx} = q_x + \frac{\partial q_x}{\partial x} dx$$

$$q_{y+dy} = q_y + \frac{\partial q_y}{\partial y} dy \quad (2.87)$$

$$q_{z+dz} = q_z + \frac{\partial q_z}{\partial z} dz$$

---

[12] Note that $q_j \sim \varepsilon_j = 1 - \rho_i$, so that $\frac{\rho_j}{1-\rho_j} \frac{\varepsilon_i F_{ij} q_j}{A_j} \sim \rho_j$.
[13] Incompressibility implies constant density.
[14] This means that there is no heat transfer due to bulk motion of the medium.

There may also be some heat generation $E_g$ in the volume $dV$, which can be expressed in terms of a volumetric heat generation rate $q_g$ [W/m$^3$] as:

$$E_g = q_g \cdot dx\, dy\, dz \quad (2.88)$$

Lastly, the internal thermal energy $U$ stored in the volume $dV$ may change by an amount $E_{\text{st}}$ which, if there are no phase changes, can be expressed as a change in *sensible energy* $U_s$ in terms of the specific heat capacity[15], $c_p$ [11]:

$$E_{\text{st}} = \frac{\partial U}{\partial t} = \frac{\partial U_s}{\partial t} = \rho c_v \frac{\partial T}{\partial t} dx\, dy\, dz = \rho c_p \frac{\partial T}{\partial t} dx\, dy\, dz \quad (2.89)$$

We can now write the heat balance equation on a rate basis for the infinitesimal volume $dV$:

$$E_{\text{in}} + E_g - E_{\text{out}} = E_{st} \quad (2.90)$$

This equation simply indicates that the change in internal energy $E_{st}$ is equal to the heat entering the volume $E_{\text{in}}$ plus the heat generated in the volume $E_g$ minus the heat leaving the volume $E_{\text{out}}$. In the case of space thermal control, the heat entering and leaving the volume is usually due to conduction and radiation. Heat generation is usually due to dissipation from electrical components or due to the action of other heating or cooling systems used for thermal control. For this derivation, we consider only conduction heat transfer and heat generation, but the result can be easily extended to include radiation heat transfer as well. For this simplified case, the heat entering and leaving the volume is entirety due to conduction, so we can write:

$$q_x + q_y + q_z + q_g \cdot dx\, dy\, dz - q_{x+dx} - q_{y+dy} - q_{z+dz} = \rho c_p \frac{\partial T}{\partial t} dx\, dy\, dz \quad (2.91)$$

---

[15] For incompressible solids, $c_v = c_p$.

Substituting the expressions from **Equation 2.87**, into the equation above, we obtain:

$$q_g \cdot dx\,dy\,dz - \frac{\partial q_x}{\partial x}dx - \frac{\partial q_y}{\partial y}dy - \frac{\partial q_z}{\partial z}dz = \rho c_p \frac{\partial T}{\partial t}dx\,dy\,dz \qquad (2.92)$$

Applying now Fourier's law from **Equation 2.1** to express the heat fluxes $q_x$, $q_y$ and $q_z$, and dividing by the volume $dV = dx\,dy\,dz$, we obtain the *heat diffusion equation*:

$$q_g + \frac{\partial}{\partial x}\left(k\frac{\partial T}{\partial x}\right) + \frac{\partial}{\partial y}\left(k\frac{\partial T}{\partial y}\right) + \frac{\partial}{\partial z}\left(k\frac{\partial T}{\partial z}\right) = \rho c_p \frac{\partial T}{\partial t} \qquad (2.93)$$

The function $k$ is the thermal conductivity of the medium, which can generally depend on the position $(x, y, z)$, the temperature $T$ and other factors. It is common to assume that $k$ is constant throughout the volume, and that the medium is isotropic (i.e. $k$ is the same in all directions), which simplifies the equation to:

$$\frac{\partial^2 T}{\partial x^2} + \frac{\partial^2 T}{\partial y^2} + \frac{\partial^2 T}{\partial z^2} + \frac{q_g}{k} = \frac{1}{\alpha}\frac{\partial T}{\partial t} \qquad (2.94)$$

The constant $\alpha = k/(\rho c_p)$ is called the thermal diffusivity of the medium, and it is a measure of how quickly heat diffuses through the medium.

For steady state conditions, the right-hand side (RHS) of the equation vanishes, and we obtain the *steady state heat diffusion equation*.

# 3
# Modelling Heat Transfer and Radiation

## 3.1 Introduction

A model is a representation of a system, often simplified or idealized, that captures the features of the system that are relevant to for a particular purpose [34]. A model should be accurate enough to provide useful insights, but not so complex that it becomes unmanageable or computationally expensive.

In this chapter, we will describe the process of creating a good model of a system for the purpose of simulating heat transfer and radiation.

## 3.2 Modelling heat transfer

In the context of heat transfer, there are two main options for modelling systems: we can use physical models, which involve creating a physical replica of the system (possibly not to scale) and measuring its behavior under different conditions of interest, or we can use mathematical models, which involve creating a mathematical representation of the system and finding solutions to the mathematical equations that describe its behavior.

Among mathematical models, a further distinction exists. Assuming a solution exists, it may be expressed analytically as a mathematical function, or numerically as an approximation obtained through numerical methods.

Unfortunately, most real-world systems do not have analytical solutions, and even when they do, the solutions are often too complex to be useful or too difficult or expen-sive to obtain. In many cases, the extra precision of an analytical solution is not worth the additional complexity and cost of obtaining it.

For these reasons, heat transfer models are frequently approached numerically, which is the method we will adopt in this work. However, there are several important considerations to keep in mind when modelling heat transfer numerically, which we will discuss in the following sections.

## 3.3  Discretization

Numerical solutions to mathematical models are obtained by dividing the system into a finite set of discrete points and approximating the solution at each of these locations. This contrasts with analytical solutions, which provide exact expressions valid at *every* point in the domain.

This process of converting a continuous problem into a discrete one is known as *discretization*, and it is a key step in numerical modelling. Choosing the right number and location of points can greatly affect the accuracy and computational efficiency of the numerical solution.

To discretize a thermal system, we divide the domain into a finite number of smaller regions, called *elements*. To each element we assign a representative point, or *node*, typically located at its center. In heat transfer problems, it is common to assume that each element is isothermal, with a uniform temperature equal to that of its corresponding node. Under this assumption, the elements are fully represented by their nodes, and the system can be modeled as a network of interconnected nodes [32]. The collection of all nodes is called the *mesh*.

Discretization can be applied in different ways. One option is a *finite element* (FE) approach, where the geometry is divided into elements and nodes are placed at their boundaries, with the elements representing the distributed material in between. An alternative is a *lumped parameter network* (LPN) approach, in which each physical component is reduced to a single node (often located at the centroid) that carries all the mass, so that the node itself represents the element. This latter strategy, in which masses are lumped at nodes, gives the method its name.

## 3.4 Discretized heat equation for spacecraft thermal systems

As mentioned in **Section 2.6**, in space applications, the heat balance equation, on a rate basis, is given by[1]:

$$Q_{\text{cond}} + Q_{\text{rad}} + Q_{\text{diss}} + Q_{\text{heat}} - Q_{\text{cool}} = E_{st} \qquad (3.1)$$

Here, $Q_{\text{cond}}$ is the heat transfer rate due to conduction, $Q_{\text{rad}}$ is the heat transfer rate due to radiation, $Q_{\text{diss}}$ is the heat generated by dissipative processes, $Q_{\text{heat}}$ is the heat added to the system by heaters, and $Q_{\text{cool}}$ is the heat removed from the system by coolers.

Using **Equation 2.89** and calling $C = \rho c_p$ the *thermal capacitance*, we can rewrite the heat balance equation as:

$$Q_{\text{cond}} + Q_{\text{rad}} + Q_{\text{diss}} + Q_{\text{heat}} - Q_{\text{cool}} = C\frac{\partial T}{\partial t} \qquad (3.2)$$

Considering now a discrete system of $N$ nodes, each of which has an associated temperature $T_i$ ($i = 1, \ldots, N$) which is uniform within the node, we can replace the partial derivative with a full derivative, and write a set of $N$ differential equations describing the heat transfer in the system:

$$Q_{\text{cond},i} + Q_{\text{rad},i} + Q_{\text{diss},i} + Q_{\text{heat},i} - Q_{\text{cool},i} = C_i\frac{dT_i}{dt} \qquad (3.3)$$

In the following sections, we will discuss how to model each of the heat transfer mechanisms in **Equation 3.3**.

### 3.4.1 Conduction

To model conduction, we assign a thermal conductivity $k_i$ to each node $i$, which is a measure of how easily heat can flow through the material of the node. Then, an effective thermal conductivity must be computed for the path between each pair of contacting nodes $i$ and $j$, since it can in general contain up to three different contributions: the conductivity of the material of the first node $i$, the conductivity of the material of the second node $j$, and the thermal contact conductance $h_c$ [W/K·m$^2$] between the two nodes $i$ and $j$, as shown in **Figure 3.1**.

To obtain the effective thermal conductivity $k_{ij}$ between nodes $i$ and $j$, the thermal resistance method introduced in **Section 2.2.2** can be employed. Considering the three

---

[1] We use $E_{\text{in}} - E_{\text{out}} = Q_{\text{cond}} + Q_{\text{rad}}$ and $E_g = Q_{\text{diss}} + Q_{\text{heat}} - Q_{\text{cool}}$ in **Equation 2.90**.

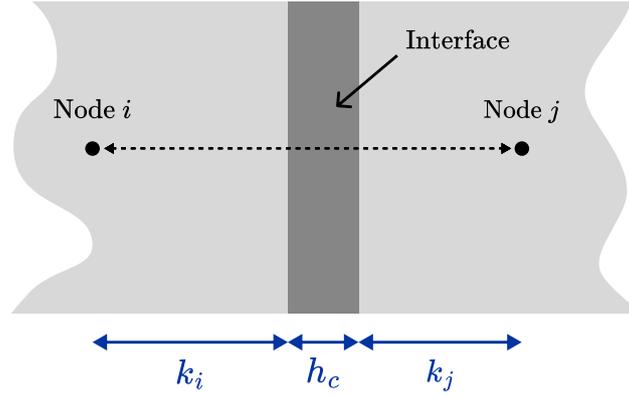

**Figure 3.1:** *Conductivity interface between two nodes i and j.*

mentioned contributions, the equivalent thermal resistance $R_{\text{eq},ij}$ is given by:

$$R_{\text{eq},ij} = R_i + R_j + R_c = \frac{l_i}{k_i A} + \frac{l_j}{k_j A} + \frac{1}{h_c A} \tag{3.4}$$

where $l_i$ and $l_j$ are the lengths travelled by the heat through the materials of nodes $i$ and $j$, respectively, and $A$ is the cross-sectional area of the interface between the two nodes.

The effective thermal conductivity $k_{ij}$ is given by:

$$k_{ij} = \frac{l_i + l_j}{R_{\text{eq},ij} A} = \frac{(l_i + l_j) k_i k_j h_c}{l_i k_j h_c + l_j k_i h_c + k_i k_j} \tag{3.5}$$

Then, the rate of thermal energy transfer per unit time due to conduction between two nodes $i$ and $j$ is given by:

$$q_{\text{cond},ij} = \frac{A k_{ij}}{L}(T_i - T_j) \tag{3.6}$$

where $L = l_i + l_j$ is the distance between the two nodes.

In the context of spacecraft thermal analysis, **Equation 3.6** is often expressed in terms of the *conductive coupling coefficient*, $GL_{ij} = A k_{ij}/L$:

$$q_{\text{cond},ij} = GL_{ij}(T_i - T_j) \tag{3.7}$$

Finally, the total heat transfer rate due to conduction at node $i$ is given by the sum of the contributions from all nodes $j$, where $GL_{ij} = 0$ for all nodes $j$ that are not in contact with node $i$:

$$Q_{\text{cond},i} = \sum_{j=1}^{N} GL_{ij}(T_i - T_j) \tag{3.8}$$

Note that $GL_{ij} = GL_{ji}$.

### 3.4.2 Radiation

In space, heat transfer primarily occurs through radiation, which includes two main types: thermal radiation, emitted by *warm* bodies due to their temperature and primarily found in the infrared range, and solar radiation, emitted by the *very hot* Sun and mainly in the visible range. Both types of radiation follow similar physical principles and can be modeled using comparable equations, although they involve different parameters.

Within a spacecraft's thermal system, heat transfer processes can be categorized into *radiative exchange between the spacecraft's surfaces* and *orbital heat fluxes* from external sources, such as direct solar flux, Earth albedo flux (reflected sunlight), and Earth infrared flux (heat emitted by the Earth). These last three are illustrated in **Figure 3.2**.

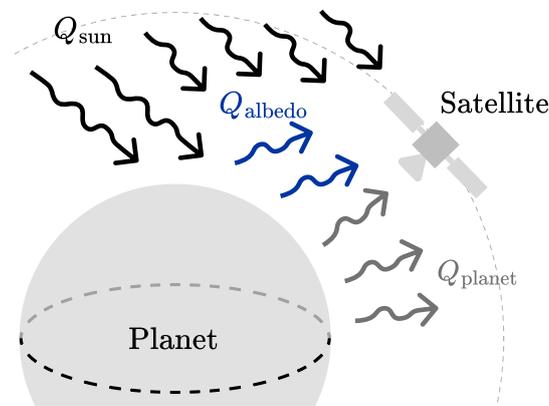

**Figure 3.2:** *Orbital heat fluxes from external sources, including direct solar flux, Earth albedo flux (reflected sunlight), and Earth infrared flux (heat emitted by the Earth). The first two are in the visible range, while the last one is in the infrared range.*

Orbital heat irradiances must be calculated based on the spacecraft's position and orientation relative to the Sun and Earth, and once established, they can be treated as constant heat sources in the thermal model. The absorbed power is then determined by multiplying the irradiance by the area of the surface that intercepts the radiation and by the surface's absorptivity in the visible (for solar and albedo heat) or IR (for planetary heat) range. In contrast, the radiative exchange between the spacecraft's surfaces is more complex and will be explored in greater detail, as it is central to the focus of this work.

**Radiative exchange between surfaces**

We already did most of the hard work towards modelling radiative exchange between surfaces in **Section 2.4.12**. In that section, we derived the expression for the net radia-tion power transferred between two real[2] surfaces $i$ and $j$:

$$q_{ij} = \varepsilon_i A_i F_{ij} \left( \sigma \left( T_i^4 - T_j^4 \right) + \left( \frac{1}{\varepsilon_j} - 1 \right) \frac{q_j}{A_j} \right) \qquad (3.9)$$

**Equation 3.9** is recursive, due to the presence of $q_j = \sum_{k=1}^{N} q_{jk}$ in the last term of its RHS, which accounts for the diffusely reflecting behavior of the intercepting surfaces $j$. Due to the complexity of this expression, a common approximation is to consider the intercepting surfaces as *black bodies*, meaning that they do not diffusely reflect radiation, and thus the last term in **Equation 3.9** can be neglected. Then, $q_{ij}$ is simplified to:

$$q_{ij} \approx \sigma \varepsilon_i A_i F_{ij} \left( T_i^4 - T_j^4 \right) \qquad (3.10)$$

Another common approximation includes the emissivity of the intercepting surface $j$ in the expression:

$$q_{ij} \approx \sigma \varepsilon_i \varepsilon_j A_i F_{ij} \left( T_i^4 - T_j^4 \right) \qquad (3.11)$$

In spacecraft thermal analysis, these expressions are often given in terms of the *radiative coupling coefficient*[3] (GR), denoted $GR_{ij}$:

$$q_{ij} = GR_{ij} \left( T_i^4 - T_j^4 \right) \qquad (3.12)$$

While it is common to find **Equation 3.12** in the literature [32, 33, 35–39], it is much more difficult to find a clear definition of the radiative coupling coefficient $GR_{ij}$, which can be different depending on whether **Equation 3.10** or **Equation 3.11** is used. For example, $GR_{ij} = \sigma \varepsilon_i A_i F_{ij}$ is used in [37, 38, 40], while $GR_{ij} = \sigma \varepsilon_i \varepsilon_j A_i F_{ij}$ is used in [39]. However, many times GRs are used in the literature without specifying their actual definition.

It is also important to note that these three expressions for $q_{ij}$ are *not* exact, but rather approximations that give reasonably good results in cases where the surfaces are not too far from being black bodies. However, when reflectivity starts to play a significant role, the accuracy of these approximations decreases dramatically, and they can no longer be used reliably, as pointed out and exemplified in [41], an unpublished but insightful manuscript.

Note that $GR_{ij} = GR_{ji}$ if we use **Equation 3.11**, but not if we use **Equation 3.10**. This may be a point in favor of the former, and may explain why it is more commonly used in the literature and in thermal analysis software [35, 37].

With these considerations in mind, we can now write the total radiative heat transfer

---

[2] With the relevant assumptions discussed already in the **Section 2.4.12**. These will be assumed henceforth when referring to "real" surfaces.

[3] Other names for $GR_{ik}$ include *radiative conductor* and *Radk*. The term *radiosity* is also sometimes used, not to be confused with $J$.

rate at node $i$ as the sum of the contributions from all nodes $j$:

$$Q_{\text{rad},i} = \sum_{j=1}^{N} GR_{ij} \left(T_i^4 - T_j^4\right) \tag{3.13}$$

### 3.4.3 Dissipation, heating, and cooling

The remaining terms in **Equation 3.3** include heat generated by dissipative processes $Q_{\text{diss},i}$, heat added by heaters $Q_{\text{heat},i}$, and heat removed by coolers $Q_{\text{cool},i}$, which are typically specified by the user. However, dissipative heat can also be calculated by integrating electrical simulations with the thermal model. In transient thermal simulations, these terms may vary with time and/or temperature (especially for heaters or coolers in the thermal control subsystem), while in steady-state simulations, they are treated as constant values.

### 3.4.4 Final discrete heat equation

Combining all the contributions discussed in the previous sections, we can write the final discrete heat equation for each node $i$ in the system:

$$\begin{aligned}
\sum_{j=1}^{N} GL_{ij}(T_i - T_j) + \sum_{j=1}^{N} GR_{ij} \left(T_i^4 - T_j^4\right) + Q_{\text{sun},i} + Q_{\text{albedo},i} + Q_{\text{earth},i} \\
+ Q_{\text{diss},i} + Q_{\text{heat},i} - Q_{\text{cool},i} = C_i \frac{dT_i}{dt}
\end{aligned} \tag{3.14}$$

## 3.5 View factor computation

In **Section 2.4.12**, we introduced the concept of the *view factor $F_{ij}$*, which is a measure of how much radiation emitted by surface $i$ is intercepted by surface $j$. We also derived the mathematical expression for the view factor integral, given in **Equation 2.68**. However, most of the time, the view factor integral is not easy to compute analytically, especially for complex geometries like those found in spacecraft thermal systems. Therefore, numerical methods are often used to compute the view factor between surfaces.

In this section, we will introduce one of the most common numerical methods for computing view factors, which is Monte Carlo Ray Tracing (MCRT).

### 3.5.1 Monte Carlo Ray Tracing

The concept of *Monte Carlo* was first introduced by Stanislaw Ulam in the 1940s, initially motivated by the challenge of calculating the probability of winning a game of solitaire [42]. Ulam recognized that determining probabilities for such a combinatorial problem was practically intractable; however, it could be approximated through statistical sampling with the emerging computers of that time [43]. The name *Monte Carlo* was suggested by Ulam's colleague, Nicholas Metropolis, inspired by the renowned casino in Monaco [44].

On the other hand, *Ray Tracing* is a computer graphics technique used for modelling light transport in a scene. It consists of tracing rays of light as they travel through the scene, simulating their emission and their interactions with the surfaces they encounter.

MCRT combines these two concepts, using Monte Carlo sampling to obtain the random initial positions and directions of the rays on the source surfaces, and using Ray Tracing to simulate the propagation of the rays through the scene. The view factor $F_{ij}$ from a source surface $i$ to a target surface $j$ can be estimated by counting the number of rays emitted from surface $i$ that reach surface $j$, and dividing this number by the total number of rays emitted from surface $i$:

$$F_{ij} = \frac{N_{ij}}{N_i} \quad (3.15)$$

where $N_{ij}$ is the number of rays emitted from surface $i$ that reach surface $j$, and $N_i$ is the total number of rays emitted from surface $i$. The accuracy of MCRT methods, their optimization and sensitivity is discussed in detail in [33, 45].

### 3.5.2 Statistical errors

MCRT is a powerful method for estimating view factors, however we must not forget that it is a statistical method, which means that it is subject to statistical errors. In particular, the view factors estimated by MCRT will usually not satisfy the *view factor relations* discussed in **Section 2.4.12**. To correct these errors, which can lead to large inaccuracies in the resulting heat transfer rates [46], several methods have been proposed for enforcing reciprocity and closure in the results obtained by MCRT [33, 47–50].

### 3.5.3 Reciprocity enforcement

Due to the statistical nature of the method, view factors computed by MCRT generally do not satisfy the reciprocity relation $A_i F_{ij} = F_{ji} A_j$. This is incompatible with the fact that $q_{ij} = -q_{ji}$, and fundamentally violates the second law of Thermodynamics. Therefore, it is necessary to enforce the reciprocity relation for the results to make physical sense. In the following sections, a few methods from the literature will be presented to enforce this relation. Unless otherwise specified, the methods presented here are valid for an enclosed system with a square $N \times N$ view factor matrix, where $N$ is the number of surfaces in the system.

**Naive and van Leersum enforcement**

The simplest way to enforce reciprocity is to discard the value of $F_{ji}$ computed by MCRT and calculate it from $F_{ij}$ using the reciprocity relation [33]:

$$F_{ji} = \frac{A_i}{A_j} F_{ij} \tag{3.16}$$

The diagonal elements $F_{ii}$ of the view factor matrix can be computed as:

$$F_{ii} = 1 - \sum_{j \neq i} F_{ij} \tag{3.17}$$

which enforces closure as a bonus. This method is simple and straightforward, but it wastefully discards half of the information obtained by MCRT, which could be used to improve the accuracy of the results. Furthermore, it does not guarantee that the view factors will be non-negative, which is a requirement for physical view factors. An iterative method proposed by J. van Leersum [50] can be used to ensure non-negativity of the view factors.

**Matrix triangulation**

A more sophisticated method for enforcing reciprocity is to use matrix triangulation. This is the method used in some of the major thermal analysis software packages like *ESARAD* (for radiative analysis) and *Thermica* [33, 51, 52].

We begin by defining a set of exchange coefficients $\eta_{ij} = A_i F_{ij}$, which will be used to perform the enforcement. In terms of a more general exchange factor $\mathcal{F}_{ij}$, we can write:

$$\eta_{ij} = \Omega_i \mathcal{F}_{ij} \tag{3.18}$$

where $\Omega_i = A_i$ and $\mathcal{F}_{ij} = F_{ij}$ for simple view factors as those we have treated so far. The reciprocity relation in terms of the exchange coefficients is given by:

$$\eta_{ij} = \eta_{ji} \tag{3.19}$$

Since we have a full set of view factors, we also have a full set of exchange coefficients which, prior to enforcement, do not satisfy **Equation 3.19**. We can then define the estimator $\hat{\eta}_{ij}$ of the real exchange coefficient $\eta_{ij}$ as:

$$\hat{\eta}_{ij} = \kappa \eta_{ij} + (1 - \kappa) \eta_{ji} \tag{3.20}$$

for some $\kappa \in [0, 1]$. This estimator is a *linear averaging* of the two exchange coefficients,

and it satisfies **Equation 3.19** by definition.

In the matrix triangulation enforcement, the weight $\kappa$ can be calculated with the following formula:

$$\kappa = \frac{1}{2}\left(1 + \text{sign}(Y) \mid Y \mid^n \right) \tag{3.21}$$

where the best value for the coefficient $n$ has been empirically determined to be $0.4$ [33, 52, 53] and $Y$ is given by:

$$Y = \frac{\frac{\Omega_j}{N_j} - \frac{\Omega_i}{N_i}}{\frac{\Omega_j}{N_j} + \frac{\Omega_i}{N_i}} \tag{3.22}$$

The constants $N_i$ and $N_j$ are the number of rays emitted from surfaces $i$ and $j$, respectively.

For open systems, the enforcement is done only with the square submatrix of the view factor matrix that corresponds to the surfaces of the system, excluding the environment. The view factors to the environment are then computed by applying the summation rule.

**Fractional variance**

Another method for enforcing reciprocity is the fractional variance method, which is based on minimizing the variance of the estimator $\hat{\eta}_{ij}$ [40]. Since $\Omega_i > 0$, the standard deviation of the exchange coefficient $\eta_{ij}$ predicted by MCRT can be estimated as [54]:

$$\sigma_{\eta_{ij}} = \text{SDV}\left(\eta_{ij}\right) = \Omega_i \cdot \text{SDV}\left(\mathcal{F}_{ij}\right) = \Omega_i \cdot z\sqrt{\frac{1 - \mathcal{F}_{ij}}{N_i \mathcal{F}_{ij}}} = z\Omega_i \sqrt{\frac{\Omega_i - \eta_{ij}}{N_i \eta_{ij}}} \tag{3.23}$$

where $z$ is a constant that depends on the desired confidence level (for instance, $z = 1.96$ for a 95% confidence level, according to the *standard normal distribution*). Then, the variance of the estimator $\hat{\eta}_{ij}$ is given by [55]:

$$\sigma^2_{\hat{\eta}_{ij}} = \text{Var}\left(\hat{\eta}_{ij}\right) = \kappa^2 \sigma^2_{\eta_{ij}} + (1 - \kappa)^2 \sigma^2_{\eta_{ji}} \tag{3.24}$$

The value of $\kappa$ that minimizes the variance of the estimator $\hat{\eta}_{ij}$ is given by:

$$\kappa = \frac{\sigma^2_{\eta_{ji}}}{\sigma^2_{\eta_{ij}} + \sigma^2_{\eta_{ji}}} \tag{3.25}$$

Again, for open systems, the enforcement is done only with the surface-to-surface view factors, and the view factors to the environment are then computed by applying the summation rule.

### 3.5.4 Closure enforcement

In general, the view factor matrix computed by MCRT does satisfy the closure relation, as all emitted rays are intercepted by some surface within the system. This ensures that the total number of emitted rays equals the total number of intercepted rays, thereby fulfilling the closure relation. The closure relation can be understood as a manifestation of the conservation of energy or the conservation of the number of rays. However, once reciprocity is enforced, the closure relation may no longer hold, as the adjustments to the view factors do not guarantee that the sum of the view factors equals 1 for each surface. Consequently, it may be necessary to enforce closure after applying reciprocity, even if the original view factor matrix computed by MCRT already satisfies the closure relation.

In what follows, we will present a few methods for enforcing closure in the view factor matrix, which can be applied after enforcing reciprocity.

**Naive and van Leersum enforcement**

The simplest way to enforce closure is to compute the diagonal elements of the view factor matrix as:

$$F_{ii} = 1 - \sum_{j \neq i} F_{ij} \tag{3.26}$$

This method is straightforward, and it maintains the reciprocity relation, as it does not change the off-diagonal elements of the view factor matrix. However, the problem of negative view factors may still arise, as the sum of the view factors may sometimes be greater than 1 after enforcing reciprocity [40]. Again, the iterative method proposed by J. van Leersum [50] may be used instead to ensure non-negativity of the diagonal view factors.

**Least squares smoothing**

Another more sophisticated method for enforcing closure is the *least squares smoothing* method, which is based on minimizing the variations of the view factors while enforcing closure in enclosed systems and maintaining reciprocity [56, 57].

In order to apply this method, we first define an objective function that includes the variations of the initial exchange coefficients $\eta_{ij}$. This objective function will then be minimized:

$$\mathcal{H} = \sum_{i=1}^{N} \sum_{j=1}^{N} \frac{(\eta_{ij} - \hat{\eta}_{ij})^2}{2\omega_{ij}} \tag{3.27}$$

The variation of the exchange coefficients is given by the difference between the initial exchange coefficients $\eta_{ij}$ and the estimated exchange coefficients $\hat{\eta}_{ij}$. The weights $\omega_{ij}$ are used to assign penalties to certain factors. The factor 2 is included for convenience.

Once the objective function is defined, we need to specify the constraints for the minimization problem. These are given by the line sums of the view factor matrix, which must equal 1 for each surface[4]:

$$\begin{cases} g_i = \Omega_i - \sum_{j=1}^{N} \hat{\eta}_{ij} = 0 \\ g_i^* = \Omega_i - \sum_{j=1}^{N} \hat{\eta}_{ji} = 0 \end{cases} \tag{3.28}$$

It is essential to have these two sets of constraints, $g_i$ for the row sums and $g_i^*$ for the column sums, to ensure that the view factors are consistent with the reciprocity relation.

To solve this constrained optimization problem, we will use *Lagrange multipliers* $\lambda_i$. Our objective function will then become[5]:

$$\mathcal{L} = \mathcal{H} + \sum_{i=1}^{N} \lambda_i g_i + \sum_{i=1}^{N} \lambda_i^* g_i^* \tag{3.29}$$

Differentiating $\mathcal{L}$ with respect to $\hat{\eta}_{ij}$ and setting the result to zero[6], we obtain the following system of equations:

$$\frac{\partial \mathcal{L}}{\partial \hat{\eta}_{ij}} = -\frac{\eta_{ij} - \hat{\eta}_{ij}}{\omega_{ij}} - \lambda_i - \lambda_j^* = 0, \quad i,j = 1, \ldots, N \tag{3.30}$$

or, equivalently:

$$\hat{\eta}_{ij} = \eta_{ij} + \omega_{ij}\left(\lambda_i + \lambda_j^*\right), \quad i,j = 1, \ldots, N \tag{3.31}$$

Substituting this expression for $\hat{\eta}_{ij}$ into the constraints $g_i$ and $g_i^*$, we obtain a system of $2N$ equations with $2N$ unknowns, which can be solved to obtain the values of $\lambda_i$ and $\lambda_i^*$:

$$\begin{cases} \Omega_i - \sum_{j=1}^{N} \eta_{ij} = \lambda_i \sum_{j=1}^{N} \omega_{ij} + \sum_{j=1}^{N} \omega_{ij}\lambda_j^* \\ \Omega_i - \sum_{j=1}^{N} \eta_{ji} = \sum_{j=1}^{N} \omega_{ji}\lambda_j + \lambda_i^* \sum_{j=1}^{N} \omega_{ji} \end{cases} \tag{3.32}$$

Once the values of $\lambda_i$ and $\lambda_i^*$ are obtained, $\hat{\eta}_{ij}$ can be computed using **Equation 3.31**.

The system in **Equation 3.32** can be expressed in matrix form $Ax = b$ where $A$ is a

---

[4] Note that the summation rule expressed in terms of the exchange coefficients is given by $\sum_{j=1}^{N} \eta_{ij} = \sum_{j=1}^{N} \Omega_i \mathcal{F}_{ij} = \Omega_i \sum_{j=1}^{N} \mathcal{F}_{ij} = \Omega_i$. Also, due to the reciprocity relation $\eta_{ij} = \eta_{ji}$, we have $\sum_{j=1}^{N} \eta_{ji} = \sum_{j=1}^{N} \eta_{ij} = \Omega_i$.

[5] In [33, 56], the constraints $g_i$ and $g_i^*$ share the same Lagrange multiplier $\lambda_i$, assuming the initial view factors satisfy reciprocity. This approach produces nearly identical results while reducing the matrix system size by a factor of four. For this discussion, the fully constrained form is adopted for generality.

[6] Care should be taken to not mix up the indices of the sums.

$2N \times 2N$ matrix given by[7]:

$$A = \left( \begin{array}{c|c} A_I & A_{II} \\ \hline A_{III} & A_{IV} \end{array} \right) \tag{3.33}$$

$$(A_I)_{ij} = \delta_{ij} \sum_{k=1}^{N} \omega_{ik}, \quad (A_{II})_{ij} = \omega_{ij}, \quad (A_{III})_{ij} = \omega_{ji}, \quad (A_{IV})_{ij} = \delta_{ij} \sum_{k=1}^{N} \omega_{kj} \tag{3.34}$$

and $b$ and $x$ are $2N$-dimensional vectors given by:

$$b = \begin{pmatrix} \Omega_1 - \sum_{j=1}^{N} \eta_{1j} \\ \vdots \\ \Omega_N - \sum_{j=1}^{N} \eta_{Nj} \\ \Omega_1 - \sum_{j=1}^{N} \eta_{j1} \\ \vdots \\ \Omega_N - \sum_{j=1}^{N} \eta_{jN} \end{pmatrix}, \quad x = \begin{pmatrix} \lambda_1 \\ \vdots \\ \lambda_N \\ \lambda_1^* \\ \vdots \\ \lambda_N^* \end{pmatrix} \tag{3.35}$$

Note that no definition of the weights $\omega_{ij}$ has been required for this derivation. In principle, any set of weights can be used, as long as they are symmetric, thus maintaining the reciprocity relation. It was determined in [33] that using the exchange coefficients themselves as weights, i.e., $\omega_{ij} = \eta_{ij}$, leads to the best results in practice, compared to other definitions of $\omega_{ij}$ based on the exchange coefficients and their variances.

This enforcement method is only valid for closed systems and, although it can be adapted to open systems (see next section below), it is done at the cost of losing the reciprocity relation, as the column constraints $g_i^*$ must be relaxed. For this reason, direct use of this method for open systems is not recommended [33].

**Least squares smoothing for open systems**

The least squares smoothing method can be adapted to open systems by suppressing the column constraints $g_i^*$, since we have no information about view factors from deep space to the model surfaces [33]. With this modification, $\mathcal{L}$ becomes:

$$\mathcal{L} = \sum_{i=1}^{N} \sum_{j=1}^{N+1} \frac{(\eta_{ij} - \hat{\eta}_{ij})^2}{2\omega_{ij}} + \sum_{i=1}^{N} \lambda_i g_i \tag{3.36}$$

where the constraint $g_i$ is given by:

$$g_i = \Omega_i - \sum_{j=1}^{N+1} \hat{\eta}_{ij} = 0 \tag{3.37}$$

---

[7] The Kronecker delta $\delta_{ij}$ is defined as $\delta_{ij} = 1$ if $i = j$ and $\delta_{ij} = 0$ otherwise.

Differentiating $\mathcal{L}$ with respect to $\hat{\eta}_{ij}$ and setting the result to zero, we obtain the following system of equations:

$$\frac{\partial \mathcal{L}}{\partial \hat{\eta}_{ij}} = -\frac{\eta_{ij} - \hat{\eta}_{ij}}{\omega_{ij}} - \lambda_i = 0, \quad i, j = 1, \ldots, N \tag{3.38}$$

or, equivalently:

$$\hat{\eta}_{ij} = \eta_{ij} + \omega_{ij}\lambda_i, \quad i, j = 1, \ldots, N \tag{3.39}$$

Substituting this expression into the closure law, $g_i = 0$, and solving for $\lambda_i$, we obtain:

$$\lambda_i = \frac{\Omega_i - \sum_{j=1}^{N+1} \hat{\eta}_{ij}}{\sum_{j=1}^{N+1} \omega_{ij}} \tag{3.40}$$

This method allows for closure enforcement in open systems, however it does not preserve reciprocity.

### 3.5.5 Simultaneous closure and reciprocity enforcement

In this thesis, two original methods for enforcing closure and reciprocity in open systems simultaneously have been developed. The first method is based on a *least squares* approximation, inspired by the *least squares smoothing* methods presented in **Section 3.5.4**, but takes a more geometrical approach. The second method is an iterative scheme combining two individual methods for enforcing reciprocity and closure, respectively.

These developments were carried out specifically for this thesis. While inspired by existing models, as indicated below, they include innovative contributions not found in the literature.

**Least squares optimum for open systems**

The *least squares smoothing* method presented in [48] and reviewed in **Section 3.5.4** enforces closure by projecting MCRT-computed view factors onto the subspace satisfying the closure relation. While this least-squares projection effectively *maintains* prior reciprocity in closed systems, the method itself cannot actively enforce it. Furthermore, this method is limited to closed systems, and a direct extension to open systems fails to preserve reciprocity. This section presents an improved method that extends this projection technique to open systems while also incorporating active reciprocity enforcement.

The problem to solve can be stated as follows: given an $N \times (N + 1)$ view factor matrix $\mathcal{F}$ computed by MCRT, where the last column corresponds to the view factors to the environment, we want to find an estimated view factor matrix $\hat{\mathcal{F}}$ that satisfies the reciprocity and closure relations while deviating as little as possible from the original view factor matrix $\mathcal{F}$.

In terms of the exchange coefficients $\eta_{ij} = \Omega_i \mathcal{F}_{ij}$, the problem can be stated as finding the estimated exchange coefficients matrix $\hat{\eta}$ such that its $i$-th row sums to $\Omega_i$ for each $i = 1, \ldots, N$ and the $N \times N$ portion of $\hat{\eta}$ is symmetric, while deviating as little as possible from the original exchange coefficients matrix $\eta$ computed by MCRT.

This can be achieved by finding the least squares approximation to the solution of a linear system $Ax = b$, where $A$ is the matrix of restrictions, $x$ is the vector of exchange coefficients, and $b$ is the vector of independent terms.

Each row in $A$ will represent one restriction, so $A$ will be an $M \times N(N-1)$ matrix, where $M$ is the number of restrictions. For a system of $N$ surfaces, we will have $N$ closure restrictions and $N(N-1)/2$ reciprocity restrictions, leading to a total of $M = N + N(N-1)/2$ restrictions. The vector $x$ will contain the exchange coefficients in row-major form, i.e., $x = (\eta_{11}, \ldots, \eta_{1N}, \eta_{21}, \ldots, \eta_{2N}, \ldots, \eta_{N1}, \ldots, \eta_{NN})$. Finally, the vector $b$ will contain the independent terms of the restrictions.

Due to the two different kinds of restrictions involved, the matrix $A$ and the vector $b$ will consist of two blocks:

$$A = \begin{pmatrix} R_C \\ R_R \end{pmatrix} \qquad b = \begin{pmatrix} c_C \\ c_R \end{pmatrix} \tag{3.41}$$

where $R_C$ is an $N \times N(N+1)$ matrix:

$$R_C = \begin{pmatrix} \overbrace{1 \cdots 1}^{N+1} & \overbrace{0 \cdots 0}^{N+1} & \cdots & \overbrace{0 \cdots 0}^{N+1} \\ 0 \cdots 0 & 1 \cdots 1 & \cdots & 0 \cdots 0 \\ \vdots & \vdots \; \vdots & \vdots \; \ddots \; \vdots & \vdots \\ 0 \cdots 0 & 0 \cdots 0 & \cdots & 1 \cdots 1 \end{pmatrix} \tag{3.42}$$

and $R_R$ is an $N(N-1)/2 \times N(N+1)$ matrix which, for the case of a system with $N = 4$ surfaces, is given as:

$$R_R = \left( \begin{array}{ccccc|ccccc|ccccc|ccccc} 0 & 1 & 0 & 0 & 0 & -1 & 0 & 0 & 0 & 0 & 0 & 0 & 0 & 0 & 0 & 0 & 0 & 0 & 0 & 0 \\ 0 & 0 & 1 & 0 & 0 & 0 & 0 & 0 & 0 & 0 & -1 & 0 & 0 & 0 & 0 & 0 & 0 & 0 & 0 & 0 \\ 0 & 0 & 0 & 1 & 0 & 0 & 0 & 0 & 0 & 0 & 0 & 0 & 0 & 0 & -1 & 0 & 0 & 0 & 0 & 0 \\ \hline & & & & & 0 & 0 & 1 & 0 & 0 & 0 & -1 & 0 & 0 & 0 & 0 & 0 & 0 & 0 & 0 \\ & & & & & 0 & 0 & 0 & 1 & 0 & 0 & 0 & 0 & 0 & 0 & 0 & -1 & 0 & 0 & 0 \\ \hline & & & & & & & & & & 0 & 0 & 0 & 1 & 0 & 0 & 0 & -1 & 0 & 0 \end{array} \right)$$

$$\tag{3.43}$$

The following procedure is proposed for constructing the matrix $R_R$ in the general case, where N is the number of surfaces in the system and Rr is initially a blank matrix of size $N(N-1)/2 \times N(N+1)$:

```
row ← 0
FOR i ← 0 TO N*(N - 1)/2 DO
    FOR j FROM (i + 1) TO (N - 1) DO
        col_i ← j + (N + 1) * i
        col_j ← i + (N + 1) * j

        Rr[row][col_i] ← 1
        Rr[row][col_j] ← -1

        row ← row + 1
    END FOR
END FOR
```

The vector $c_C$ is of size $N$ and contains the factors $\Omega_i$. On the other hand, the vector $c_R$ is of size $N(N-1)/2$ and contains all zeros.

Once the system of equations $Ax = b$ is constructed, the least squares solution can be obtained. When $A$ is full rank[8], the least squares solution is unique, and it is generally obtained by solving the *normal equation*, where $A^T A$ is also full rank [58]:

$$A^T A \, \hat{x} = A^T b \tag{3.44}$$

This yields the formula:

$$\hat{x} = (A^T A)^{-1} A^T b \tag{3.45}$$

This solution $\hat{x}$ minimizes the error $\|Ax - b\|_2$ when $A$ is full rank. If $A$ is not full rank, the least squares solution is not unique, and it can be factored out into two orthogonal components: the **row space** component $x_r$ (the particular solution, which is the same for all solutions $x$ to the system) and the **null space** component $x_n$ (the homogeneous solution, which varies for different solutions and satisfies $Ax_n = 0$) [59]:

$$x = x_r + x_n \tag{3.46}$$

Then, $Ax = Ax_r + Ax_n = Ax_r = b$, so we are left with a system $Ax_r = b$ for the particular solution $x_r$; along with the freedom to choose any $x_n$ in the null space of $A$.

*Least squares solution*

The exchange coefficient system under consideration will, in general, have infinitely many solutions; in other words, $A$ will not be full rank. This is intuitive: not all systems with $N$ surfaces of equal area but different spatial arrangements will share the same view factor matrix. Consequently, the system $Ax = b$ must be supplemented with information about the geometric configuration of the surfaces in order to fully determine its solution. Moreover, among the possible least squares solutions to the system, we are interested in the one that deviates the least from the original exchange coefficients $\eta_{ij}$ computed by MCRT.

---

[8] In this case, we mean full *column* rank, meaning that the rank is equal to the number of columns.

Thus, we must employ the second least squares solution discussed above, which treats $A$ as rank-deficient, and incorporate geometric information to determine the correct homogeneous solution $x_n$. This geometric information is obtained indirectly from the initial exchange coefficients $\eta_{ij}$ computed by MCRT.

The row space component $x_r$ can be computed as[9]:

$$x_r = A^T(AA^T)^{-1}b \tag{3.47}$$

This is the particular least squares solution to the system and, as well as minimizing the error $\|Ax_r - b\|_2$, it is always the solution of minimum norm. Adding a null space component maintains the error minimization, but it increases the norm of the solution. However, we are not interested in minimizing the norm of the solution, rather we are interested in minimizing the deviation from the original exchange coefficients $\eta_{ij}$ computed by MCRT.

To this end, we will express the null space component $x_n$ as a linear combination of the basis vectors of the null space of $A$:

$$x_n = N_b w \tag{3.48}$$

where $N_b$ is the matrix of basis vectors of the null space of $A$ and $w$ is a vector containing the coefficients of the linear combination [48].

If the exchange coefficients computed by MCRT were exact, we would expect $x_n = x - x_r$, where $x$ is the vector of exchange coefficients computed by MCRT and $x_r$ is the row space component. We would then be able to solve the following system for the coefficients $w$:

$$N_b w = x - x_r \tag{3.49}$$

However, since the exchange coefficients computed by MCRT are not exact, the system in **Equation 3.49** is incompatible, so we resort now to the direct least squares solution from **Equation 3.45**:

$$w = (N_b^T N_b)^{-1} N_b^T (x - x_r) \tag{3.50}$$

---

[9] When $A$ has full row rank, which is usually the case for underdetermined systems, it can be shown that $AA^T$ is invertible. Then, $Ax_r = b \to Ax_r = \mathbb{1}b \to Ax_r = (AA^T)(AA^T)^{-1}b \to x_r = A^T(AA^T)^{-1}b$.

Then:
$$x_n = N_b w = N_b(N_b^T N_b)^{-1} N_b^T (x - x_r) \qquad (3.51)$$

Finally, the least squares solution to the system $Ax = b$ is given by:

$$\hat{x} = x_r + x_n = A^T(AA^T)^{-1}b + N_b(N_b^T N_b)^{-1} N_b^T (x - x_r) \qquad (3.52)$$

*Note on non-negativity*: The least squares solution presented above does not guarantee non-negativity of the estimated exchange coefficients $\hat{\eta}_{ij}$. However, non-negativity can be ensured by combining this method with the simple two-step procedure presented in [48] and outlined in **Appendix A.1**, which we will refer to going forward as *non-negativity rectification* (NNR).

*Note on zero values*: From experiment (see **Section 6.2**), it has been seen that this enforcer, combined with non-negativity, tends to make zero values turn slightly positive. An improvement to this method, named *small positive value avoidance* (SPVA), has also been developed in this work for avoiding this issue, assuming that the zero view factors calculated by MCRT are considered exact. In this case, the zero exchange coefficients can be removed from the problem, avoiding the introduction of small positive values and also reducing the order of the problem. See **Appendix A.2** for the details on SPVA.

**Iterative closure and reciprocity enforcer**

The least squares method presented above is a powerful approach for enforcing closure and reciprocity in open systems, providing a solid theoretical tool to enforce view factor closure and reciprocity in the most general case. However, its main practical limitation is the high computational cost, as it involves working with very large matrices that grow rapidly in size. To address this, an iterative method has been developed that enforces closure and reciprocity without requiring large matrices, making it far more efficient in practice. This method builds on the idea introduced by J. van Leersum to avoid negative view factors in his naive enforcer [50], but uses it instead to enforce both closure and reciprocity simultaneously.

Each iteration of this method comprises two main steps: enforcing closure and enforcing reciprocity. Closure enforcement can be performed using any appropriate method described in **Section 3.5.4**, while reciprocity enforcement can be applied using any appropriate method outlined in **Section 3.5.3**. In this work, closure and reciprocity are enforced using a combination of the *least squares smoothing* (*for open systems*) method and the *fractional variance* method, respectively. Since the fractional variance method is designed for enclosed systems (and does not account for environment view factors), reciprocity is enforced only on the square part of the view factor matrix. The environment view factors are taken from the previous iteration, prior to enforcing closure with the least squares smoothing method. The schematic representation of the iterative enforcer implemented in this work is shown in **Figure 3.3**.

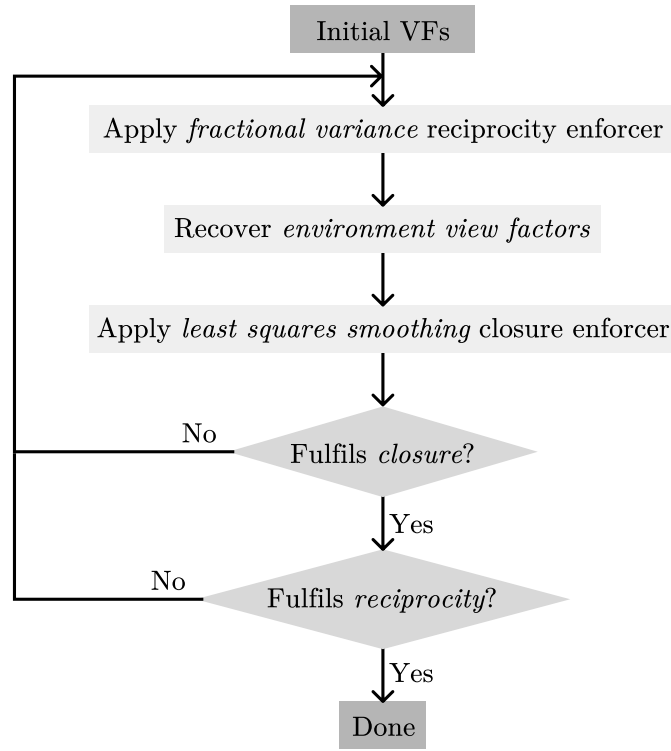

**Figure 3.3:** *Schematic representation of the iterative enforcer method for view factor closure and reciprocity.*

## 3.6 Single-node and multi-node models

An important consideration in thermal simulations is how the system's physical model (e.g., a satellite) is represented and discretized. Models are typically composed of simple flat or curved surfaces forming the system geometry. In *lumped parameter* models, such as those used in this work, calculations are performed on nodes associated with these surfaces.

There are two main ways to relate surfaces and nodes, referred to here as *single-node* and *multi-node* models. In *single-node* models, each surface corresponds to a single node. In *multi-node* models, multiple nodes can be linked to one surface, enabling more detailed representations of geometry and thermal behavior. The main advantages and challenges of *multi-node* models are outlined below.

### 3.6.1 Advantages of multi-node models

*Multi-node* models offer a balance between resolution and computational efficiency, providing the benefits of increased nodal detail without the need to define one surface per node. Compared to models where each surface corresponds to a single node, a *multi-node* representation reduces the total number of surfaces for a given nodal resolution. This leads to smaller view factor matrices, lower memory requirements, and reduced computational overhead, while still enabling fine spatial resolution.

Defining relationships such as shared properties, boundary conditions, or constraints is also more efficient in a *multi-node* framework, as these can be applied at the surface level. In contrast, when using many *single-node* surfaces, such relationships must be managed individually for each surface, increasing modelling complexity and the potential for errors.

From an accuracy standpoint, a single node per surface may fail to capture variations in geometry or radiative properties across a surface, especially for irregular or complex shapes. By associating multiple nodes with a single surface, local variations can be better represented and resolved. *Multi-node* models also enable smoother interpolation of spatially varying properties, such as temperature, emissivity, and heat flux, preserving continuity that is often lost when breaking a surface into multiple independent *single-node* entities.

For radiative interactions, multiple nodes per surface improve the resolution of shadowing effects, occlusions, and energy transfer variations. They also mitigate artificial edge effects introduced when a surface is divided into separate *single-node* surfaces, where discontinuities or unrealistic shading may appear at boundaries. This approach supports the treatment of self-interactions within a surface, which can be important for complex geometries where one part of a surface may exchange radiation with another.

Finally, in large systems, *multi-node* discretization can improve the numerical stability and convergence of the solution. By providing a more accurate distribution of view factors and radiative exchange terms, the conditioning of the resulting system of equations is often enhanced. Iterative solvers, in particular, benefit from this improved geometric representation, which can accelerate convergence and reduce solver instability.

### 3.6.2 Challenges of multi-node models

One drawback of *multi-node* models is the higher computational cost from the increased number of nodes, which may lead to larger matrices and more complex integrations. Nevertheless, for a given nodal resolution, *multi-node* models remain more efficient than an equivalent number of *single-node* surfaces, as fewer total surfaces are required.

A greater challenge lies in handling view factors, particularly when applying closure, reciprocity enforcers, or similar modifications. Maintaining consistency between node-to-node and surface-to-surface view factors is non-trivial, even in *single-node* models. For instance, reciprocity can only be enforced or checked when each "portion" of the model (surface or node) has a clearly defined area—something not straightforward for nodes[10]. What is the area of a node representing a surface potentially exposed on both sides? In *multi-node* models, where nodes may represent only part of a surface, the question becomes even more complex. Clearly, reciprocity is more easily enforced at the surface level, but this requires a method to derive corrected node-to-node view factors from corrected surface-to-surface values, which is challenging when multiple nodes share a surface.

As a result, *multi-node* models require more elaborate processing to maintain physical and mathematical consistency, even though their finer granularity can significantly improve accuracy.

In this work, a multi-node model has been employed for solving the thermal problem, so **Chapter 5** will be dedicated to addressing some of these challenges.

---

[10] The same could be said for optical properties as is said here for area.

# 4
# Modelling Diffuse Reflectivity

## 4.1 Introduction

We discussed the directional nature of light in **Section 2.4.3** and introduced *diffuse* and *specular* reflection in **Section 2.4.10**, noting their key role in radiative heat transfer and the need to model them for accurate simulations. In **Section 3.4.2**, we outlined the challenges of accurately modelling these reflection processes and noted that approximations are often required to limit complexity. For critical applications, such as space thermal control, more precise modelling is essential to ensure reliable thermal predictions. This section presents a widely used approach to modelling diffuse reflectivity which has been implemented in the *Radian* software package as part of this work.

## 4.2 The Gebhart method

There are two main approaches to solving the thermal problem with diffuse reflection. The first uses view factors and direct heat fluxes (DHFs), as in **Section 2.4.12**. The second, simpler in practice, uses *radiative exchange factors* (REFs) and *absorbed heat fluxes*[1] (AHFs). These replace view factors[2] and DHFs[3] in **Equation 3.14**, respectively. The Gebhart method belongs to this second family.

### 4.2.1 Radiative exchange factors

The concept of REF presented in **Definition 39** was first introduced by Benjamin Gebhart in 1957 [60], and applications were discussed more in detail a few years later, also by Gebhart, in [61].

---

[1] The AHFs are the net heat fluxes absorbed by the surfaces of the model, accounting for multi reflections.
[2] That is, GRs are obtained by replacing each view factor by the corresponding REF.
[3] That is, the DHFs (solar, albedo and planetary) are replaced by the corresponding AHF.

> **Definition 39: Radiative Exchange Factor**
>
> The radiative exchange factor[a] (REF) from surface $i$ to surface $j$, denoted $B_{ij}$, is the fraction of the energy leaving surface $i$ that is finally absorbed by surface $j$, after any number of diffuse reflections [60].
>
> ---
> [a] Also commonly known as *Gebhart's Factor*.

Gebhart does not define the emission process, however we will consider only diffuse emission, since that is a reasonable assumption for most problems. In any case, the emission process would be determined by the MCRT, and would not alter this definition.

**Radiative exchange between surfaces**

Radiative exchange between surfaces can now be exactly described using REFs. The net radiation power transferred between two real surfaces $i$ and $j$ is given by:

$$q_{ij} = \sigma \varepsilon_i A_i B_{ij} \left( T_i^4 - T_j^4 \right) \tag{4.1}$$

which is similar to **Equation 3.10**, but now exact. We can also identify the radiative coupling coefficient $GR_{ij}$ as:

$$GR_{ij} = \sigma \varepsilon_i A_i B_{ij} \tag{4.2}$$

**Closure and reciprocity relations**

Analogously to view factors, REFs fulfill the following closure and reciprocity relations:

$$\text{Closure:} \quad \sum_j B_{ij} = 1 \quad \forall i \tag{4.3}$$

$$\text{Reciprocity:} \quad \varepsilon_i A_i B_{ij} = \varepsilon_j A_j B_{ji} \quad \forall i, j \tag{4.4}$$

As with view factors, the closure relation comes from conservation of energy, and the reciprocity relation comes from setting $GR_{ij} = GR_{ji}$ so that $q_{ij} = -q_{ji}$.

### 4.2.2 Computing radiative exchange factors and absorbed heat fluxes

There are two main approaches to computing REFs and AHFs. The first is to obtain them directly from MCRT, by considering diffuse reflectivity in the interaction between rays and surfaces. When a ray intersects a surface, a random process related to the diffuse reflectivity determines if the ray is reflected or absorbed. If the ray is reflected, a random direction is chosen and propagation of the ray continues. In other words, multiple diffuse reflections are incorporated to the MCRT routine directly. The resulting exchange factors are the REFs, and the heat fluxes are the AHFes. The drawback of this method is that it requires many rays to accurately capture the effects of multiple reflections, which becomes increasingly expensive as reflectivity rises[4].

The second method is Gebhart's method, which computes the REFs from the non-diffuse view factors. It avoids the full MCRT, and it is much less computationally demanding. This method is outlined in the sections below.

### 4.2.3 Gebhart's formulation

As mentioned at the beginning of **Section 4.2**, the Gebhart method replaces view factors by radiative exchange factors and DHFs by AHFs in **Equation 3.14**. The rest of the resolution procedure is identical to that of the original problem, except for the fact that we now need to compute the REFs and AHFs.

Gebhart presented in [62] a recursive expression relating the REFs to the view factors, which can be solved by an iterative process or by *Gebhart's matrix method*, which consists of inverting a matrix based on the view factors (see **Appendix B**). This expression is **Equation 4.5**.

$$B_{ij} = \varepsilon_j F_{ij} + \sum_{k=1}^{N} F_{ik} \rho_k B_{kj} \tag{4.5}$$

The first term in the RHS of this equation is the fraction of the energy that leaves surface $i$ and is directly absorbed by surface $j$. The second term accounts for the fraction of energy that leaves surface $i$ which is diffusely reflected by surface $k$ and which is finally absorbed by surface $j$. The surfaces $k$ represent all possible intermediary surfaces.

Gebhart's formulation is originally valid only for closed systems. A simple extension of the formulation to open systems, as well as the necessary adaptions to *Gebhart's matrix method*, have been developed in this work and are presented in **Appendix B.2**. This extension has not been found in the current literature.

Closure and reciprocity can be enforced on REFs using the same methods as in **Section 3.5.3** and **Section 3.5.4**, simply by setting $\Omega_i = \varepsilon_i A_i$ instead of $\Omega_i = A_i$.

Gebhart also presented in [62] a simple formula to obtain the AHFs from the REFs and the DHFs:

$$Q_i^{(a)} = A_i \varepsilon_i Q_i^{(d)} + \sum_{j=1}^{N} B_{ji} A_j \rho_j Q_j^{(d)} \tag{4.6}$$

Here, the superscript $(a)$ has been used to denote *absorbed* heat flux, and the superscript $(d)$ has been used to denote *direct* heat flux. The first term in the RHS of this equation is the portion of the heat flux that is directly incident on surface *i* that is absorbed. The second term is the heat flux that is diffusely reflected by each intermediary surface *j* and finally absorbed by surface *i*.

---
[4] Higher reflectivity increases the number of ray reflections.

Note that the REFs depend on the optical properties of the surfaces, as well as on the view factors. This makes them also dependent on the wavelength of the considered radiation. For example, different REFs would be required when considering the solar heat flux (visible range) and the planetary heat flux (IR range), as the optical properties in these two ranges differ. For this reason, and for others discussed in [33], it has been chosen not to use AHFs in this work, and to use only REFs instead. Using AHFs is left for future work. **Figure 4.1** summarizes the sequence of steps used in this work for solving the thermal problem with diffuse reflectivity.

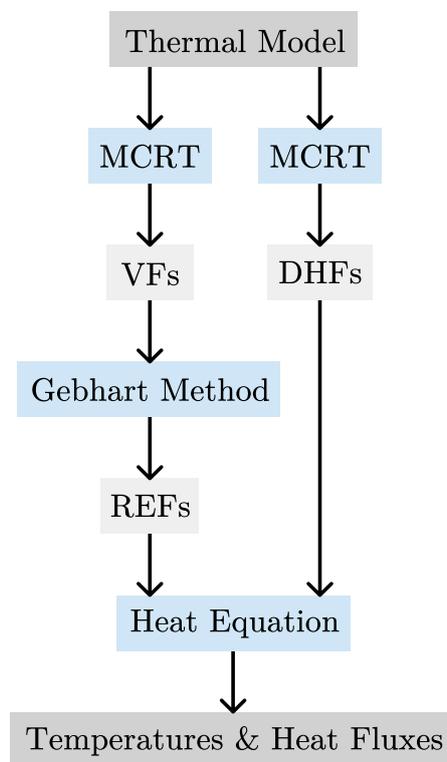

**Figure 4.1:** *Proposed sequence of steps to follow for solving the thermal problem with diffuse reflectivity.*

# 5
# Multi-Node Surface Model Relations

## 5.1 Introduction

In **Section 3.6**, we compared single-node and multi-node models, highlighting the advantages and challenges of the latter, which are used in this work. Here, we present an original method for efficiently handling and switching between different levels of subdivision in the multi-node surface model. The concept of *view fractions*, necessary for averaging optical properties among the two faces of a node, will also be introduced. This method is essential for applying view factor corrections—including enforcement, and even computation of REFs, which can be interpreted as a correction to view factors.

The developments in this chapter are fully original and have not been previously published or suggested in the literature.

## 5.2 Derivation

To derive the relations between view factors at different subdivision levels, we start with a single pair of multi-node surfaces and extend the result to all pairs.

For example, if a geometrical model of a system has $T$ surfaces, then we can look, in particular, at a pair of surfaces $n$ and $m$, like those on **Figure 5.1**, that have $N_n$ and $N_m$ sub-surfaces, respectively. Sub-surfaces are the areas around each node in the surface. A surface can have multiple nodes, but there is only one node per sub-surface. We will use this simple model as a basis for the derivations that follow.

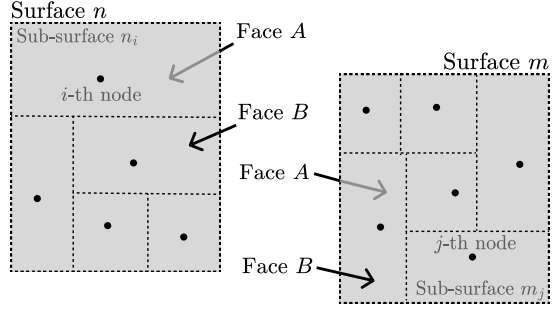

**Figure 5.1:** *Surfaces n and m, each with their sub-surfaces indicated.*

### 5.2.1 Assumptions

We make the following assumptions, which are necessary but do not limit the generality of the results in most cases:

1. Heat is emitted uniformly in each face of surfaces *n* and *m*. This is a reasonable assumption since, in the ray tracing calculation of the view factors, rays have been emitted from random positions on each surface.
2. If an arbitrary surface has two faces *A* and *B*, the heat emitted by face *A* of the surface is equal to the heat emitted by face *B* of the same surface. This is a reasonable assumption as long as, in ray tracing calculation of the view factors, the same number of rays is emitted from each face of all surfaces that have two faces.
3. If an arbitrary emitting surface has only one face, it is always labelled as face *A*, and it emits all the heat emitted by the surface. This is a logical assumption, as rays will only be emitted by faces of the surface that exist.

### 5.2.2 Notation

We denote $F_{xy}^{X \to Y}$ as the view factor from face *X* of surface *x* to face *Y* of surface *y*, where *X* and *Y* can take the values *A* and *B* and where *x* and *y* can take the values of the name of any sub-surface or surface. For example, we denote $F_{n_i m_j}^{A \to B}$ as the view factor from face *A* of sub-surface $n_i$ to face *B* of sub-surface $m_j$; and we denote the view factor $F_{nm}^{B \to B}$ as the view factor from face *B* of surface *n* to face *B* of surface *m*.

We denote $F_{ij}$ as the view factor from node *i* to node *j*, that is, the nodes of sub-surfaces $n_i$ and $m_j$.

Analogous notation will be used for heat, where we will use $H_{xy}^{X \to Y}$ and $H_{ij}$. The notation $H_i$ will be used to denote the total heat emitted by a node *i*, which is the same as the total heat emitted by sub-surface $n_i$, denoted $H_{n_i}$. The total heat emitted by a full surface *n* will be denoted $H_n$.

We denote $f_{n_i m_j}^{X \to Y}$, as the fraction of the total heat emitted from sub-surface $n_i$ that arrives at sub-surface $m_j$ that was emitted from face *X* of the source sub-surface and received at face $Y$ of the target surface.

### 5.2.3 View factors of surfaces and view factors of sub-surfaces

In this section, we will derive the relation between the view factor of a full surface and the view factors of its constituent sub-surfaces. From the definition of view factor, we can write the following expression of the view factor between face $X$ of surface $n$ and face $Y$ of surface $m$:

$$F_{nm}^{X \to Y} = \frac{\text{Heat from surface } n \text{ face } X \text{ to surface } m \text{ face } Y}{\text{Heat from surface } n \text{ face } X} = \frac{H_{nm}^{X \to Y}}{H_n^X} \tag{5.1}$$

From the same definition, we can also write the expression of the view factor between face $X$ of sub-surface $n_i$ and face $Y$ of sub-surface $Y$:

$$F_{n_i m_j}^{X \to Y} = \frac{\text{Heat from surface } n_i \text{ face } X \text{ to surface } m_j \text{ face } Y}{\text{Heat from surface } n_i \text{ face } X} = \frac{H_{n_i m_j}^{X \to Y}}{H_{n_i}^X} \tag{5.2}$$

From **Equation 5.2**, we obtain that:

$$H_{n_i m_j}^{X \to Y} = H_{n_i}^X F_{n_i m_j}^{X \to Y} \tag{5.3}$$

If we now apply conservation of energy, we can express the total heat going from face $X$ of surface $n$ to face $Y$ of surface $m$ as:

$$H_{nm}^{X \to Y} = \sum_{i=1}^{N_n} H_{n_i m}^{X \to Y} = \sum_{i=1}^{N_n} \sum_{j=1}^{N_m} H_{n_i m_j}^{X \to Y} \tag{5.4}$$

Then, combining this with **Equation 5.1** and **Equation 5.3**, we obtain:

$$F_{nm}^{X \to Y} = \frac{H_{nm}^{X \to Y}}{H_n^X} = \sum_{i=1}^{N_n} \sum_{j=1}^{N_m} \frac{H_{n_i m_j}^{X \to Y}}{H_n^X} = \sum_{i=1}^{N_n} \sum_{j=1}^{N_m} \frac{H_{n_i}^X F_{n_i m_j}^{X \to Y}}{H_n^X} \tag{5.5}$$

Here, the first assumption is used to find that $H_{n_i}^X / H_n^X = S_{n_i} / S_n$, where $S_{n_i}$ is the area of sub-surface $n_i$ and $S_n$ is the area of the full surface $n$. Then:

$$F_{nm}^{X \to Y} = \sum_{i=1}^{N_n} \sum_{j=1}^{N_m} \left( F_{n_i m_j}^{X \to Y} \cdot \frac{S_{n_i}}{S_n} \right) \tag{5.6}$$

For view factors to deep space, we can set $m_j = \infty$ and use the fact that there is only ever a single deep space node; and it does not have faces:

$$F_{n\infty}^{X\to\infty} = \sum_{i=1}^{N_n} \left( F_{n_i\infty}^{X\to\infty} \cdot \frac{S_{n_i}}{S_n} \right) \tag{5.7}$$

### 5.2.4 View factors of nodes and view factors of sub-surfaces

In this section, we will derive the relation between the view factor of a node and the view factors of its corresponding sub-surfaces (faces $A$ and $B$). From the definition of view factor, we have that the view factor between nodes $i$ and $j$ is:

$$F_{ij} = \frac{\text{Heat from node } i \text{ to node } j}{\text{Heat from node } i} = \frac{H_{ij}}{H_i} \tag{5.8}$$

From conservation of energy, we have:

$$H_{ij} = \sum_{\substack{X,Y \\ \in \{A,B\}}} H_{n_i m_j}^{X\to Y} = H_{n_i m_j}^{A\to A} + H_{n_i m_j}^{A\to B} + H_{n_i m_j}^{B\to A} + H_{n_i m_j}^{B\to B} \tag{5.9}$$

Using **Equation 5.3**, we obtain:

$$\begin{aligned} H_{ij} &= H_{n_i}^A F_{n_i m_j}^{A\to A} + H_{n_i}^A F_{n_i m_j}^{A\to B} + H_{n_i}^B F_{n_i m_j}^{B\to A} + H_{n_i}^B F_{n_i m_j}^{B\to B} \\ &= H_{n_i}^A \left( F_{n_i m_j}^{A\to A} + F_{n_i m_j}^{A\to B} \right) + H_{n_i}^B \left( F_{n_i m_j}^{B\to A} + F_{n_i m_j}^{B\to B} \right) \end{aligned} \tag{5.10}$$

To proceed, we need to consider two possible cases, depending on whether the source face has one or two existing faces ($A$ and $B$ or just $A$).

**Source surfaces with two faces**

If the source surface has two faces $A$ and $B$, the second assumption is used to find that $H_{n_i}^A = H_{n_i}^B = H_{n_i}/2 = H_i/2$, where $H_{n_i}$ is the heat emitted by surface $n_i$ (considering both faces), which is equal to $H_i$, the heat emitted by the corresponding node. Then:

$$H_{ij} = \frac{H_i}{2} \left( F_{n_i m_j}^{A\to A} + F_{n_i m_j}^{A\to B} + F_{n_i m_j}^{B\to A} + F_{n_i m_j}^{B\to B} \right) \tag{5.11}$$

This can be combined with **Equation 5.8** to obtain:

$$F_{ij} = \frac{1}{2} \left( F_{n_i m_j}^{A\to A} + F_{n_i m_j}^{A\to B} + F_{n_i m_j}^{B\to A} + F_{n_i m_j}^{B\to B} \right) \tag{5.12}$$

For the special case of view factors to deep space, we have $H_{i\infty} = H_{n_i\infty}^{A\to\infty} + H_{n_i\infty}^{B\to\infty}$, so:

$$F_{i\infty} = \frac{1}{2}\left(F_{n_i\infty}^{A\to\infty} + F_{n_i\infty}^{B\to\infty}\right) \tag{5.13}$$

**Source surfaces with one face**

If the source surface has only one face, $A$, the third assumption is used to find that $H_{n_i}^B = 0$, so $H_{n_i}^A = H_{n_i} = H_i$, where $H_{n_i}$ is the heat emitted by surface $n_i$, which is equal to $H_i$, the heat emitted by the corresponding node. Also, it is clear that $F_{n_im_j}^{B\to Y} = 0$ for all $Y$ and for all $m_j$, and also $F_{n_i\infty}^{B\to\infty} = 0$. Then:

$$H_{ij} = H_i \left(F_{n_im_j}^{A\to A} + F_{n_im_j}^{A\to B}\right) \tag{5.14}$$

This can be combined with **Equation 5.8** to obtain:

$$F_{ij} = F_{n_im_j}^{A\to A} + F_{n_im_j}^{A\to B} \tag{5.15}$$

For the special case of view factors to deep space when we only have face $A$, we have $H_{i\infty} = H_{n_i\infty}^{A\to\infty}$, so:

$$F_{i\infty} = F_{n_i\infty}^{A\to\infty} \tag{5.16}$$

**General case**

We can combine the results of the previous two sections to obtain a general result, applicable to source faces with either one or two faces. This is done by defining a new parameter, $C_n$, which is equal to the number of faces of surface $n$ (and of sub-surfaces $n_i$, $\forall i$). Then, we can write these general expressions:

$$\begin{cases} F_{ij} = \dfrac{1}{C_n}\left(F_{n_im_j}^{A\to A} + F_{n_im_j}^{A\to B} + F_{n_im_j}^{B\to A} + F_{n_im_j}^{B\to B}\right) \\[2mm] F_{i\infty} = \dfrac{1}{C_n}\left(F_{n_i\infty}^{A\to\infty} + F_{n_i\infty}^{B\to\infty}\right) \end{cases} \tag{5.17}$$

Since $C_n = 1$ and $F_{n_im_j}^{B\to Y} = F_{n_i\infty}^{B\to\infty} = 0$ for all $Y$, $m_j$ and $i$ when the source surface $n$ has only one face, then under such conditions these expressions reduce to **Equation 5.15** and **Equation 5.16**.

### 5.2.5 View fractions and view factors of sub-surfaces

Many times, nodes have two associated faces, which may not have the same optical properties, such as emissivity. This poses an obvious question: which emissivity should I use when performing radiative calculations on a node with two optically different faces? The answer is that a weighed average of the two emissivities should be used. As proven in **Appendix C**, it turns out that the weights are obtained precisely using the view fractions that we are about to discuss.

According to the definition from **Section 5.2.2**, we can express $f_{n_i m_j}^{X \to Y}$ as:

$$f_{n_i m_j}^{X \to Y} = \frac{\text{Heat emitted from face } X \text{ of surface } n_i \text{ that arrives at face } Y \text{ of surface } m_j}{\text{Total heat from surface } n_i \text{ that arrives at surface } m_j}$$

$$= \frac{H_{n_i m_j}^{X \to Y}}{H_{n_i m_j}} \tag{5.18}$$

Again, to proceed we need to distinguish between nodes with one and nodes with two faces.

**Source surfaces with two faces**

Using **Equation 5.3**, and using the second assumption as before, we can write:

$$H_{n_i m_j}^{X \to Y} = H_{n_i}^X F_{n_i m_j}^{X \to Y} = \frac{H_{n_i}}{2} \cdot F_{n_i m_j}^{X \to Y} \tag{5.19}$$

Then, we have:

$$f_{n_i m_j}^{X \to Y} = \frac{\frac{H_{n_i}}{2} \cdot F_{n_i m_j}^{X \to Y}}{H_{n_i m_j}} = \frac{H_i \cdot F_{n_i m_j}^{X \to Y}}{2 H_{ij}} \tag{5.20}$$

where we have used that the heat transfer from two sub-surfaces is the same as the heat transfer between their two respective nodes, so $H_{n_i} = H_i$ and $H_{n_i m_j} = H_{ij}$. Noting from **Equation 5.8** that $H_{ij} = H_i F_{ij}$, we finally obtain:

$$f_{n_i m_j}^{X \to Y} = \frac{F_{n_i m_j}^{X \to Y}}{2 F_{ij}} \tag{5.21}$$

Similarly, for the special case of deep space, we have:

$$f_{n_i \infty}^{X \to \infty} = \frac{\text{Heat emitted from face } X \text{ of surface } n_i \text{ to deep space}}{\text{Total heat from surface } n_i \text{ that goes to deep space}} = \frac{H_{n_i \infty}^{X \to \infty}}{H_{n_i \infty}}$$

$$= \frac{H_{n_i}^X F_{n_i \infty}^{X \to \infty}}{H_{i \infty}} = \frac{(H_{n_i}/2) \cdot F_{n_i \infty}^{X \to \infty}}{H_i F_{i \infty}} = \frac{F_{n_i \infty}^{X \to \infty}}{2 F_{i \infty}} \tag{5.22}$$

**Source surfaces with one face**

When there is only one face for source sub-surface $n_i$, clearly $f_{n_i m_j}^{B \to Y} = f_{n_i \infty}^{B \to \infty} = F_{n_i m_j}^{B \to Y} = F_{n_i \infty}^{B \to \infty} = 0$ for any $Y$ and for any $m_j$. Using **Equation 5.3**, and using the third assumption as before, we can write:

$$H_{n_i m_j}^{A \to Y} = H_{n_i}^A F_{n_i m_j}^{A \to Y} = H_{n_i} \cdot F_{n_i m_j}^{A \to Y} \tag{5.23}$$

Then, we have:

$$f_{n_i m_j}^{A \to Y} = \frac{H_{n_i} \cdot F_{n_i m_j}^{A \to Y}}{H_{n_i m_j}} = \frac{H_i \cdot F_{n_i m_j}^{A \to Y}}{H_{ij}} \tag{5.24}$$

where, again, we have used that the heat transfer from two sub-surfaces is the same as the heat transfer between their two respective nodes, so $H_{n_i} = H_i$ and $H_{n_i m_j} = H_{ij}$. Noting from **Equation 5.8** that $H_{ij} = H_i F_{ij}$, we finally obtain:

$$f_{n_i m_j}^{A \to Y} = \frac{F_{n_i m_j}^{A \to Y}}{F_{ij}} \tag{5.25}$$

Similarly, for the special case of deep space, we have:

$$f_{n_i \infty}^{A \to \infty} = \frac{\text{Heat emitted from face } A \text{ of surface } n_i \text{ to deep space}}{\text{Total heat from surface } n_i \text{ that goes to deep space}} = \frac{H_{n_i \infty}^{A \to \infty}}{H_{n_i \infty}}$$
$$= \frac{H_{n_i}^A F_{n_i \infty}^{A \to \infty}}{H_{i \infty}} = \frac{H_{n_i} F_{n_i \infty}^{A \to \infty}}{H_i F_{i \infty}} = \frac{F_{n_i \infty}^{A \to \infty}}{F_{i \infty}} \tag{5.26}$$

**General case**

As we can see from the previous sections, and using the expressions in **Equation 5.17** to express the fractions as a function of only the sub-surface view factors, we obtain the following relations that are valid independently of the number of faces that the source surface has:

$$f_{n_i m_j}^{X \to Y} = \frac{F_{n_i m_j}^{X \to Y}}{F_{n_i m_j}^{A \to A} + F_{n_i m_j}^{A \to B} + F_{n_i m_j}^{B \to A} + F_{n_i m_j}^{B \to B}} \quad \text{and} \quad f_{n_i \infty}^{X \to \infty} = \frac{F_{n_i \infty}^{X \to \infty}}{F_{n_i \infty}^{A \to \infty} + F_{n_i \infty}^{B \to \infty}} \tag{5.27}$$

If we count on $F_{n_i m_j}^{B \to Y}$ being zero $\forall Y$ when the source sub-surface $n_i$ has only one face (face $A$), then these expressions are valid for surfaces with one or two faces.

### 5.2.6 Multi-node surface model relations

In the previous sections, we derived the relations in **Equation 5.28** and **Equation 5.29**, where $C_n$ is the number of faces of surface $n$ (and $n_i$, $\forall i$), and $S_{n_i}$ and $S_n$ are the areas of the sub-surfaces and the surfaces, respectively. These expressions allow us to obtain the surface view factors and the node view factors from a common set of sub-surface view factors, which can be computed by MCRT. Any enforcement or correction, including the computation of REFs, can be applied at the sub-surface level, and then the surface and node view factors can be computed from the corrected set of sub-surface view factors.

$$\begin{cases} F_{nm}^{X \to Y} = \sum_{i=1}^{N_n} \sum_{j=1}^{N_m} \left( F_{n_i m_j}^{X \to Y} \cdot \dfrac{S_{n_i}}{S_n} \right) & \text{for } X, Y \in \{A, B\} \\[1em] F_{n\infty}^{X \to \infty} = \sum_{i=1}^{N_n} \left( F_{n_i \infty}^{X \to \infty} \cdot \dfrac{S_{n_i}}{S_n} \right) & \text{for } X \in \{A, B\} \\[1em] C_n F_{ij} = F_{n_i m_j}^{A \to A} + F_{n_i m_j}^{A \to B} + F_{n_i m_j}^{B \to A} + F_{n_i m_j}^{B \to B} \\[1em] C_n F_{i\infty} = F_{n_i \infty}^{A \to \infty} + F_{n_i \infty}^{B \to \infty} \end{cases} \quad (5.28)$$

$$\begin{cases} f_{n_i m_j}^{X \to Y} = \dfrac{F_{n_i m_j}^{X \to Y}}{F_{n_i m_j}^{A \to A} + F_{n_i m_j}^{A \to B} + F_{n_i m_j}^{B \to A} + F_{n_i m_j}^{B \to B}} & \text{for } X, Y \in \{A, B\} \\[1em] f_{n_i \infty}^{X \to \infty} = \dfrac{F_{n_i \infty}^{X \to \infty}}{F_{n_i \infty}^{A \to \infty} + F_{n_i \infty}^{B \to \infty}} & \text{for } X \in \{A, B\} \end{cases} \quad (5.29)$$

Note that these expressions for the node view factors $F_{ij}$ and $F_{i\infty}$ account for both single-face and two-face surfaces. In case the surface $n$ only has one face (face $A$), all factors $F_{n_i m_j}^{B \to A}$, $F_{n_i m_j}^{B \to B}$ and $F_{n_i m_j}^{B \to \infty}$ for all sub-surfaces $n_i$ and $m_j$ are equal to zero, and also $C_n = 1$, so we recover **Equation 5.15** and **Equation 5.16**.

## 5.3 Analysis

In this section, we will analyze the way in which the relations presented above modify or preserve closure and reciprocity.

### 5.3.1 Preservation of closure

To see how the multi-node surface relations affect closure, we will start by assuming that closure has been enforced on the initial sub-surface view factors $F_{n_i m_j}^{X \to Y}$ computed

by MCRT. This means:

$$F_{n_i\infty}^{X\to\infty} + \sum_{\substack{Y\in \\ \{A,B\}}} \sum_m \sum_{j=1}^{N_m} F_{n_i m_j}^{X\to Y} = 1, \quad \forall i, \quad X \in \{A, B\} \tag{5.30}$$

Form this, we want to obtain the closure relation for the surfaces and nodes view factors. For surface $n$:

$$\begin{aligned}
\sum_{\substack{Y\in \\ \{A,B\}}} \sum_m F_{nm}^{X\to Y} &= \sum_{\substack{Y\in \\ \{A,B\}}} \sum_m \sum_{i=1}^{N_n} \sum_{j=1}^{N_m} \left( F_{n_i m_j}^{X\to Y} \cdot \frac{S_{n_i}}{S_n} \right) \\
&= \sum_{i=1}^{N_n} \sum_{\substack{Y\in \\ \{A,B\}}} \sum_m \sum_{j=1}^{N_m} \left( F_{n_i m_j}^{X\to Y} \cdot \frac{S_{n_i}}{S_n} \right) \\
&= \frac{1}{S_n} \sum_{i=1}^{N_n} \left( S_{n_i} \sum_{\substack{Y\in \\ \{A,B\}}} \sum_m \sum_{j=1}^{N_m} F_{n_i m_j}^{X\to Y} \right) \\
&= \frac{1}{S_n} \sum_{i=1}^{N_n} S_{n_i} \left( 1 - F_{n_i\infty}^{X\to\infty} \right) \\
&= \frac{1}{S_n} \cdot \left( S_n - \sum_{i=1}^{N_n} S_{n_i} F_{n_i\infty}^{X\to\infty} \right) \\
&= 1 - \frac{1}{S_n} \cdot \sum_{i=1}^{N_n} S_{n_i} F_{n_i\infty}^{X\to\infty} = 1 - F_{n\infty}^{X\to\infty}
\end{aligned} \tag{5.31}$$

Considering the view factors to space:

$$F_{n\infty}^{X\to\infty} + \sum_{\substack{Y\in \\ \{A,B\}}} \sum_m F_{nm}^{X\to Y} = 1 \tag{5.32}$$

Therefore, the *closure is preserved for surfaces*. Applying now the closure to node $i$, which corresponds to sub-surface $n_i$ of surface $n$:

$$
\begin{aligned}
F_{n\infty}^{X\to\infty} + \sum_j F_{ij} &= \frac{F_{n_i\infty}^{A\to\infty} + F_{n_i\infty}^{B\to\infty}}{C_n} + \sum_{\substack{X\in\\\{A,B\}}}\sum_{\substack{Y\in\\\{A,B\}}}\sum_m \sum_{j=1}^{N_m} \frac{F_{n_i m_j}^{X\to Y}}{C_n} \\
&= \frac{1}{C_n}\left(F_{n_i\infty}^{A\to\infty} + F_{n_i\infty}^{B\to\infty} + \sum_{\substack{X\in\\\{A,B\}}}\sum_{\substack{Y\in\\\{A,B\}}}\sum_m \sum_{j=1}^{N_m} F_{n_i m_j}^{X\to Y}\right) \\
&= \frac{1}{C_n}\sum_{\substack{X\in\\\{A,B\}}}\left(F_{n_i\infty}^{X\to\infty} + \sum_{\substack{Y\in\\\{A,B\}}}\sum_m \sum_{j=1}^{N_m} F_{n_i m_j}^{X\to Y}\right) \\
&= \frac{1}{C_n}\sum_{\substack{X\in\\\{A,B\}}} 1 = \frac{1}{C_n}\cdot C_n = 1
\end{aligned}
\tag{5.33}
$$

Therefore, the *closure is preserved for nodes also*.

### 5.3.2 Preservation of reciprocity

We proceed similarly to closure, by assuming that reciprocity has been enforced on the initial sub-surface view factors $F_{n_i m_j}^{X\to Y}$ computed by MCRT. This means:

$$
S_{n_i} F_{n_i m_j}^{X\to Y} = S_{m_j} F_{m_j n_i}^{Y\to X}, \quad \forall n, m, i, j, \quad X, Y \in \{A, B\} \tag{5.34}
$$

Considering surfaces $n$ and $m$:

$$
\begin{aligned}
S_n F_{nm}^{X\to Y} &= S_n \sum_{i=1}^{N_n}\sum_{j=1}^{N_m}\left(F_{n_i m_j}^{X\to Y}\cdot \frac{S_{n_i}}{S_n}\right) \\
&= \sum_{i=1}^{N_n}\sum_{j=1}^{N_m}\left(F_{n_i m_j}^{X\to Y} S_{n_i}\right) \\
&= \frac{S_m}{S_m}\sum_{j=1}^{N_m}\sum_{i=1}^{N_n}\left(F_{m_j n_i}^{Y\to X} S_{m_j}\right) \\
&= S_m \sum_{j=1}^{N_m}\sum_{i=1}^{N_n}\left(F_{m_j n_i}^{Y\to X}\cdot \frac{S_{m_j}}{S_m}\right) = S_m F_{mn}^{Y\to X}
\end{aligned}
\tag{5.35}
$$

So, *reciprocity is preserved for surfaces*. Considering node $i$ of sub-surface $n_i$ and node $j$

of sub-surface $m_j$:

$$
\begin{aligned}
C_n S_{n_i} F_{ij} &= C_n S_{n_i} \cdot \frac{1}{C_n} \sum_{\substack{X \in \\ \{A,B\}}} \sum_{\substack{Y \in \\ \{A,B\}}} \sum_{m} \sum_{j=1}^{N_m} F_{n_i m_j}^{X \to Y} \\
&= \sum_{\substack{X \in \\ \{A,B\}}} \sum_{\substack{Y \in \\ \{A,B\}}} \sum_{m} \sum_{j=1}^{N_m} S_{n_i} F_{n_i m_j}^{X \to Y} \\
&= \sum_{\substack{X \in \\ \{A,B\}}} \sum_{\substack{Y \in \\ \{A,B\}}} \sum_{m} \sum_{j=1}^{N_m} S_{m_j} F_{m_j n_i}^{Y \to X} \\
&= C_m S_{m_j} \cdot \frac{1}{C_m} \sum_{\substack{X \in \\ \{A,B\}}} \sum_{\substack{Y \in \\ \{A,B\}}} \sum_{m} \sum_{j=1}^{N_m} F_{m_j n_i}^{Y \to X} = C_m S_{m_j} F_{ji}
\end{aligned}
\tag{5.36}
$$

Finally, the *reciprocity is preserved for nodes also*. Note that the area of a node used in the reciprocity relation (and in all other computations) is the total area of its sub-surface, accounting for both faces when present.

## 5.4 Comments and alternative approaches

In the previous sections, we have developed some expressions to work in multi-surface models, and we have proven the following important facts.

1. Closure is preserved between surfaces and sub-surfaces (**Section 5.3.1**).
2. Closure is preserved between nodes and sub-surfaces (**Section 5.3.1**).
3. Reciprocity is preserved between surfaces and sub-surfaces (**Section 5.3.2**).
4. Reciprocity is preserved between nodes (for which it is written as $C_n S_{n_i} F_{ij} = C_m S_{m_j} F_{ji}$) and sub-surfaces (**Section 5.3.2**).
5. Optical properties of nodes can be obtained by averaging the optical properties of the two faces of their corresponding sub-surfaces, by means of the view fractions (**Appendix C**).

With these facts—especially the last two—we can revisit the challenges of multi-node surface models discussed in **Section 3.6**. For instance, the question of the area of a node representing a surface exposed on both sides can be resolved using the fourth point: the area of a node $i$ can be taken as its total exposed area, $C_i S_i$, for radiative and REF computations as well as for enforcement. This result applies to both single- and multi-node surface models. Consequently, reciprocity (and closure, trivially) could be enforced directly on node view factors without resorting to sub-surface view factors. Incorporating our knowledge obtained in the fifth point about node optical properties, REFs could even be computed directly on nodes using view fractions. This would simplify calculations, although in some cases sub-surface view factors may still be preferred— for example, to avoid duplicate enforcement at both surface and node levels.

Moreover, the multi-node surface model relations provide a conceptually powerful framework for related tasks. For example, the author has applied it to compare view factor (and REF) results computed in *Radian* with those computed in *ESATAN-TMS*, where nodes are grouped differently, preventing direct comparison.

# 6
# Applied Examples

## 6.1 Introduction

As most developments in this work were implemented in the commercial software package *Radian* for space thermal analysis, the proprietary code cannot be disclosed. Instead, this section presents simple results obtained with the software to demonstrate the usefulness of the developed methods.

## 6.2 Enforcers

In this section, both the least squares and the iterative closure and reciprocity enforcers for open systems developed in this work will be demonstrated with the simple geometrical model shown in **Figure 6.1**. Some of the other enforcers discussed in **Section 3.5.3** and **Section 3.5.4** have also been implemented in *Radian*, so they will be demonstrated as well.

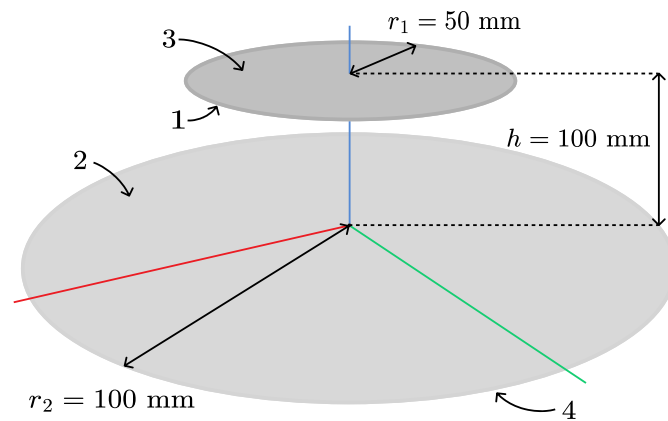

**Figure 6.1:** *Geometrical model used to demonstrate the enforcers.*

The model consists of two parallel discs of different radii $r_1 = 50$ mm and $r_2 = 100$ mm, separated by a distance of $h = 100$ mm. This model has been chosen because the analytical expressions for the view factors between the discs are known, allowing for good validation of the numerical results. The analytical expression for the view factor from disc 1 to disc 2 is [63]:

$$F_{12} = \frac{1}{2}\left(X - \sqrt{X^2 - 4\left(\frac{R_2}{R_1}\right)^2}\right) \tag{6.1}$$

where

$$X = 1 + \frac{1 + R_2^2}{R_1^2}, \quad R_1 = \frac{r_1}{h}, \quad R_2 = \frac{r_2}{h} \tag{6.2}$$

A similar expression can be found for the view factor $F_{21}$. Using these formulas we can compute the complete analytical view factor matrix for this open model, considering that surfaces 3 and 4 only see the environment:

$$F_{\text{analytical}} = \begin{pmatrix} F_{11} & F_{12} & F_{13} & F_{14} & F_{1\infty} \\ F_{21} & F_{22} & F_{23} & F_{24} & F_{2\infty} \\ F_{31} & F_{32} & F_{33} & F_{34} & F_{3\infty} \\ F_{41} & F_{42} & F_{43} & F_{44} & F_{4\infty} \end{pmatrix}$$
$$\approx \begin{pmatrix} 0 & 0.763932 & 0 & 0 & 0.236068 \\ 0.190983 & 0 & 0 & 0 & 0.809017 \\ 0 & 0 & 0 & 0 & 1 \\ 0 & 0 & 0 & 0 & 1 \end{pmatrix} \tag{6.3}$$

To demonstrate the enforcers, the view factor matrix $F$ of the system has been obtained using the software *Radian*, with an MCRT routine using $10^7$ rays. The following corrections have then been applied:

1. No correction.

$$F^{(1)} = \begin{pmatrix} 0 & 0.764011 & 0 & 0 & 0.235989 \\ 0.190823 & 0 & 0 & 0 & 0.809177 \\ 0 & 0 & 0 & 0 & 1 \\ 0 & 0 & 0 & 0 & 1 \end{pmatrix} \tag{6.4}$$

2. Least squares smoothing closure enforcer for open systems (**Section 3.5.4**).

$$F^{(2)} = \begin{pmatrix} 0 & 0.764011 & 0 & 0 & 0.235989 \\ 0.190823 & 0 & 0 & 0 & 0.809177 \\ 0 & 0 & 0 & 0 & 1 \\ 0 & 0 & 0 & 0 & 1 \end{pmatrix} \tag{6.5}$$

3. Fractional variance reciprocity enforcer (**Section 3.5.3**).

$$F^{(3)} = \begin{pmatrix} 0 & 0.763651 & 0 & 0 & 0.235989 \\ 0.190913 & 0 & 0 & 0 & 0.809177 \\ 0 & 0 & 0 & 0 & 1 \\ 0 & 0 & 0 & 0 & 1 \end{pmatrix} \quad (6.6)$$

4. Least squares optimum for open systems with NNR (**Section 3.5.5** and **Appendix A.1**).

$$F^{(4)} = \begin{pmatrix} 0.000019 & 0.763948 & 0.000006 & 0.000017 & 0.236009 \\ 0.190987 & 0 & 0 & 0 & 0.809013 \\ 0.000006 & 0 & 0 & 0 & 0.999994 \\ 0.000004 & 0 & 0 & 0 & 0.999996 \end{pmatrix} \quad (6.7)$$

5. Least squares optimum for open systems with NNR and SPVA (**Section 3.5.5** and **Appendix A.2**).

$$F^{(5)} = \begin{pmatrix} 0 & 0.763969 & 0 & 0 & 0.236031 \\ 0.190992 & 0 & 0 & 0 & 0.809008 \\ 0 & 0 & 0 & 0 & 1 \\ 0 & 0 & 0 & 0 & 1 \end{pmatrix} \quad (6.8)$$

6. Iterative closure and reciprocity enforcer for open systems (**Section 3.5.5**).

$$F^{(6)} = \begin{pmatrix} 0 & 0.763848 & 0 & 0 & 0.236152 \\ 0.190962 & 0 & 0 & 0 & 0.809038 \\ 0 & 0 & 0 & 0 & 1 \\ 0 & 0 & 0 & 0 & 1 \end{pmatrix} \quad (6.9)$$

From **Table 6.1**, we can clearly see the impact of the different enforcers on the mean absolute error (MAE) of the view factors computed by MCRT. We shall take this metric as an indicator of accuracy.

**Table 6.1:** *Mean absolute error of the view factor matrix obtained using different enforcement methods.*

| Enforcement | Mean Absolute Error |
| --- | --- |
| None | 0.0000239 |
| Closure | 0.0000239 |
| Reciprocity | 0.0000295 |
| Closure + reciprocity (least squares + NNR) | 0.0000073 |
| Closure + reciprocity (least squares + NNR + SPVA) | 0.0000046 |
| Closure + reciprocity (iterative) | 0.0000105 |

It is interesting to note that closure enforcement has no impact on the MAE or on any individual view factor. As explained in **Section 3.5.4**, this is because view factors computed by MCRT already satisfy closure: all rays emitted in the process are accounted for and reach some target surface.

Reciprocity enforcement, in this case, has a slightly negative impact on the MAE. This is expected for an open system, since enforcing reciprocity usually introduces violations of closure. Indeed, the line sums of the first and second rows of $F^{(3)}$ are $0.99964$ and $1.00009$, respectively.

Simultaneous closure and reciprocity enforcement significantly improves the MAE in all cases, with the full least squares optimum method combined with NNR and SPVA yielding the best results. The iterative closure and reciprocity enforcer also reduces the MAE, though less effectively than either least squares method. However, its simplicity and lower computational cost make it more practical for larger models. In this case, convergence is reached after $13$ iterations.

## 6.3 Diffuse Reflectivity

In this section, we demonstrate the application of the proposed methods to the computation of diffuse reflectivity in the simple parallel plates open model shown in **Figure 6.2**. The top plate has a diffuse emissivity of $0.2$, and the bottom plate has a diffuse emissivity of $0.5$.

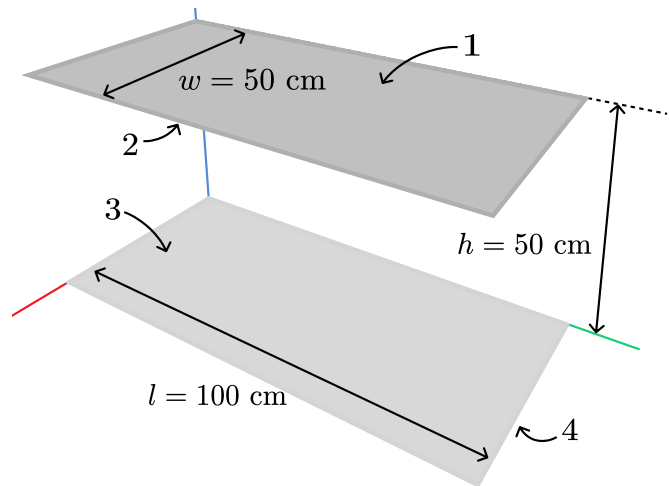

**Figure 6.2:** *Geometrical model used to demonstrate diffuse reflectivity.*

The view factor and REF matrices for this model are, respectively:

$$F = \begin{pmatrix} 0. & 0. & 0. & 0. & 1. \\ 0. & 0. & 0.285913 & 0. & 0.714087 \\ 0. & 0.285927 & 0. & 0. & 0.714073 \\ 0. & 0. & 0. & 0. & 1. \end{pmatrix} \quad (6.10)$$

$$B = \begin{pmatrix} 0 & 0 & 0 & 0 & 1 \\ 0 & 0.020864 & 0.145939 & 0 & 0.833197 \\ 0 & 0.145946 & 0.020864 & 0 & 0.833190 \\ 0 & 0 & 0 & 0 & 1 \end{pmatrix} \qquad (6.11)$$

The view factors were computed using an MCRT routine with $10^7$ rays, and the REFs were computed using Gebhart's matrix method extended to open systems (see **Section 4.2.3** and **Appendix B.2**). No enforcement was applied to neither view factors nor REFs.

From looking at the view factors and the REFs, it is clearly perceived how, when not considering diffuse reflectivity, each plate only sees the other plate and the environment. However, when diffuse reflectivity is taken into account, each plate also partly sees itself, thanks to the radiation that is reflected from the other plate.

But we can go a step further to see the impact of diffuse reflectivity. If we fix the temperature of the bottom plate to $500°C$ and consider the environment temperature to be $27°C$, we can compute the steady state temperature of the top plate in both scenarios: with and without consideration of diffuse reflectivity. The results are shown in **Table 6.2**.

**Table 6.2:** *Temperatures of the top plate with and without diffuse reflectivity.*

| Factors used | Temperature of the top plate ($°C$) |
|---|---|
| View factors | 159.37 |
| REFs | 155.83 |

These results show that the top plate reaches a lower temperature when diffuse reflectivity is included. This aligns with physical intuition: if incident radiation is partly reflected away, the surface absorbs less heat and stabilizes at a lower steady-state temperature.

Already in a simple example such as this one, the effect of diffuse reflectivity can be clearly observed, which is a testament to the importance of considering it in thermal analyses. The methods presented and developed in this work make it possible to include diffuse reflectivity in a computationally efficient manner, without the need for more complex MCRT techniques, even for open systems.

# 7
# Conclusions

The motivation for this work arose from the increasing complexity of satellite missions and the growing demand for advanced thermal control systems, which require highly accurate modeling of radiative heat transfer. Strict regulatory standards, such as the ESA guidelines for thermal control, further highlight the importance of rigorous simulations capable of reliably predicting spacecraft temperature behavior under diverse orbital and environmental conditions.

In this chapter, the key contributions and findings of the thesis are summarized, their implications are discussed, the original objectives are revisited, and the main conclusions are presented.

## 7.1 Key contributions and findings

After establishing the theoretical foundations of radiative heat transfer, this thesis developed a comprehensive framework for modeling diffuse reflectivity in space thermal analysis. The framework encompasses methods for enforcing reciprocity and closure, efficient computation of REFs to account for diffuse reflection, and a systematic approach for handling *multi-node surface* representations of thermal systems.

The primary novel contributions of this work—representing developments not previously available in the literature—are as follows:

1. Development and demonstration of a simultaneous closure and reciprocity enforcer for open systems based on the least squares optimum (**Section 3.5.5**).
2. Introduction of *small positive value avoidance* (SPVA), a method designed to eliminate spurious small positive values arising in the least squares optimum enforcer (**Appendix A.2**).

3. Development and demonstration of an iterative approach for enforcing closure and reciprocity in open systems (**Section 3.5.5**).
4. Design and validation of a framework for handling multi-node surface representations of thermal systems, addressing both efficiency and consistency issues (**Chapter 5**).
5. Extension and formalization of the Gebhart formulation and Gebhart's matrix method to open systems (**Appendix B.2**).

These contributions were demonstrated through two applied examples in **Chapter 6**. The first example showed that simultaneous enforcement significantly reduced the mean absolute error (MAE) of the view factor matrix. Among the tested methods, the least squares optimum enforcer with non-negativity rectification (NNR) and small positive value avoidance (SPVA) achieved the highest accuracy, with a MAE of $4.6 \cdot 10^{-6}$. The iterative method, while slightly less accurate (MAE of $1.05 \cdot 10^{-5}$), offered superior computational efficiency, making it more suitable for large-scale models. An additional observation was that view factors computed using MCRT typically satisfied closure initially; however, reciprocity enforcement could disrupt it, underscoring the need for combined enforcement.

The second example demonstrated the impact of diffuse reflectivity on thermal predictions. In a parallel-plates model, including REFs resulted in a lower steady-state temperature for the top plate (155.83°C) compared to calculations without diffuse reflectivity (159.37°C). This difference, attributable to the reduction in absorbed heat due to diffuse reflection, is physically intuitive and emphasizes the importance of incorporating diffuse reflectivity for accurate thermal analysis.

Beyond the examples in **Chapter 6**, the developed methods were also implemented in practice, leading to significant enhancements in the thermal analysis capabilities of *Radian*, a cloud-based platform for space thermal modeling. Although detailed figures cannot be disclosed due to confidentiality, substantial improvements in both accuracy and computational efficiency were demonstrated across multiple internal test cases and benchmarks against established tools such as *ESATAN-TMS*.

## 7.2 Alignment with objectives

This thesis successfully met all its stated objectives:

1. To enhance the modeling of surface interactions in *Radian* by incorporating diffuse reflectivity into the radiative heat transfer calculations: This has been achieved through the implementation the Gebhart formulation and matrix method for diffuse reflectivity and its extension to open systems.

2. To ensure the physical consistency of view factors and REFs by implementing methods that enforce closure and reciprocity conditions in radiative exchange modeling: This has been accomplished by developing and implementing two novel simultaneous enforcement methods for open systems.
3. To develop and implement original methods for simultaneously enforcing closure and reciprocity for view factors in open systems: This has been fulfilled by the least squares optimum method for open systems (**Section 3.5.5**) with NNR (**Appendix A.1**) and SPVA (**Appendix A.2**) and by the iterative enforcer (**Section 3.5.5**).
4. To formally extend Gebhart's formulation and matrix method for computing REFs to open systems: This has been successfully developed and detailed in **Appendix B.2**.
5. To develop a framework for efficiently handling and switching between different levels of subdivision in a multi-node surface model: This has been successfully achieved through the derivation of global relations and the concept of view fractions described in **Chapter 5**.
6. To ensure the preservation of closure and reciprocity relations when applying these transformations across different levels of the multi-node model: This has been analytically proven in **Section 5.3**.
7. To develop applied examples that demonstrate the practical relevance and effectiveness of the proposed methods: This has been accomplished through the examples in **Chapter 6** and through implementation of internal test cases and benchmarks in *Radian*.

## 7.3 Conclusion

In conclusion, the innovations presented in this thesis represent a substantial enhancement of the radiative heat transfer modeling capabilities in *Radian*. By addressing critical aspects such as diffuse reflectivity and view factor consistency in open systems, this work directly contributes to increasing the accuracy and reliability of thermal simulations for satellite applications, thereby supporting compliance with evolving regulatory frameworks and enabling robust mission performance in the increasingly complex small satellite sector.

*This page intentionally left blank.*

# Appendices

*This page intentionally left blank.*

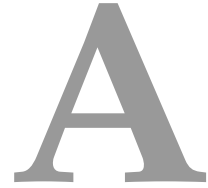

# A
# Improvements to the least squares optimum enforcer

## A.1 Ensuring non-negativity

In this section, we outline the *non-negativity rectification* (NNR) algorithm for the least squares optimum method. This rectification adapted from [48]. The idea behind it is to consider negative view factors to be equal to zero, removing them from the view factor matrix and re-applying the least squares enforcement with the reduced order system. This procedure is repeated until no negative view factors remain.

The steps of the process are:

1. Enforce closure and reciprocity using the chosen method (that may not ensure non-negativity).
2. If there are no negative view factors, the enforcement is complete and complies with non-negativity. Otherwise, set all negative view factors to zero and remove them from consideration. For the least squares closure and reciprocity enforcer, this includes removing the corresponding restrictions from the matrices $R_C$ and $R_R$, and removing the corresponding elements from vectors $c_C$ and $c_R$ (see **Section 3.5.5**).
3. Return to step 1 with the reduced set of factors.

This procedure has proven to give the exact same results as other more complex and computationally expensive methods, such as nonlinear programming approaches, in 90% of the cases, and very similar ones in the other 10% of the cases [48].

## A.2 Avoiding small positive values

It has been seen from experiment (see **Section 6.2**) that the least squares optimum enforcer, combined with non-negativity, tends to make originally zero view factors turn slightly positive. In this section, we present *small positive value avoidance* (SPVA), an extension to the method that can be used alongside non-negativity rectification[1] and prevents this issue. This extension has been developed fully in this work and cannot be found in the existing literature at this point.

The extension of the least squares optimum enforcer involves two steps, to be done at the beginning of each iteration[2]. Once we have constructed the restriction matrix $A$, the initial MCRT guess of the exchange coefficients $x$, and the vector of independent terms $b$, we proceed to the following steps:

1. Remove from $x$ all the zero exchange factors.
2. Remove all the corresponding columns from $A$.
3. Remove all the rows from $A$ that have become all zeros as a consequence of the previous step.
4. Remove all the corresponding elements from $b$.
5. We have now obtained a new system $A'x' = b'$ of reduced order, which concerns only the exchange factors that were originally non-zero. We can now apply the least squares optimum enforcer to this reduced system, finally replacing in $x$ the subset of the original exchange factors that have been corrected.

---

[1] Although it can also be used separately from non-negativity rectification.
[2] If applying the non-negativity rectification, these steps should be performed before.

# B

# Gebhart's formulation and matrix method

## B.1 Closed systems

In this section, we present Gebhart's original matrix method for computing REFs from view factors using **Equation 4.5**, which is valid for closed systems. It consists of writing the mentioned equation in matrix form. This can be done as follows:

$$\begin{aligned}
\varepsilon_j F_{ij} + \sum_{k=1}^{N} \rho_k F_{ik} B_{kj} &= B_{ij} \\
\Rightarrow \varepsilon_j F_{ij} &= B_{ij} - \sum_{k=1}^{N} \rho_k F_{ik} B_{kj} \\
\Rightarrow \varepsilon_j F_{ij} &= \sum_{k=1}^{N} \delta_{ik} B_{kj} - \sum_{k=1}^{N} \rho_k F_{ik} B_{kj} \\
\Rightarrow \varepsilon_j F_{ij} &= \sum_{k=1}^{N} \left( \delta_{ik} - \rho_k F_{ik} \right) B_{kj} \\
\Rightarrow E &= K \cdot B
\end{aligned} \tag{B.1}$$

where:

$$E = \begin{bmatrix} \varepsilon_1 F_{11} & \varepsilon_2 F_{12} & \cdots & \varepsilon_N F_{1N} \\ \varepsilon_1 F_{21} & \varepsilon_2 F_{22} & \cdots & \varepsilon_N F_{2N} \\ \vdots & \vdots & \ddots & \vdots \\ \varepsilon_1 F_{N1} & \varepsilon_2 F_{N2} & \cdots & \varepsilon_N F_{NN} \end{bmatrix} \tag{B.2}$$

$$K = \begin{bmatrix} 1 - \rho_1 F_{11} & -\rho_2 F_{12} & \cdots & -\rho_N F_{1N} \\ -\rho_1 F_{21} & 1 - \rho_2 F_{22} & \cdots & -\rho_N F_{2N} \\ \vdots & \vdots & \ddots & \vdots \\ -\rho_1 F_{N1} & -\rho_2 F_{2N} & \cdots & 1 - \rho_N F_{NN} \end{bmatrix} \quad (B.3)$$

$$B = \begin{bmatrix} B_{11} & B_{12} & \cdots & B_{1N} \\ B_{21} & B_{22} & \cdots & B_{2N} \\ \vdots & \vdots & \ddots & \vdots \\ B_{N1} & B_{N2} & \cdots & B_{NN} \end{bmatrix} \quad (B.4)$$

Then, the matrix $B$ of REFs can be computed by solving the system $K \cdot B = E$, for example by inverting $K$:

$$B = K^{-1} E \quad (B.5)$$

## B.2 Open systems

The original Gebhart formulation method presented above is valid only for closed systems, although the extension to open systems is trivial once the following formula is considered for the REFs to outer space:

$$B_{i\infty} = F_{i\infty} + \sum_{k=1}^{N} F_{ik} \rho_k B_{k\infty} \quad (B.6)$$

The first term in the RHS of this equation is the fraction of the energy that leaves surface $i$ and is directly lost to space. The second term accounts for the fraction of energy that leaves surface $i$ which is diffusely reflected by surface $k$ and which is finally lost to space. The surfaces $k$ represent all possible intermediary surfaces.

Although simple, this formula has not been found by the author of this thesis in the literature. It is suspected by the author that the current practice when computing the REFs to outer space in an open system is simply to assume that closure is satisfied, and then to obtain $B_{i\infty} = 1 - \sum_{j=1}^{N} B_{ij}$. However, **Equation B.6** allows us to compute the REFs to outer space in an open system even when the closure assumption is not satisfied.

In matrix form, we obtain a similar expression to the one for closed systems:

$$F_{i\infty} = \sum_{k=1}^{N} \left( \delta_{ik} - \rho_k F_{ik} \right) B_{k\infty} \Rightarrow E_\infty = K \cdot B_\infty \quad (B.7)$$

where now $E_\infty$ is the column vector of view factors to outer space, $K$ is the same matrix as in the closed system case, and $B_\infty$ is the column vector of REFs to outer space.

This result can be combined with the closed system case to obtain both the REFs between surfaces and the REFs to outer space in a single system of equations $E = K \cdot B$, where:

$$E = \begin{bmatrix} \varepsilon_1 F_{11} & \varepsilon_2 F_{12} & \cdots & \varepsilon_N F_{1N} & F_{1\infty} \\ \varepsilon_1 F_{21} & \varepsilon_2 F_{22} & \cdots & \varepsilon_N F_{2N} & F_{2\infty} \\ \vdots & \vdots & \ddots & \vdots & \vdots \\ \varepsilon_1 F_{N1} & \varepsilon_2 F_{N2} & \cdots & \varepsilon_N F_{NN} & F_{N\infty} \end{bmatrix} \quad (B.8)$$

$$K = \begin{bmatrix} 1 - \rho_1 F_{11} & -\rho_2 F_{12} & \cdots & -\rho_N F_{1N} \\ -\rho_1 F_{21} & 1 - \rho_2 F_{22} & \cdots & -\rho_N F_{2N} \\ \vdots & \vdots & \ddots & \vdots \\ -\rho_1 F_{N1} & -\rho_2 F_{2N} & \cdots & 1 - \rho_N F_{NN} \end{bmatrix} \quad (B.9)$$

$$B = \begin{bmatrix} B_{11} & B_{12} & \cdots & B_{1N} & B_{1\infty} \\ B_{21} & B_{22} & \cdots & B_{2N} & B_{2\infty} \\ \vdots & \vdots & \ddots & \vdots & \vdots \\ B_{N1} & B_{N2} & \cdots & B_{NN} & B_{N\infty} \end{bmatrix} \quad (B.10)$$

Then, the matrix $B$ of REFs can be computed by solving the system $K \cdot B = E$, for example by inverting $K$:

$$B = K^{-1} E \quad (B.11)$$

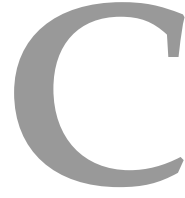

# Optical properties of nodes

When computing radiative couplings between nodes $i$ and $j$, we find the following expressions:

$$GR_{ij} = \varepsilon_i A_i B_{ij} \tag{C.1}$$

However, since a node $i$ may belong to a sub-surface with two faces having different optical properties, the question is which emissivity should be used in this formula. The answer is found when we break this GR into the contributions from each face $A$ and $B$ of the source and target nodes. Since the transmission between each combination of those faces happens in parallel, the corresponding GRs can be added up to obtain the final GR:

$$GR_{ij} = GR_{ij}^{A \to B} + GR_{ij}^{B \to A} + GR_{ij}^{A \to A} + GR_{ij}^{B \to B} \tag{C.2}$$

Then, substituting $GR_{ij}^{X \to Y} = \varepsilon_i^X S_i B_{n_i m_j}^{X \to Y}$, where $S_i = A_i / C_i$ is the area of each of the $C_i$ faces of node $i$, we obtain:

$$\begin{aligned}
GR_{ij} &= \varepsilon_i C_i S_i B_{ij} \\
&= \varepsilon_i^A S_i B_{n_i m_j}^{A \to B} + \varepsilon_i^B S_i B_{n_i m_j}^{B \to A} + \varepsilon_i^A S_i B_{n_i m_j}^{A \to A} + \varepsilon_i^B S_i B_{n_i m_j}^{B \to B} \\
&= \varepsilon_i^A S_i \left( B_{n_i m_j}^{A \to B} + B_{n_i m_j}^{A \to A} \right) + \varepsilon_i^B S_i \left( B_{n_i m_j}^{B \to A} + B_{n_i m_j}^{B \to B} \right)
\end{aligned} \tag{C.3}$$

Then, we have:

$$\begin{aligned}
\varepsilon_i &= \varepsilon_i^A \cdot \frac{B_{n_i m_j}^{A \to A} + B_{n_i m_j}^{A \to B}}{C_i B_{ij}} + \varepsilon_i^B \cdot \frac{B_{n_i m_j}^{B \to A} + B_{n_i m_j}^{B \to B}}{C_i B_{ij}} \\
&= \varepsilon_i^A \cdot \left( f_{n_i m_j}^{A \to A} + f_{n_i m_j}^{A \to B} \right) + \varepsilon_i^B \cdot \left( f_{n_i m_j}^{B \to A} + f_{n_i m_j}^{B \to B} \right)
\end{aligned} \tag{C.4}$$

The $f_{n_i m_j}^{X \to Y}$ are the view fractions derived in **Section 5.2.5**.